\documentclass[a4paper,12pt,%
  DIV13,%
  smallheadings,%
  headsepline,%
  headinclude,%
  footexclude,%
  bibtotoc,%
  BCOR14mm,%
  pointlessnumbers%
]{scrbook}

\usepackage{graphicx,color}

\usepackage[pagebackref]{hyperref}
\hypersetup{%
  bookmarksopen=true, %
  bookmarksnumbered=true, %
  pdftitle={Renormalization, Conservation Laws and Transport in
    Correlated Electron Systems},
  pdfsubject={Doctoral dissertation},
  pdfauthor={Tilman Enss},
  pdfkeywords={correlated electrons}
}

\usepackage{feynmp}
\ifpdfoutput{\DeclareGraphicsRule{*}{mps}{*}{}}{} %
\let\comma=, %

\usepackage[fleqn]{amsmath}
\allowdisplaybreaks %
\ifx\undefined\gtrsim %
  \usepackage{amssymb} %
\fi

\providecommand{\abb}[1]{{\texorpdfstring{\textsc{\lowercase{#1}}}{#1}}}
\def\RG{\abb{RG}}
\def\fRG{f\kern1pt\abb{RG}}
\def\WI{\abb{WI}}
\def\mWI{m\WI}
\def\onePI{\abb{1PI}}
\def\DMRG{\abb{DMRG}}
\def\LL{\abb{TLL}}
\def\QMC{\abb{QMC}}
\def\CST{\abb{CST}}
\def\UST{\abb{UST}}
\def\oneD{\abb{1D}}
\def\twoD{\abb{2D}}
\def\LHS{\abb{LHS}}
\def\RHS{\abb{RHS}}
\def\ie{i.\kern.5pt e.}
\def\zB{z.\kern1pt B.}
\def\PhD{Ph.\kern1pt D.}
\newcommand{\lat}[1]{\textit{#1}} %

\DeclareMathOperator{\Tr}{Tr}
\DeclareMathOperator{\tr}{tr}
\DeclareMathOperator{\sgn}{sgn}
\providecommand{\vev}[1]{\langle#1\rangle}
\providecommand{\inner}[2]{\left(#1,#2\right)}
\providecommand{\abs}[1]{\lvert#1\rvert}
\newcommand{\ket}[1]{\lvert#1\rangle}
\newcommand{\bra}[1]{\langle#1\rvert}
\newcommand{\id}{\boldsymbol 1}
\renewcommand{\Re}{\mathrm{Re}}
\renewcommand{\Im}{\mathrm{Im}}
\newcommand{\at}{\Big\vert}
\newcommand{\dmux}[3]{d\mu_{#1}[#2,#3]}
\newcommand{\dmu}[2]{\dmux{#1}{#2}{\bar #2}}
\renewcommand{\vec}[1]{\boldsymbol{#1}}
\newcommand{\Laplace}[1]{\Delta_{#1}}
\newcommand{\ddd}[1]{\frac{\delta}{\delta #1}}
\newcommand{\dl}{\partial_\Lambda}
\newcommand{\ds}{\partial_s}
\newcommand{\dT}{\frac{d}{dT}}
\newcommand{\F}{\mathcal{F}}
\newcommand{\G}{\mathcal{G}}
\newcommand{\Ord}{\mathcal{O}}
\newcommand{\T}{\mathcal{T}}
\newcommand{\V}{\mathcal{V}}
\newcommand{\W}{\mathcal{W}}
\newcommand{\Z}{\mathcal{Z}}
\newcommand{\imp}{\text{imp}}
\newcommand{\ret}{\text{ret}}
\newcommand{\leads}{\text{leads}}
\newcommand{\tint}{\text{int}}
\newcommand{\ab}{{\alpha_B}}
\newcommand{\kf}{\ensuremath{k_{\text F}}}
\newcommand{\vf}{\ensuremath{v_{\text F}}}
\newcommand{\etaf}{\ensuremath{\eta_{\text F}}}
\newcommand{\tdot}{\text{dot}}
\newcommand{\p}{\text{p}}

\newcommand{\fig}[4][clip]{\begin{figure}[ht!]%
    \centerline{\includegraphics*[#1]{#2}}
    \caption{\small\label{#3}#4}\end{figure}}

\providecommand{\SetFigFont}[5]{} %

\newcommand{\setsize}[1]{\linespread{#1}\normalsize}

\begin{document}

\setkomafont{caption}{\itshape}
\setkomafont{pagehead}{\itshape}

\pagestyle{empty}

\mainmatter

\begin{flushleft}
  \large\sffamily
  {\LARGE\bfseries
    Renormalization, Conservation Laws and \\[1ex]
    Transport in Correlated Electron Systems} \\[5ex]
  {\Large\bfseries Tilman Enss} \\[4ex]
  Dissertation accepted by the University of Stuttgart%
  \footnote{\raggedright\sffamily Version without the German summary; full
    version available online at \\
    \href{http://elib.uni-stuttgart.de/opus/volltexte/2005/2258/}
    {\ttfamily http://elib.uni-stuttgart.de/opus/volltexte/2005/2258/}} \\
  for the degree of Doctor of Natural Sciences \\[2ex]
  Examiner: Prof.\ Dr.\ Walter Metzner \\
  Co-examiner: Prof.\ Dr.\ Siegfried Dietrich \\[2ex]
  Max-Planck-Institut f\"ur Festk\"orperforschung \\
  Stuttgart, Germany 2005
\end{flushleft}
\vfill
{\Large\sffamily\bfseries Abstract} \\[.7ex]


\noindent This thesis comprises two parts centered around the functional
renormalization-group framework: in the first part, I study the role
of symmetries and conservation laws in approximate solutions, while in
the second part I analyze Friedel oscillations and transport in
Luttinger liquids with impurities.

The functional renormalization group (\fRG) has been developed as a
new computational tool in the theory of interacting Fermi systems.
The effective behavior of a given microscopic model is calculated by
solving coupled differential flow equations for the Green functions
with an energy scale as the flow parameter.  The symmetries of the
microscopic model imply Ward identities between Green and response
functions.  It is shown that solutions of truncated flow-equation
hierarchies satisfy Ward identities if the cutoff bare action is gauge
invariant.  However, truncations are generally not self-consistent
approximations in the sense of Baym and Kadanoff.

The \fRG\ is then applied to study Luttinger liquids.  By computing
the full spatial effective potential of a single impurity, long-range
Friedel oscillations are observed in the density profile with the
expected power laws for systems with up to $10^7$ lattice sites.  For
a double barrier enclosing a dot region we find temperature regimes in
which the conductance follows power laws with universal exponents, as
well as non-universal crossover regimes in intermediate parameter
regions.


\clearpage

\begin{center}
  \vspace*{35mm}
  \large F\"ur Carmen
\end{center}
\cleardoubleplainpage

\setsize{1.1}
\pagestyle{plain}
\tableofcontents
\cleardoubleplainpage

\pagestyle{headings}


\chapter{Introduction}
\label{sec:intro}

In one-dimensional metals electrons can move freely in one direction
but are confined in the two transverse directions.  The interaction
between the electrons leads to \emph{Luttinger-liquid\kern1pt}
behavior with unusual properties different from conventional
(Fermi-liquid) metals.  In particular, the low-energy behavior of
Luttinger liquids is strongly affected by impurities.  Already a
single static impurity has a dramatic effect: for a repulsive
interaction, the backscattering amplitude grows as the energy scale is
lowered, until at $T=0$ transport is inhibited and the chain is
effectively cut into two pieces.  The local density of states near an
impurity, as well as the spatial density profile away from the
impurity, obey characteristic power laws depending only on the bulk
parameters.  The conductance through a single impurity with varying
parameters can be collapsed onto a single curve by a one-parameter
scaling ansatz.  A double barrier shows particularly rich behavior: it
can be tuned to resonance, and additional scales are introduced by the
separation of the two barriers and the detuning from the resonance.
The conductance as a function of temperature is non-monotonous,
exhibiting several distinct power laws, as well as a complex
non-universal crossover behavior for intermediate parameter ranges.

In recent years experiments on carbon nanotubes have allowed to
measure the effect of one or two impurities in an otherwise perfectly
clean one-dimensional metal.  While several field-theoretical
predictions were confirmed, transport through a double barrier did not
obey the expected asymptotic power laws.  This led to a renewed
theoretical interest to understand the behavior in intermediate
parameter ranges accessible in experiments.  Different analytical and
computational methods applied to a spinless double-barrier model
either supported or disagreed with the experimental data.  This
prompted us to investigate the problem with the functional
renormalization-group method, which we have already used to treat
complex multi-scale problems, such as Luttinger liquids with a single
impurity.

The functional renormalization group (\fRG) has been developed in
recent years as a new computational tool to study interacting Fermi
systems.  It is particularly efficient in low dimensions.  Starting
from a specific microscopic model, high-energy modes are successively
integrated out to obtain the effective behavior on all energy scales.
The method captures universal scaling laws in certain limits, as well
as non-universal crossover phenomena at intermediate scales.

Formally, the \fRG\ flow equations constitute an infinite hierarchy of
coupled differential equations which describe the change of all Green
functions as the energy scale is lowered.  This hierarchy of flow
equations produces the exact solution to all orders of perturbation
theory at the end of the flow.  In practice, however, the full
hierarchy has to be truncated by neglecting the flow of some higher
Green functions, which is justified perturbatively for weak
renormalized interactions.  In contrast to other renormalization-group
methods we not only follow the flow of a few running couplings but of
whole functions, such as the impurity potential.

For the \oneD\ problems with one or two impurities, we approximate the
interaction by an effective nearest-neighbor coupling but retain the
full effective impurity potential.  Our method is thus perturbative in
the renormalized interaction but non-perturbative in the impurity
strength.  This already yields the expected universal scaling of the
local density of states.  In order to obtain the spatial density
profile it is necessary to follow the flow of the density-response
vertex.  We treat the simpler case of spinless fermions; the more
realistic modeling of electrons by spinful fermions is currently being
considered.  Our \fRG\ results have been checked against numerically
exact density-matrix renormalization group (\DMRG) data for systems
with up to 1000 sites.  The computation of the conductance as a
function of temperature requires several extensions of the method.  We
develop the flow of the full impurity potential at fixed,
\emph{finite} temperature, as well as the temperature-flow scheme with
self-energy feedback.  Moreover, it is shown that on the level of our
approximation, no corrections to the current vertex appear in the Kubo
formula for the conductance.  We are thus able to compute the
conductance consistently within one approximation over several orders
of magnitude in temperature, for arbitrary impurity strength.

Besides the formal developments, the practical feasibility of the
method depends crucially on the required computation time.  Using a
little-known mathematical theorem, an algorithm has been developed
which scales linearly in the system size, instead of quadratically.
For a lattice of 60.000 sites, the zero-temperature flow now takes
minutes instead of days, and systems of up to $10^7$ sites have been
computed.  This allows to find interesting regions in a large
parameter space much more quickly.

The vanishing of current-vertex corrections to the conductance is an
example of a more general topic: the role of \emph{symmetries} and
\emph{conservation laws} in the \fRG\ formalism.  The microscopic
model considered above, for instance, has a local $U(1)$ gauge
symmetry which implies charge conservation.  As a consequence, the
exact Green functions are related by Ward identities.  In particular
for transport calculations it is crucial to respect these identities
exactly even in approximate calculations.  This raises the question
whether typical approximations in the \fRG, especially the truncation
of the infinite hierarchy of flow equations, satisfy---or can be made
to satisfy---Ward identities.  This problem has been addressed ten
years ago in the context of gauge theories with a fluctuating gauge
field, using either modified Ward identities or the background-field
method, but simple gauge-invariant truncations remained elusive.  In
those cases where a gauge-invariant flow is possible, for instance if
the model is regularized not by a momentum cutoff but by a finite
temperature, we show that even truncated flows satisfy the Ward
identities.  On the other hand, we find that ``self-consistency''
between Green functions of different degree, a feature of the
conserving approximations by Baym and Kadanoff, is generally not
satisfied by common truncations of the \fRG\ flow equations.
\\

This thesis is organized as follows:
\begin{itemize}
\item In Chapter~\ref{sec:rg} the \fRG\ formalism is introduced.
  After a brief review of generating functionals and their expansion
  in terms of Green or vertex functions, an infrared cutoff is defined
  which introduces a scale dependence in the generating functionals.
  A derivative with respect to this scale leads to functional flow
  equations.  These are then expanded in terms of their constituent
  Green functions to obtain an infinite hierarchy of coupled
  differential flow equations for the Green functions and a
  diagrammatic representation of the flow equations.  The merits of
  different schemes are compared.
\item In Chapter~\ref{sec:ward} Ward identities are derived expressing
  the symmetry of the bare action in the functional formalism.  A
  momentum cutoff generally modifies the Ward identities.  For other
  flow schemes which preserve Ward identities we show that they hold
  even in truncated flows.  Conserving approximations are reviewed as
  an example of self-consistent approximations.  It is then shown that
  common truncated \fRG\ flow equations are generally not
  self-consistent.
\item In Chapter~\ref{sec:lutt} the general \fRG\ formalism is applied
  to study one-dimensional correlated fermion systems (Luttinger
  liquids) with impurities, in particular their single-particle and
  transport properties.  In this technical part the precise form of
  the flow equations on the lattice is derived, as well as the details
  of the finite-temperature cutoff procedure, truncations of the
  flow-equation hierarchy and parametrizations of the flowing
  vertices.  At the end of the flow we obtain the effective impurity
  potential (self energy) and the renormalized density profile.  In
  order to compute transport in our approximation, we then have to
  solve the scattering problem of non-interacting electrons in this
  effective potential.  It is also shown that current-vertex
  corrections to the conductance vanish in our approximation, in
  accordance with the Ward identities.  The loop algorithm for
  nearest-neighbor interaction, which scales linearly in the system
  size, is derived in appendix~\ref{sec:app:loop}.
\item In Chapter~\ref{sec:results} new results are reported for the
  Friedel oscillations of the spatial density profile generated by a
  boundary or impurity in one dimension, and the temperature-dependent
  conductance through a double barrier.  For appropriate parameter
  ranges universal scaling is observed with several distinct power
  laws in temperature.  In intermediate regions the full non-universal
  crossover behavior is obtained, suggesting an interpretation of
  recent measurements on carbon nanotubes.
\end{itemize}

The publications based on this thesis are listed on page
\pageref{sec:publ}.


\begin{fmffile}{diag} %
\fmfcmd{%
  vardef slashbaro (expr p, len, ang) =
    ((-len/2,0)--(len/2,0))
    rotated (ang + angle direction length(p)/2 of p)
    shifted point length(p)*0.3 of p
  enddef;
  vardef slashbarm (expr p, len, ang) =
    ((-len/2,0)--(len/2,0))
    rotated (ang + angle direction length(p)/2 of p)
    shifted point length(p)*0.5 of p
  enddef;
  vardef slashbar (expr p, len, ang) =
    ((-len/2,0)--(len/2,0))
    rotated (ang + angle direction length(p)/2 of p)
    shifted point length(p)*0.7 of p
  enddef;
  vardef blob (expr z_arg, diameter) =
    save p,currentpen; path p; pen currentpen;
    pickup pencircle scaled thick;
    p = fullcircle scaled diameter shifted z_arg;
    cfill p;
  enddef;
  style_def slplaino expr p =
    cdraw p;
    ccutdraw slashbaro (p, 5mm, 45)
  enddef;
  style_def slplainm expr p =
    cdraw p;
    ccutdraw slashbarm (p, 5mm, 45)
  enddef;
  style_def slplain expr p =
    cdraw p;
    ccutdraw slashbar (p, 5mm, 45)
  enddef;
  style_def sldasheso expr p =
    draw_dashes p;
    ccutdraw slashbaro (p, 5mm, 45)
  enddef;
  style_def sldashesm expr p =
    draw_dashes p;
    ccutdraw slashbarm (p, 5mm, 45)
  enddef;
  style_def dashesp expr p =
    draw_dashes p;
    blob (point length(p)*0.5 of p, 1.6mm)
  enddef;
}


\chapter{Functional renormalization group}
\label{sec:rg}

Challenging many-body problems often involve effects on many energy
scales.  In perturbation theory one has to perform loop integrals over
all energy scales, which may lead to infrared or ultraviolet
divergences.  Some of these divergences have a physical origin
indicating for instance a phase transition, while others are an
artefact of perturbation theory.  Wilson's exact renormalization group
(\RG) \cite{Wil71,WK74} provides a method to deal with such problems:
the different energy scales are successively taken into account by
integrating out momentum shells.  This can be done by introducing for
example an infrared cutoff in the bare propagator which suppresses all
modes with an energy below the cutoff scale $\Lambda$.  Then, all
correlation functions depend on the scale $\Lambda$.  One follows the
change (\kern1pt\emph{flow}) of the correlation functions as the cutoff
scale is lowered until finally the cutoff is removed and the original
theory is recovered.  An important advantage of this procedure is that
the right-hand side (\RHS) of the flow equation remains regular even
if perturbation theory leads to unphysical divergences.

There are several variants of the exact functional \RG\ flow
equations.  After Wilson's early review \cite{WK74} on the exact \RG,
\cite{Pol84} derived equivalent continuum flow equations with a smooth
cutoff in order to prove perturbative renormalizability of massive
Euclidean $\varphi^4$ theory in $D=4$.  Keller, Kopper, and Salmhofer
\cite{KKS92} simplified and extended the proof and showed that these
flow equations determine the connected amputated Green functions (cf.\ 
section \ref{sec:rg:flow:v}).  \cite{WH73} derived flow equations for
a sharp cutoff, however there were ambiguities which they avoided by
assuming discrete momenta.  \cite{Wei76} took the continuum limit and
found that the flow could be formulated without ambiguity by expanding
the connected amputated Green functions in trees, \ie, in terms of
one-particle irreducible (\onePI) vertex functions.  The \onePI\ flow
equation of the Legendre effective action, the generating functional
of the \onePI\ vertex functions, was derived by
\cite{NC77,Wet93,BDM93,Mor94,SH01,Kopietz:0103633}.  Another scheme,
obtained from the Polchinski scheme by Wick-ordering
\cite{Wie88,Salmhofer:9706188,Salmhofer:1999}, is particularly suited
for rigorous proofs because it allows strong bounds on the growth of
correlation functions, even near the Fermi surface and to all orders
in the renormalized interaction.

By then, several people had started to use the functional \RG\ (\fRG)
for fermionic lattice models.  This introduces several problems not
present in $\varphi^4$ theory, for example the determination of the
Fermi surface, whose shape and position is not known \lat{a priori}.
Another problem is that even at low energies, when only momenta close
to the Fermi surface are important, the two-electron interaction is a
complicated function of momenta.  Therefore, it needs to be
parametrized by many discrete couplings, in contrast to the single
renormalized coupling $\lambda$ at zero external momenta for the
$\varphi^4$ theory.  On the other hand, the lattice provides a natural
ultraviolet cutoff which leads to significant simplifications as
compared to the continuum field-theoretical models.  Important
applications of the \fRG\ in condensed-matter physics include the
\twoD\ Hubbard model using the Polchinski scheme
\cite{Zanchi:9703189,Zanchi:9812303}, the Wick-ordered scheme
\cite{HM00} and also the \onePI\ scheme \cite{HSFR01}, while \oneD\ 
impurity problems and Luttinger-liquid physics are conveniently
investigated in the \onePI\ scheme \cite{MMSS02a,MMSS02b,AEMMSS04}.

Initially, many of the flow-equation schemes have not been derived in
the most straightforward way, and I found it worthwhile to derive them
again in a simple and uniform notation, highlighting the relation
between different schemes.  While the results are not new, some
derivations are much easier than those found in the literature, and I
hope the reader new to this method will find them helpful.  After
introducing the functional formalism and several types of correlation
functions in section \ref{sec:rg:form}, the most commonly used flow
equations are derived in section \ref{sec:rg:flow}.  I compare
important features of the different schemes in section
\ref{sec:rg:sum}.


\section{Functional formalism}
\label{sec:rg:form}


\subsection{Bare action}
\label{sec:rg:form:bare}

A system of interacting spinless\footnote{Ultimately we aim to
  describe electrons which are fermions with spin; presently in the
  applications in Chapter~\ref{sec:lutt}, however, we only consider
  fermions without spin and, therefore, specialize to this case.}
fermions is described by the action
\begin{align}
  \label{eq:action}
  S[\psi,\bar\psi] = \inner{\bar\psi}{Q\psi} - V_0[\psi,\bar\psi]
\end{align}
with the kinetic (quadratic) term defined as a scalar product
\begin{align}
  \label{eq:kintermreal}
  \inner{\bar\psi}{Q\psi}
  & = \int dx\, dy\, \bar\psi(y) Q(y,x) \psi(x) \\
  \label{eq:kintermmom}
  & = \sum_K (i\omega-\xi_{\vec{k}}) \bar\psi_{K} \psi_K
  \quad \text{(for translational invariance)}
\end{align}
where for a translationally invariant system \eqref{eq:kintermmom},
the multi-index $K=(\omega,\vec{k})$ contains space-time indices
(frequency, momentum) and could be extended by internal degrees of
freedom like the spin projection $\sigma$.  The $\psi_K$, $\bar\psi_K$
are Grassmann variables, and $Q(K) = i\omega-\xi_{\vec{k}}$ is the
inverse bare propagator with $\xi_{\vec{k}} = \varepsilon_{\vec{k}} -
\mu$ the dispersion around the chemical potential $\mu$.  If, however,
one includes an impurity potential in the bare propagator that depends
specifically on space and not just differences of positions,
translational invariance is broken: then one has to use the general
expression \eqref{eq:kintermreal}, cf.\ section \ref{sec:lutt:model}.
The bare propagator is the inverse operator of $Q$,
\begin{align}
  \label{eq:bareprop}
  C := Q^{-1}.
\end{align}
The functional $V_0[\psi,\bar\psi]$ is the bare many-body interaction,
for example in the typical case of a density-density interaction (with
frequency conservation implicit):
\begin{align}
  \label{eq:densdensint}
  V_0[\psi,\bar\psi]
  & = \int dx\, dy\, V_0(x-y)\; n(x)\, n(y) \\
  & = \sum_{k_1,k_2,q} \tilde V_0(q)\; \bar\psi_{k_1} \bar\psi_{k_2+q}
  \psi_{k_2} \psi_{k_1+q}. \notag
\end{align}


\subsection{Partition function and connected Green functions}
\label{sec:rg:form:z}

All information about the physical system with action
\eqref{eq:action} is encoded in the normalized partition function
\begin{align}
  \Z[\eta,\bar\eta]
  := \frac{1}{Z_0} \int [d\psi\bar\psi]\, e^{S[\psi,\bar\psi]}\,
  e^{-\inner{\bar\psi}{\eta}-\inner{\bar\eta}{\psi}} \; ,
\end{align}
a functional integral with weight $e^S$ and coupled to Grassmann
source fields $\eta_K$, $\bar\eta_K$.  The integration measure is
abbreviated as $[d\psi\bar\psi] := \prod_K d\psi_K \, d\bar\psi_K$.
The partition function $\Z[\eta,\bar\eta]$ is the generating
functional for the Green functions (connected and disconnected).  The
normalization constant $Z_0$ is the non-interacting partition
function,
\begin{align}
  \label{eq:Z0}
  Z_0 := \int [d\psi\bar\psi]\, e^{\sum_K \bar\psi_K Q_K \psi_K}
  = \prod_K \int d\psi_K\, d\bar\psi_K\, e^{\bar\psi_K Q_K \psi_K}
  = \prod_K Q_K
  = \det(Q).
\end{align}
It is convenient to absorb the quadratic part of the action as well as
the normalization factor into the measure,
\begin{align}
  \label{eq:Z}
  \Z[\eta,\bar\eta]
  = \int \dmu{Q}{\psi}\, e^{-V_0[\psi,\bar\psi]}\,
  e^{-\inner{\bar\psi}{\eta}-\inner{\bar\eta}{\psi}}
\end{align}
with the normalized Gaussian path-integral measure
\begin{align}
  \label{eq:measure}
  \dmu{Q}{\psi} & := \frac{1}{Z_0} [d\psi\bar\psi]\,
  e^{\inner{\bar\psi}{Q\psi}}
  & \text{such that } \int \dmu{Q}{\psi} & = 1.
\end{align}
If there is no interaction, $V_0=0$, the integral is quadratic
(Gaussian) and can be performed analytically by completing the square:
\begin{equation*}
  \begin{split}
    \Z^{\text{nonint}}[\eta,\bar\eta]
    & = \int \dmu{Q}{\psi}\,
    e^{-\inner{\bar\psi}{\eta}-\inner{\bar\eta}{\psi}} \\
    & = \frac{1}{Z_0} \int [d\psi\bar\psi]\,
    e^{\inner{\bar\psi}{Q\psi}
      -\inner{\bar\psi}{\eta}-\inner{\bar\eta}{\psi}} \\
    & = \frac{1}{Z_0} \int [d\psi\bar\psi]\,
    e^{\inner{[\bar\psi-C^t\bar\eta]}{Q[\psi-C\eta]}
      -\inner{\bar\eta}{C\eta}} \\
    & = e^{-\inner{\bar\eta}{C\eta}}
    \int \dmux{Q}{\psi-C\eta}{\bar\psi-C^t\bar\eta} \\
    & = e^{-\inner{\bar\eta}{C\eta}}
  \end{split}
\end{equation*}
where $C^t$ is the transposed propagator.

The \emph{connected} Green functions are generated by the functional
\begin{align*}
  \G[\eta,\bar\eta] := - \ln\Z[\eta,\bar\eta]
\end{align*}
as
\begin{align*}
  \G[\eta,\bar\eta] = \sum_{m=0}^\infty \frac{1}{(m!)^2}
  \sum_{K_1\dots K_m} \sum_{K_1'\dots K_m'} 
  G_m(K_1',\dots,K_m';K_1,\dots,K_m) 
  \prod_{j=1}^m \bar\eta_{K_j'} \eta_{K_j}  
\end{align*}
and
\begin{align*}
  G_m(K_1',\dots,K_m';K_1,\dots,K_m)
  & = -\vev{\psi_{K_1'}\dots\psi_{K_m'}
    \bar\psi_{K_m}\dots\bar\psi_{K_1}}_{\text{conn}} \\
  & = \frac{\delta^m}{\delta\eta_{K_1}\dots\delta\eta_{K_m}}
  \frac{\delta^m}{\delta\bar\eta_{K_m'}\dots\delta\bar\eta_{K_1'}}
  \G[\eta,\bar\eta]\at_{\eta=\bar\eta=0} \;,
\end{align*}
respectively.  For the non-interacting system follows
\begin{align}
  \label{eq:Gnonint}
  \G^{\text{nonint}} = \inner{\bar\eta}{C\eta}
\end{align}
such that $G_1^{\text{nonint}}(K) = C(K)$ is the bare propagator, and
all other $G_m^{\text{nonint}}$ vanish.


\subsection[\onePI\ vertex functions]{1PI vertex functions}
\label{sec:rg:form:gamma}

There is another set of correlation functions that is particularly
useful for describing phase transitions and fields whose expectation
value does not always vanish, such as the order parameter in a
symmetry-broken phase: the \emph{one-particle irreducible} (\onePI)
vertex functions $\gamma_m$ generated by the functional
$\Gamma[\phi,\bar\phi]$.  $\Gamma$ is obtained via Legendre
transformation from the connected Green functions \cite[section
7.8]{Zin02},
\begin{align}
  \label{eq:gammadef}
  \Gamma[\phi,\bar\phi] + \inner{\bar\phi}{Q\phi}
  := \G[\eta,\bar\eta] +\inner{\bar\phi}{\eta} -\inner{\bar\eta}{\phi}.
\end{align}
This differs from the textbook definition by taking the inverse bare
propagator $Q$ out of the one-particle vertex function.  The
transformations between $\phi$ and $\eta$ are
\begin{equation}
  \label{eq:legendre}
  \begin{split}
    \delta_\phi \Gamma -Q^t \bar\phi & = \bar\eta, & \delta_\eta \G
    & = \bar\phi, & \delta_\eta \delta_{\bar\eta} \G
    & = (\delta_\phi \delta_{\bar\phi} \Gamma + Q)^{-1} \\
    \delta_{\bar\phi} \Gamma + Q\phi & = \eta, & \delta_{\bar\eta} \G & = \phi.
  \end{split}
\end{equation}
$\Gamma$ is expanded in $\bar\phi_K$, $\phi_K$ as
\begin{align*}
  \Gamma[\phi,\bar\phi] = \sum_{m=0}^\infty \frac{1}{(m!)^2}
  \sum_{K_1\dots K_m} \sum_{K_1'\dots K_m'} 
  \gamma_m(K_1',\dots,K_m';K_1,\dots,K_m) 
  \prod_{j=1}^m \bar\phi_{K_j'} \phi_{K_j}.
\end{align*}
Each \onePI\ vertex function $\gamma_m$ is made up of those diagrams
of $G_m$ which cannot be split into two disconnected parts by cutting
a single line.  As all self-energy contributions on external legs are
one-particle \emph{reducible} with respect to the main part of the
diagram, full propagators are amputated from all external legs.  In
the special case without interaction with
$\G^{\text{nonint}}[\eta,\bar\eta] = \inner{\bar\eta}{C\eta}$ we
obtain $\Gamma^{\text{nonint}}[\phi,\bar\phi] = 0$.


\subsection{Connected amputated Green functions}
\label{sec:rg:form:v}

Another way of looking at a system is by considering the generating
functional of the connected amputated Green functions, the
\emph{effective interaction} $\V[\chi,\bar\chi]$:
\begin{align}
  \label{eq:vdef}
  e^{-\V[\chi,\bar\chi]}
  & := \int \dmu{Q}{\psi}\, e^{-V_0[\psi+\chi,\bar\psi+\bar\chi]} \\
  & = e^{-V_0[\delta_{\bar\varphi},\delta_\varphi]}\, \int \dmu{Q}{\psi}\,
  e^{\inner{\bar\varphi}{\psi+\chi}-\inner{\bar\psi+\bar\chi}{\varphi}}
  \at_{\varphi=\bar\varphi=0} \notag \\
  & = e^{-V_0[\delta_{\bar\varphi},\delta_\varphi]}\;
  e^{\inner{\bar\varphi}{C\varphi}}\;
  e^{\inner{\bar\varphi}{\chi}-\inner{\bar\chi}{\varphi}}
  \at_{\varphi=\bar\varphi=0} \notag \\
  & = e^{-V_0[\delta_{\bar\varphi},\delta_\varphi]}\;
  e^{\inner{\delta_\chi}{C\delta_{\bar\chi}}}\;
  e^{\inner{\bar\varphi}{\chi}-\inner{\bar\chi}{\varphi}}
  \at_{\varphi=\bar\varphi=0} \notag \\
  & = e^{\Laplace{C}}\; e^{-V_0[\chi,\bar\chi]} \notag
\end{align}
where the \emph{functional Laplace operator} is defined as
\begin{align*}
  \Laplace{C}
  := \inner{\ddd\chi}{C\ddd{\bar\chi}}
  = \int dx \, dy \; \frac{\delta}{\delta\chi(y)} C(y,x)
  \frac{\delta}{\delta\bar\chi(x)} 
  = \sum_K \ddd{\chi_K} C_K \ddd{\bar\chi_K}.
\end{align*}
I shall use the shorthand notation $\Laplace{C}$ only if it is
unambiguous on which Grassmann variable the derivatives act and use
the explicit notation otherwise.  $\Laplace{C}$ acts on a functional
$\F[\chi,\bar\chi]$ in the following way: the derivatives $\ddd\chi$
and $\ddd{\bar\chi}$ pick an ingoing and outgoing leg from each
diagram in $\F$ and connect them by a bare propagator $C$.  In this
picture, perturbation theory for $e^{-\V}$ may be visualized as
follows: $e^{-V_0}$ is a collection of any number of disconnected bare
interaction vertices, and $\Laplace{C}\, e^{-V_0}$ contains all
diagrams where either $V_0$ is closed by one $C$ loop to create a
tadpole diagram, or two $V_0$'s are connected by a $C$ line to form a
tree diagram.  Repeating this procedure indefinitely, higher tree and
loop diagrams appear, and finally $e^{-\V}=e^{\Laplace{C}}\, e^{-V_0}$
contains all Feynman diagrams.  Taking the logarithm to obtain $\V$,
only the connected diagrams are retained \cite[section 1.2.1]{Zin02}.

What is the relation between $\G$ and $\V$ as both generate all
connected Green functions?  We observe that by the substitution
$\chi:=C\eta$, $\bar\chi:=C^t\bar\eta$,
\begin{align*}
  e^{-\V[C\eta,C^t\bar\eta]}
  & = \int \dmu{Q}{\psi}\, e^{-V_0[\psi+C\eta,\bar\psi+C^t\bar\eta]} \\
  & = \int \dmux{Q}{\psi'-C\eta}{\bar\psi'-C^t\bar\eta}\,
  e^{-V_0[\psi',\bar\psi']} \\
  & = \int \dmu{Q}{\psi'}\,
  e^{-\inner{\bar\psi'}{\eta} -\inner{\bar\eta}{\psi'}
    +\inner{\bar\eta}{C\eta}}\, e^{-V_0[\psi',\bar\psi']} \\
  & = e^{\inner{\bar\eta}{C\eta}-\G[\eta,\bar\eta]}
\end{align*}
such that
\begin{align}
  \label{eq:VGrel}
  \V[C\eta,C^t\bar\eta] = \G[\eta,\bar\eta] - \inner{\bar\eta}{C\eta}.
\end{align}
In the non-interacting case, $\V=0$ because $\inner{\bar\eta}{C\eta}$
cancels the non-interacting part \eqref{eq:Gnonint} of $\G$, in
accordance with the name effective \emph{interaction}.  Generally,
$\V$ generates connected amputated Green functions $V_m$,
\begin{align}
  \V[\chi,\bar\chi] = \sum_{m=0}^\infty \frac{1}{(m!)^2}
  \sum_{K_1\dots K_m} \sum_{K_1'\dots K_m'} 
  V_m(K_1',\dots,K_m';K_1,\dots,K_m) 
  \prod_{j=1}^m \bar\chi_{K_j'} \chi_{K_j}\,,
\end{align}
\ie, connected Green functions $G_m$ with \emph{bare} propagators
$C(K)$ amputated from all external legs:
\begin{align*}
  V_m(K_1',\dots,K_m';K_1,\dots,K_m)
  & = \frac{\delta^m}{\delta\chi_{K_1}\dotsm\delta\chi_{K_m}}
  \frac{\delta^m}{\delta\bar\chi_{K_m'}\dotsm\delta\bar\chi_{K_1'}}
  \V[\chi,\bar\chi]\at_{\chi=\bar\chi=0} \\
  & = \frac{\delta\eta_{K_1}}{\delta\chi_{K_1}} 
  \frac{\delta}{\delta\eta_{K_1}} \dotsm
  \frac{\delta\bar\eta_{K_1'}}{\delta\bar\chi_{K_1'}} 
  \frac{\delta}{\delta\bar\eta_{K_1'}}
  \bigl[ \G[\eta,\bar\eta] - \inner{\bar\eta}{C\eta} \bigr]
  _{\eta=\bar\eta=0} \notag \\
  & = \frac{G_m(K_1',\dotsc,K_m';K_1,\dotsc,K_m)}
  {C^t(K_1')\dotsm C(K_1)} - \frac{\delta_{m,1}}{C(K_1)}\,.
\end{align*}


\section{Renormalization-group flow equations}
\label{sec:rg:flow}

When computing a Green function perturbatively by summing the
contributions of certain Feynman diagrams, there can be infrared
divergences both as an artefact of perturbation theory and physically
as an indication of a phase transition.  These can be regularized by
an infrared cutoff $\Lambda$ in the bare propagator which suppresses
modes of low frequency or momentum close to the Fermi surface.  The
change of the Green functions as the cutoff scale is changed is
governed by the \emph{renormalization-group flow equation}.

Alternatively, one can regularize the problem by going to sufficiently
high temperatures $T>T_c$, weak coupling $g^2 U < U$, and/or a finite
system size $N$; then one can consider the flow of the Green functions
as the parameters $T$ or $g$ are changed.


\subsection{Regularization and flow parameters}
\label{sec:rg:flow:regu}

The flow schemes used for the functional \RG\ are constructed by
making \emph{only} the quadratic part of the bare action depend on the
cutoff scale or flow parameter.  In all cases the regularization is
done with respect to \emph{energy} scales.

\subsubsection*{Frequency and momentum cutoff}

The \emph{frequency cutoff} is defined by multiplying the bare
propagator in the action with a cutoff function
$\chi^\Lambda(\omega)$,
\begin{align}
  \label{eq:Ccutoff}
  C^\Lambda(K) = \chi^\Lambda(\omega) C(K),
\end{align}
where $\chi^\Lambda(\omega)$ cuts out modes with frequency $\abs\omega
< \Lambda$ in the frequency basis where $C(K)$ is diagonal.  It may be
either a sharp cutoff function using the step function $\Theta(x)$,
\begin{align*}
  \chi^\Lambda(\omega) = \Theta(\abs\omega-\Lambda),
\end{align*}
or a smooth cutoff.  Generally one would think that a smooth cutoff
with a differentiable cutoff function is easier to handle.  However,
especially at zero temperature, a sharp cutoff can in fact be
favorable (cf.\ section \ref{sec:rg:flow:gamma}).

The \emph{momentum cutoff} is defined analogously with
\begin{align}
  \label{eq:momcutoff}
  \chi^\Lambda(\vec k) = \Theta(\abs{\xi_{\vec k}}-\Lambda).
\end{align}
Generally, the propagator is split as
\begin{align}
  C(K) = C^\Lambda(K) + D^\Lambda(K)
\end{align}
where $C^\Lambda(K)$ is the propagator for the high-energy
(\emph{hard}) modes, and $D^\Lambda(K)$ is the propagator for the
remaining low-energy (\emph{soft}) modes that are yet to be integrated
out.  As $\Lambda\to 0$, the cutoff is removed and $C^\Lambda\to C$,
$D^\Lambda\to 0$.

It is simplest to define the cutoff in the basis where $C$ is
diagonal, as a function multiplying the kinetic energy eigenvalues
$\epsilon$ with $\chi^\Lambda(\epsilon)$.  We will later (section
\ref{sec:ward:cutoff:bg}) see an example where in the presence of an
external field the propagator is not diagonal in momentum space: then
the cutoff can be defined by diagonalizing first.

Note that at finite $\Lambda$, the frequency cutoff leads to
non-analyticities in certain correlation functions, for example in the
current-current response function.  This disadvantage is absent in the
temperature flow.

\subsubsection*{Temperature flow}

In order to use the temperature $T$ as a flow parameter in the \fRG\ 
formalism, only the quadratic part of the action may depend on $T$.
Initially, however, the action has powers of $T$ both in the kinetic
and in the interaction part:
\begin{align*}
  S[\psi,\bar\psi] & = T\sum_{i\omega_n} \sum_k \bar\psi_K
  (i\omega_n-\xi_{\vec{k}}) \psi_K \\
  & \quad + \frac{1}{2} T^3
  \sum_{K_1,K_2,K_1'} V(K_1';K_1,K_2)\, \bar\psi_{K_1'}
  \bar\psi_{K_1+K_2-K_1'} \psi_{K_2} \psi_{K_1} \,.
\end{align*}
By rescaling the fields $\tilde\psi := T^{3/4}\, \psi$, one can shift
the explicit $T$ dependence of a \emph{quartic} interaction into the
quadratic part \cite{HS01}:
\begin{align*}
  \tilde S[\tilde\psi,\bar{\tilde\psi}]
  & = T^{-1/2} \sum_{i\omega_n} \sum_k \bar{\tilde\psi}_K
  (i\omega_n-\xi_{\vec{k}}) \tilde\psi_K \\
  & \quad + \frac{1}{2}
  \sum_{K_1,K_2,K_1'} V(K_1';K_1,K_2)\, \bar{\tilde\psi}_{K_1'}
  \bar{\tilde\psi}_{K_1+K_2-K_1'} \tilde\psi_{K_2} \tilde\psi_{K_1} \,.
\end{align*}
We can now perform the \fRG\ flow on this action to obtain Green
functions $G_m^T$ in terms of the rescaled fields $\tilde\psi$ for a
whole temperature range and rescale them back to obtain the usual
Green functions\footnote{This is in contrast to \cite{HS01} where the
  powers of $T$ in the original action are considered part of the
  Green functions.}
\begin{align}
  \label{eq:Tscaling}
  G_m(\dotsc) = T^{(2-m)/2}\, G_m^T(\dotsc).
\end{align}

\subsubsection*{Interaction flow}

There are systems which can be treated also by a much simpler method
\cite{HRAE04}, namely by simply rescaling the propagator with a number
$g\in [0,1]$.  Then, all Feynman diagrams become dressed by a global
factor of $g$ for each internal line.  This seemingly trivial change
to weight each diagram by a power of $g$ has the advantage that the
\fRG\ flow equation for the Green functions from $g=0$ to $g=1$ resums
infinite subclasses of Feynman diagrams which would otherwise be more
tedious to do.  At the same time we can rescale the fields and observe
that for $g<1$ the model is the same as one with full propagators but
reduced interaction strength $g^2 \abs U < \abs U$:
\begin{align*}
  S & \sim \inner{\bar\psi}{[Q/g]\psi}
  -U\, \bar\psi \bar\psi \psi \psi \\
  & \sim \inner{\bar{\tilde\psi}}{Q\tilde\psi}
  -g^2 U\, \bar{\tilde\psi} \bar{\tilde\psi} \tilde\psi \tilde\psi
\end{align*}
where $\tilde\psi = g^{-1/2}\, \psi$.  Integrating from $g=0$ to $g=1$
thus yields a flow in the bare interaction.


\subsection{Connected Green function flow}
\label{sec:rg:flow:g}

The above regularization and flow schemes lead to an action with a
quadratic part depending on a scale parameter $\Lambda$, which for
notational simplicity shall include $T$ for the temperature flow and
$g$ for the interaction flow,
\begin{align}
  \label{eq:cutoffaction}
  S^\Lambda[\psi,\bar\psi]
  := \inner{\bar\psi}{Q^\Lambda\psi} - V_0[\psi,\bar\psi] \,,
\end{align}
where the inverse bare propagator with cutoff is defined in analogy to
equation \eqref{eq:bareprop},
\begin{align}
  \label{eq:cutoffbareprop}
  Q^\Lambda := (C^\Lambda)^{-1}.
\end{align}
Generally, if $\chi^\Lambda(K)=1$ then $C^\Lambda(K) = \chi^\Lambda(K)
C(K)$ is just the original propagator while for $\chi^\Lambda=0$,
$C^\Lambda(K)=0$ and the kinetic term $\inner{\bar\psi}
{(C^\Lambda)^{-1}\psi}$ in the action becomes infinite, giving the
cutoff modes of the fermion fields an infinite mass such that they are
frozen out.  By the action $S^\Lambda$ all generating functionals, and
thus their Green function components, depend on $\Lambda$.  The
generating functional $\G^\Lambda$ for the connected Green functions
in the presence of the cutoff is defined as
\begin{align}
  \label{eq:glambda1}
  e^{-\G^\Lambda[\eta,\bar\eta]}
  & := \int \dmu{Q^\Lambda}{\psi}\, e^{-V_0[\psi,\bar\psi]}\,
  e^{-\inner{\bar\psi}{\eta}-\inner{\bar\eta}{\psi}} \\
  \label{eq:glambda}
  & = \frac{1}{Z_0^\Lambda} \int [d\psi\bar\psi]\,
  e^{\inner{\bar\psi}{Q^\Lambda\psi} - V_0[\psi,\bar\psi]}\,
  e^{-\inner{\bar\psi}{\eta}-\inner{\bar\eta}{\psi}}
\end{align}
where $Q$ is replaced by $Q^\Lambda$, and the normalization factor is
changed accordingly to $Z_0^\Lambda=\det(Q^\Lambda)$.  The flow
equation for $e^{-\G^\Lambda}$ is obtained by taking the $\Lambda$
derivative, denoted by the dot, on both sides of equation
\eqref{eq:glambda},
\begin{align*}
  -\left(\dl\G^\Lambda\right) e^{-\G^\Lambda[\eta,\bar\eta]}
  & = - \, \frac{\dl \det Q^\Lambda}{\det Q^\Lambda} \,
  e^{-\G^\Lambda} + \\
  & \quad + \int \dmu{Q^\Lambda}{\psi} \,
  \inner{\bar\psi}{\dot Q^\Lambda\psi} \, e^{-V_0[\psi,\bar\psi]} \,
  e^{-\inner{\bar\psi}{\eta}-\inner{\bar\eta}{\psi}} \\
  & = \left( -\Tr(\dot Q^\Lambda C^\Lambda) -\Laplace{\dot
      Q^\Lambda} \right) e^{-\G^\Lambda[\eta,\bar\eta]}
\end{align*}
where the first term comes from the derivative of the normalization
factor, $\dl \ln \det(Q^\Lambda) = \Tr \dl \ln(Q^\Lambda) = \Tr(\dot
Q^\Lambda C^\Lambda)$.  Here, $\Tr$ denotes a sum over all space-time
indices.  Thus, the flow of $\G^\Lambda$ is
\begin{align}
  \label{eq:gflow}
  \boxed{
    \dl \G^\Lambda[\eta,\bar\eta]
    = \Tr(\dot Q^\Lambda\, C^\Lambda)
    -\Tr\left(\dot Q^\Lambda\,
    \frac{\delta^2 \G^\Lambda[\eta,\bar\eta]}
    {\delta\eta\, \delta\bar\eta}\right)
    +\inner{\frac{\delta\G^\Lambda[\eta,\bar\eta]}{\delta\eta}} {\dot
      Q^\Lambda \frac{\delta\G^\Lambda[\eta,\bar\eta]}{\delta\bar\eta}}}\,.
\end{align}
As a check, in the non-interacting case
\begin{align*}
  \G^\Lambda[\eta,\bar\eta]
  & = \inner{\bar\eta}{C^\Lambda\eta} \\
  G_1^\Lambda
  & = \frac{\delta^2 \G^\Lambda} {\delta\eta\, \delta\bar\eta}
  = C^\Lambda \\
  \dot\G^\Lambda
  & = \Tr(\dot Q^\Lambda\, C^\Lambda)
  -\Tr(\dot Q^\Lambda\, C^\Lambda) 
  +\inner{-\bar\eta}{C^\Lambda\, \dot Q^\Lambda\, C^\Lambda \eta}
  = \inner{\bar\eta}{\dot C^\Lambda\eta}.
\end{align*}
Two technical notes are in order.  Naively, the $\dot Q^\Lambda$
appearing in the flow equation \eqref{eq:gflow} for $\G^\Lambda$ looks
ill-defined:
\begin{align*}
  \dot Q_K^\Lambda
  = -\frac{\dot\chi^\Lambda(K)}{[\chi^\Lambda(K)]^2} Q(K)
\end{align*}
contains a division by zero for the cutoff modes where
$\chi^\Lambda(K)=0$.  But all Green functions $G_m^\Lambda$ have
$C^\Lambda$ on their external legs so only the combination
\begin{align}
  \label{eq:singlescaleconn}
  S^\Lambda(K) & := C^\Lambda\, \dot Q^\Lambda\, C^\Lambda
  = -\dot C^\Lambda
  = -\dot\chi^\Lambda(K) C(K)
  \quad \text{(\emph{single-scale propagator})}
\end{align}
appears.  This quantity is well-defined and, for a sharp cutoff, has
support only on the scale $\Lambda$ since $\dot\chi^\Lambda$ is a
$\delta$ function.  The other seemingly ill-defined contribution is
\begin{align}
  \Tr(\dot Q^\Lambda C^\Lambda)
  = -\sum_K \frac{\dot\chi^\Lambda(K)}{\chi^\Lambda(K)} ,
\end{align}
but this is canceled by the bare part of the second term in the flow
equation \eqref{eq:gflow},
\begin{align*}
  \Tr(\dot Q^\Lambda C^\Lambda)
  -\Tr \Bigl( \dot Q^\Lambda\, \frac{\delta^2 \G^\Lambda}
  {\delta\eta\, \delta\bar\eta} \Bigr) \at_{\eta=\bar\eta=0}
  & = \Tr(\dot Q^\Lambda [C^\Lambda - G^\Lambda]) \\
  = -\sum_{k=1}^\infty \Tr(\dot Q^\Lambda\, C^\Lambda
  [\Sigma^\Lambda\, C^\Lambda]^k)
\end{align*}
which again contains only the well-defined combination $S^\Lambda =
C^\Lambda\, \dot Q^\Lambda\, C^\Lambda$.  On the second line, the full
propagator $G^\Lambda$ has been expanded into a geometric series via
the Dyson equation $(G^\Lambda)^{-1} = (C^\Lambda)^{-1} -
\Sigma^\Lambda$.


\subsection[\onePI\ vertex function flow]{1PI vertex function flow}
\label{sec:rg:flow:gamma}

From the flow equation \eqref{eq:gflow} of the connected Green
functions it is simple to derive the flow of the \onePI\ generating
functional $\Gamma^\Lambda$.  We use the Legendre transformation
\eqref{eq:legendre} but all functionals and $Q$ are taken at scale
$\Lambda$,
\begin{align}
  \label{eq:leglambda}
  \Gamma^\Lambda[\phi,\bar\phi] + \inner{\bar\phi}{Q^\Lambda\phi}
  := \G^\Lambda[\eta,\bar\eta]
  + \inner{\bar\phi}{\eta} - \inner{\bar\eta}{\phi}.
\end{align}
We express $\G^\Lambda$ and its derivatives by $\Gamma^\Lambda$,
\begin{align*}
  \dl \Gamma^\Lambda[\phi,\bar\phi]
  & = \dl \G^\Lambda[\eta,\bar\eta]
  - \inner{\bar\phi}{\dot Q^\Lambda\phi} \\[1ex]
  & = \Bigl(\frac{\delta\G^\Lambda[\eta,\bar\eta]}{\delta\eta},
  \dot Q^\Lambda \frac{\delta\G^\Lambda[\eta,\bar\eta]}{\delta\bar\eta}\Bigr)
  -\Tr \dot Q^\Lambda\,
  \Bigl[\frac{\delta^2 \G^\Lambda[\eta,\bar\eta]} {\delta\eta\,\delta\bar\eta}
  -C^\Lambda\Bigr] - \inner{\bar\phi}{\dot Q^\Lambda\phi} \\
  & = \inner{\bar\phi}{\dot Q^\Lambda\phi}
  -\Tr \dot Q^\Lambda \Bigl[ \Bigl(\frac{\delta^2
    \Gamma^\Lambda[\phi,\bar\phi]}{\delta\phi\,\delta\bar\phi}
  + Q^\Lambda \Bigr)^{-1}-C^\Lambda \Bigr]
  - \inner{\bar\phi}{\dot Q^\Lambda\phi},
\end{align*}
to arrive at the \emph{\onePI\ flow equation}
\begin{align}
  \label{eq:gammaflow}
  \boxed{
    \dl\Gamma^\Lambda = \Tr \dot Q^\Lambda
    \Bigl[ C^\Lambda - \Bigl(\frac{\delta^2
      \Gamma^\Lambda[\phi,\bar\phi]}{\delta\phi\,\delta\bar\phi}
    + Q^\Lambda \Bigr)^{-1} \Bigr]}\,.
\end{align}
If we define the abbreviation $\Gamma^{(2)} := \delta^2 \Gamma^\Lambda
/ \delta\phi \, \delta\bar\phi$, the flow equation can be written in a
more compact form as
\begin{align}
  \label{eq:gammaflow2}
  \dl\Gamma^\Lambda
  & = \dl \Tr \ln Q^\Lambda - \dl \Tr \ln (Q^\Lambda + \Gamma^{(2)})
  = \dl \Tr \ln \frac{Q^\Lambda}{Q^\Lambda + \Gamma^{(2)}},
\end{align}
where in equation \eqref{eq:gammaflow2}, the $\Lambda$ derivative acts
only on the cutoff functions inside $Q^\Lambda$.

The inverse of the functional $(Q^\Lambda+\Gamma^{(2)})$ is the
inverse full propagator in the presence of the fields $\phi$,
$\bar\phi$ and can be safely defined via a geometric series, as in the
Dyson equation.  To see this, split $(Q^\Lambda+\Gamma^{(2)})$ into a
part independent of $\phi$, $\bar\phi$ which is the usual inverse full
cutoff propagator $[G^\Lambda]^{-1} = Q^\Lambda - \Sigma^\Lambda$, and
a remaining functional $\tilde\Gamma^\Lambda[\phi,\bar\phi]$,
\begin{align*}
  Q^\Lambda + \frac{\delta^2 \Gamma^\Lambda[\phi,\bar\phi]}
  {\delta\phi\,\delta\bar\phi} 
  & = (G^\Lambda)^{-1} + \tilde\Gamma^\Lambda[\phi,\bar\phi],
  & -\Sigma^\Lambda
  & = \gamma_1^\Lambda
  = \frac{\delta^2 \Gamma^\Lambda[\phi,\bar\phi]}
  {\delta\phi\,\delta\bar\phi} \at_{\phi=\bar\phi=0} \;.
\end{align*}
Thus,
\begin{align*}
  \Bigl( Q^\Lambda + \frac{\delta^2 \Gamma^\Lambda[\phi,\bar\phi]}
    {\delta\phi\,\delta\bar\phi} \Bigr)^{-1}
  & = \bigl( 1 + G^\Lambda\, \tilde\Gamma^\Lambda \bigr)^{-1}
  G^\Lambda 
  = \Bigl( 1 - G^\Lambda\, \tilde\Gamma^\Lambda 
    + [G^\Lambda\, \tilde\Gamma^\Lambda]^2 - \dotsb
  \Bigr) G^\Lambda.
\end{align*}
Together with the definition of the single-scale propagator
$S^\Lambda$ in analogy to \eqref{eq:singlescaleconn},
\begin{align*}
  S^\Lambda & := G^\Lambda\, \dot Q^\Lambda\, G^\Lambda
  \quad \text{(\emph{single-scale propagator})},
\end{align*}
the flow equation becomes
\begin{align}
  \label{eq:gammaflow3}
  \dl \Gamma^\Lambda = \Tr(\dot Q^\Lambda [C^\Lambda-G^\Lambda])
  +\Tr(S^\Lambda [\tilde\Gamma^\Lambda - \tilde\Gamma^\Lambda\,
  G^\Lambda\, \tilde\Gamma^\Lambda\, + \dotsb])
\end{align}
where the last trace consists of a one-loop term with any number of
vertices $\tilde\Gamma^\Lambda$, which contribute at least two
external legs each, connected by one single-scale propagator
$S^\Lambda$ and several full propagators $G^\Lambda$.  Writing the
flow equation in terms of the components $\gamma_m^\Lambda$, we
obtain the following diagrams for the general hierarchy of flow
equations and for the first few levels of this infinite hierarchy:
\begin{align}
  \label{eq:1piflow}
  \frac{\partial}{\partial\Lambda}
  \parbox{23mm}{\unitlength=1mm\fmfframe(3,2)(2,2){
      \begin{fmfgraph*}(18,18)
        \fmfsurroundn{e}{16}
        \fmfv{d.sh=circle,d.f=30,d.si=10mm,
          label.angle=-90,label.dist=11mm,label=$\gamma_m^\Lambda$}{v}
        \fmfv{d.sh=circle,d.f=full,d.si=0.5mm}{e1,e2,e8,e9,e10,e16}
        \fmf{plain}{v,e3}
        \fmf{plain}{v,e5}
        \fmf{plain}{v,e7}
        \fmf{plain}{v,e11}
        \fmf{plain}{v,e13}
        \fmf{plain}{v,e15}
      \end{fmfgraph*}}}
  & =
  \parbox{23mm}{\unitlength=1mm\fmfframe(3,12)(2,12){
      \begin{fmfgraph*}(18,18)
        \fmfsurroundn{e}{16}
        \fmfv{d.sh=circle,d.f=30,d.si=10mm,
          label.angle=-90,label.dist=11mm,label=$\gamma_{m+1}^\Lambda$}{v}
        \fmfv{d.sh=circle,d.f=full,d.si=0.5mm}{e1,e2,e8,e9,e10,e16}
        \fmfleft{i}\fmfright{o}\fmf{phantom}{i,v,o}\fmffreeze
        \fmf{plain}{v,e3}
        \fmf{plain}{v,e7}
        \fmf{plain}{v,e11}
        \fmf{plain}{v,e13}
        \fmf{plain}{v,e15}
        \fmf{slplaino,right=100,tension=0.4,label=$S^\Lambda$}{v,v}
      \end{fmfgraph*}}}
  + \sum
  \parbox{55mm}{\unitlength=1mm\fmfframe(3,0)(2,0){
      \begin{fmfgraph*}(50,18)
        \fmfleftn{l}{7}
        \fmfrightn{r}{7}
        \fmfv{d.sh=circle,d.f=30,d.si=10mm,
          label.angle=-90,label.dist=11mm,label=$\gamma_{m'}^\Lambda$}{vl}
        \fmfv{d.sh=circle,d.f=30,d.si=10mm,
          label.angle=-70,label.dist=12mm,label=$\gamma_{m''}^\Lambda$}{vr}
        \fmffixed{(.55w,0)}{vl,vr}
        \fmfleft{i}\fmfright{o}\fmf{phantom}{i,vl,vr,o}\fmffreeze
        \fmf{plain}{vl,l1}
        \fmf{plain}{vl,l3}
        \fmfv{d.sh=circle,d.f=full,d.si=0.5mm}{l4}
        \fmfv{d.sh=circle,d.f=full,d.si=0.5mm}{l5}
        \fmfv{d.sh=circle,d.f=full,d.si=0.5mm}{l6}
        \fmf{plain}{vl,l7}
        \fmf{plain}{vr,r1}
        \fmfv{d.sh=circle,d.f=full,d.si=0.5mm}{r2}
        \fmfv{d.sh=circle,d.f=full,d.si=0.5mm}{r3}
        \fmfv{d.sh=circle,d.f=full,d.si=0.5mm}{r4}
        \fmf{plain}{vr,r5}
        \fmf{plain}{vr,r7}
        \fmf{slplain,left=0.5,label=$S^\Lambda$}{vl,vr}
        \fmf{wiggly,right=0.5,label.side=left,label=expand}{vl,vr}
      \end{fmfgraph*}}}
  \\
  \frac{\partial}{\partial\Lambda}
  \parbox{23mm}{\unitlength=1mm\fmfframe(3,8)(2,8){
      \begin{fmfgraph*}(18,18)
        \fmfv{d.sh=circle,d.f=30,d.si=10mm,
          label.angle=-90,label.dist=7mm,label=$\gamma_1^\Lambda$}{v}
        \fmfleft{i}
        \fmfright{o}
        \fmf{plain}{i,v,o}
      \end{fmfgraph*}}}
  & =
  \parbox{18mm}{\unitlength=1mm\fmfframe(2,10)(2,10){
      \begin{fmfgraph*}(18,18)
        \fmfsurroundn{e}{8}
        \fmfv{d.sh=circle,d.f=30,d.si=10mm,
          label.angle=-90,label.dist=7mm,label=$\gamma_2^\Lambda$}{v}
        \fmfleft{i}\fmfright{o}\fmf{phantom}{i,v,o}\fmffreeze
        \fmf{plain}{v,e6}
        \fmf{plain}{v,e8}
        \fmf{slplaino,right=100,tension=0.4,label=$S^\Lambda$}{v,v}
      \end{fmfgraph*}}}
  \notag \\
  \frac{\partial}{\partial\Lambda}
  \parbox{23mm}{\unitlength=1mm\fmfframe(3,8)(2,8){
      \begin{fmfgraph*}(18,18)
        \fmfsurroundn{e}{8}
        \fmfv{d.sh=circle,d.f=30,d.si=10mm,
          label.angle=-90,label.dist=7mm,label=$\gamma_2^\Lambda$}{v}
        \fmfleft{i}\fmfright{o}\fmf{phantom}{i,v,o}\fmffreeze
        \fmf{plain}{v,e2}
        \fmf{plain}{v,e4}
        \fmf{plain}{v,e6}
        \fmf{plain}{v,e8}
      \end{fmfgraph*}}}
  & =
  \parbox{23mm}{\unitlength=1mm\fmfframe(3,10)(2,10){
      \begin{fmfgraph*}(18,18)
        \fmfsurroundn{e}{6}
        \fmfv{d.sh=circle,d.f=30,d.si=10mm,
          label.angle=-93,label.dist=9mm,label=$\gamma_3^\Lambda$}{v}
        \fmfleft{i}\fmfright{o}\fmf{phantom}{i,v,o}\fmffreeze
        \fmf{plain}{v,e1}
        \fmf{plain}{v,e4}
        \fmf{plain}{v,e5}
        \fmf{plain}{v,e6}
        \fmf{slplaino,right=100,tension=0.4,label=$S^\Lambda$}{v,v}
      \end{fmfgraph*}}}
  +
  \parbox{47mm}{\unitlength=1mm\fmfframe(-4,6)(1,6){
      \begin{fmfgraph*}(50,18)
        \fmfleftn{l}{3}
        \fmfrightn{r}{3}
        \fmfv{d.sh=circle,d.f=30,d.si=10mm,
          label.angle=-90,label.dist=7mm,label=$\gamma_2^\Lambda$}{vl}
        \fmfv{d.sh=circle,d.f=30,d.si=10mm,
          label.angle=-90,label.dist=7mm,label=$\gamma_2^\Lambda$}{vr}
        \fmffixed{(.55w,0)}{vl,vr}
        \fmfleft{i}\fmfright{o}\fmf{phantom}{i,vl,vr,o}\fmffreeze
        \fmf{plain}{vl,l1}
        \fmf{plain}{vl,l3}
        \fmf{plain}{vr,r1}
        \fmf{plain}{vr,r3}
        \fmf{slplain,left=0.5,label=$S^\Lambda$}{vl,vr}
        \fmf{plain,right=0.5,label.side=left,label=$G^\Lambda$}{vl,vr}
      \end{fmfgraph*}}}
  \notag \\
  \frac{\partial}{\partial\Lambda}
  \parbox{23mm}{\unitlength=1mm\fmfframe(3,8)(2,8){
      \begin{fmfgraph*}(18,18)
        \fmfsurroundn{e}{12}
        \fmfv{d.sh=circle,d.f=30,d.si=10mm,
          label.angle=-85,label.dist=9mm,label=$\gamma_3^\Lambda$}{v}
        \fmfleft{i}\fmfright{o}\fmf{phantom}{i,v,o}\fmffreeze
        \fmf{plain}{v,e2}
        \fmf{plain}{v,e4}
        \fmf{plain}{v,e6}
        \fmf{plain}{v,e8}
        \fmf{plain}{v,e10}
        \fmf{plain}{v,e12}
      \end{fmfgraph*}}}
  & =
  \parbox{23mm}{\unitlength=1mm\fmfframe(3,10)(2,10){
      \begin{fmfgraph*}(18,18)
        \fmfsurroundn{e}{10}
        \fmfv{d.sh=circle,d.f=30,d.si=10mm,
          label.angle=-97,label.dist=9mm,label=$\gamma_4^\Lambda$}{v}
        \fmfleft{i}\fmfright{o}\fmf{phantom}{i,v,o}\fmffreeze
        \fmf{plain}{v,e1}
        \fmf{plain}{v,e6}
        \fmf{plain}{v,e7}
        \fmf{plain}{v,e8}
        \fmf{plain}{v,e9}
        \fmf{plain}{v,e10}
        \fmf{slplaino,right=100,tension=0.4,label=$S^\Lambda$}{v,v}
      \end{fmfgraph*}}}
  +
  \parbox{47mm}{\unitlength=1mm\fmfframe(-4,6)(1,6){
      \begin{fmfgraph*}(50,18)
        \fmfleftn{l}{3}
        \fmfrightn{r}{4}
        \fmfv{d.sh=circle,d.f=30,d.si=10mm,
          label.angle=-90,label.dist=7mm,label=$\gamma_2^\Lambda$}{vl}
        \fmfv{d.sh=circle,d.f=30,d.si=10mm,
          label.angle=-90,label.dist=7mm,label=$\gamma_3^\Lambda$}{vr}
        \fmffixed{(.55w,0)}{vl,vr}
        \fmfleft{i}\fmfright{o}\fmf{phantom}{i,vl,vr,o}\fmffreeze
        \fmf{plain}{vl,l1}
        \fmf{plain}{vl,l3}
        \fmf{plain}{vr,r1}
        \fmf{plain}{vr,r2}
        \fmf{plain}{vr,r3}
        \fmf{plain}{vr,r4}
        \fmf{slplain,left=0.5,label=$S^\Lambda$}{vl,vr}
        \fmf{plain,right=0.5,label.side=left,label=$G^\Lambda$}{vl,vr}
      \end{fmfgraph*}}}
  +
  \parbox{30mm}{\unitlength=1mm\fmfframe(-3,0)(-2,0){
      \begin{fmfgraph*}(35,35)
        \fmfsurroundn{e}{24}
        \fmfv{d.sh=circle,d.f=30,d.si=10mm}{vl,vr,vu}
        \fmffixed{(.5w,0)}{vl,vr}
        \fmffixed{(.25w,.43h)}{vl,vu}
        \fmf{plain}{vu,e6}
        \fmf{plain}{vu,e8}
        \fmf{plain}{vl,e14}
        \fmf{plain}{vl,e16}
        \fmf{plain}{vr,e22}
        \fmf{plain}{vr,e24}
        \fmf{slplain,left=0.5,label=$S^\Lambda$}{vl,vu}
        \fmf{plain,left=0.5,label=$G^\Lambda$}{vu,vr}
        \fmf{plain,left=0.5,label=$G^\Lambda$}{vr,vl}
      \end{fmfgraph*}}} \notag
\end{align}
In the general form of equation \eqref{eq:1piflow}, the wiggly line in
the last diagram denotes that the expansion of the inverse second
derivative yields further 1-loop diagrams with an appropriate number
of higher vertices insertions, such that the number of external legs
on the \LHS\ and \RHS\ matches.  This flow scheme is derived in
\cite{Wei76,NC77,Wet93,BDM93,Mor94,SH01}.

\subsubsection*{Sharp-cutoff flow equations}

Consider the flow equation \eqref{eq:gammaflow2} in a basis where $Q$
is diagonal and the cutoff function $\chi^\Lambda$ is multiplicative
so that it can be taken out of $Q^\Lambda$,
\begin{align}
  \label{eq:gammachideriv}
  \dl\Gamma^\Lambda
  & = \Tr \, \inner{\dl\chi^\Lambda}{\frac{\delta}{\delta\chi^\Lambda}} 
  \ln \frac{Q}{Q+\chi^\Lambda \Gamma^{(2)}}.
\end{align}
Assume that all $\chi^\Lambda$ along the loop have the same value.
This is clearly not the general case but is sufficient in our
application in Chapter~\ref{sec:lutt} where $\chi^\Lambda$ is a sharp
frequency cutoff and all vertices are evaluated at zero frequencies,
such that all propagators in the loop have the same frequency and
hence the same value of the cutoff function.  Then $\dl\chi^\Lambda =
-\delta(\abs\omega-\Lambda)$ restricts the frequency to the $\Lambda$
shell but the \RHS\ contains $\Theta(\abs\omega-\Lambda)$ which has a
step right at the shell.  Such expressions are unique if the sharp
cutoff is implemented as the limit of increasingly sharp, broadened
cutoff functions $\Theta_\epsilon$ with broadening parameter
$\epsilon\to 0$.  This is demonstrated by a lemma due to Morris
\cite{Mor94}: for an arbitrary continuous function $f(t)$,
\begin{align}
  \label{eq:lemmamorris}
  \delta_\epsilon(x-\Lambda) \, f(\Theta_\epsilon(x-\Lambda)) 
  \to \delta_\epsilon(x-\Lambda) \, \int_0^1 f(t) \, dt
\end{align}
where $\delta_\epsilon(x) = \Theta_\epsilon'(x)$.  Then, equation
\eqref{eq:gammachideriv} reads
\begin{align}
  \dl \Gamma^\Lambda 
  & = \Tr \dot\chi^\Lambda \int_0^1 dt \, \frac{d}{dt}
  \ln \frac{Q}{Q+t\Gamma^{(2)}} \notag \\
  & = \Tr \dot\chi^\Lambda \ln \frac{Q}{Q+\Gamma^{(2)}} \notag \\
  & = \Tr \dot\chi^\Lambda \ln \frac{Q}{Q-\Sigma^\Lambda} \,
  \frac{Q-\Sigma^\Lambda}{Q-\Sigma^\Lambda+\tilde\Gamma^\Lambda} \notag \\
  & = \Tr -\dot\chi^\Lambda \left[ \ln(1-C\Sigma^\Lambda) +
    \ln(1+\tilde G^\Lambda \tilde\Gamma^\Lambda) \right] \notag \\
  \label{eq:Gshcutoff}
  & = \frac{1}{2\pi} \sum_{\omega=\pm\Lambda} \tr \Bigl[
    \ln(1-C\Sigma^\Lambda) - \sum_{k=1}^\infty \frac{(-1)^k}{k}
    (\tilde G^\Lambda \tilde\Gamma^\Lambda)^k \Bigr],
\end{align}
where the $T=0$ Matsubara sum $\frac{1}{2\pi} \int d\omega$ has been
performed in the last line such that $\tr$ denotes the sum over the
remaining spatial indices.  The new sharp-cutoff propagator is defined
as
\begin{align}
  \label{eq:sharpsscale}
  \tilde G^\Lambda := [Q - \Sigma^\Lambda]^{-1}.
\end{align}
This propagator has no step at $\abs\omega=\Lambda$ as opposed to
$G^\Lambda$, hence the \RHS\ of the flow equation which contains
one-loop terms built from powers of $\tilde G^\Lambda
\tilde\Gamma^\Lambda$ is a smooth function of $\Lambda$ and the
vertices $\tilde\Gamma^\Lambda$.

However, the sharp frequency cutoff is only possible if the loop
integral and hence the vertices $\tilde\Gamma^\Lambda$ are continuous:
at $T>0$, an imaginary frequency integral is restricted to discrete
Matsubara frequencies $\omega=\omega_n$ at which all vertices have a
step as a function of $\Lambda$, hence the condition that $f(t)$ be a
continuous function in Morris' lemma \eqref{eq:lemmamorris} is not
satisfied any more.


\subsection{Connected amputated Green function flow}
\label{sec:rg:flow:v}

The Polchinski scheme \cite{Pol84} has been introduced to prove the
renormalizability of the $\varphi^4$ theory via the flow of the
effective interaction, the generating functional of the connected
amputated Green functions (cf.\ section \ref{sec:rg:form:v}).  In the
presence of a cutoff, the effective interaction
$\V^\Lambda[\chi,\bar\chi]$ is defined as
\begin{equation}
  \label{eq:defvlambda}
  \begin{split}
    e^{-\V^\Lambda[\chi,\bar\chi]} & := 
    \int \dmu{Q^\Lambda}{\psi} \,
    e^{-V_0[\psi+\chi,\bar\psi+\bar\chi]} \\
    & = e^{\inner{\ddd\chi}{C^\Lambda \ddd{\bar\chi}}}\,
    e^{-V_0[\chi,\bar\chi]} \\
    & = e^{\Laplace{C^\Lambda}}\,
    e^{-V_0[\chi,\bar\chi]}\,.
  \end{split}
\end{equation}
Thus, all Feynman diagrams contain not $C$ but $C^\Lambda$ on the
internal lines.  In order to obtain the full $\V$ without cutoff, one
has to apply the missing $e^{\Laplace{D^\Lambda}}$ on
$e^{-\V^\Lambda}$:
\begin{align}
  e^{-\V[\chi,\bar\chi]}
  & = e^{\Laplace{C}}\,
  e^{-V_0[\chi,\bar\chi]} \notag \\
  & = e^{\Laplace{D^\Lambda+C^\Lambda}}\,
  e^{-V_0[\chi,\bar\chi]} \notag \\
  \label{eq:hardsoft}
  & = e^{\Laplace{D^\Lambda}}\,
  e^{\Laplace{C^\Lambda}}\,
  e^{-V_0[\chi,\bar\chi]} \\
  & = e^{\Laplace{D^\Lambda}}\,
  e^{-\V^\Lambda[\chi,\bar\chi]}\,. \notag
\end{align}
Equation \eqref{eq:hardsoft} can be interpreted as follows: if all
Feynman diagrams (connected and disconnected) with hard internal lines
are reconnected again in all possible ways with soft lines, one
obtains all diagrams with full lines.

The flow equation for $\V^\Lambda$ is derived by taking the $\Lambda$
derivative of equation \eqref{eq:defvlambda},
\begin{align*}
  \dl\V^\Lambda & = -e^{\V^\Lambda} \dl e^{-\V^\Lambda} \\
  & = -e^{\V^\Lambda} \dl (e^{\Laplace{C^\Lambda}} \, e^{-V_0}) \\
  & = -e^{\V^\Lambda} \Laplace{\dot C^\Lambda} e^{-\V^\Lambda},
\end{align*}
to arrive at the \emph{Polchinski flow equation}
\begin{align}
  \label{eq:vflow1}
  \boxed{
    \dl\V^\Lambda = \Tr \Bigl( \dot C^\Lambda\,
    \frac{\delta^2 \V^\Lambda} {\delta\chi\, \delta\bar\chi} \Bigr)
    - \inner{\frac{\delta\V^\Lambda}{\delta\chi}}
    {\dot C^\Lambda \frac{\delta\V^\Lambda}{\delta\bar\chi}}}\,.
\end{align}
In a more compact notation,
\begin{align}
  \label{eq:vflow}
  \dl\V^\Lambda & = \Laplace{\dot C^\Lambda} \V^\Lambda
  - \tfrac{1}{2} \Laplace{\dot C^\Lambda}^{12}
  \V^\Lambda[\chi_1,\bar\chi_1] \, \V^\Lambda[\chi_2,\bar\chi_2]
  \at_{\chi_1=\chi_2=\chi,\bar\chi_1=\bar\chi_2=\bar\chi}
\end{align}
where $\Laplace{C}^{12}=\inner{\ddd{\chi_1}}{C\ddd{\bar\chi_2}} +
\inner{\ddd{\chi_2}}{C\ddd{\bar\chi_1}}$ connects two \emph{different}
vertices.  The initial condition is the bare interaction:
\begin{align*}
  \V^{\Lambda_0}[\chi,\bar\chi]
  = V_0[\chi,\bar\chi]
  = \parbox{6mm}{\unitlength=1mm\begin{fmfgraph}(6,6)
      \fmfleft{i1,i2}
      \fmfright{i4,i3}
      \fmfdot{v}
      \fmf{plain}{i1,v}
      \fmf{plain}{i2,v}
      \fmf{plain}{i3,v}
      \fmf{plain}{i4,v}
    \end{fmfgraph}} \;\; \text{(for a two-particle interaction)}.
\end{align*}
The graphical representation of the Polchinski equation for the
connected amputated Green functions $V_m^\Lambda$ features both
tadpole and tree diagrams:
\begin{align*}
  \frac{\partial}{\partial\Lambda}
  \parbox{23mm}{\unitlength=1mm\fmfframe(3,0)(2,0){
      \begin{fmfgraph*}(18,18)
        \fmfsurroundn{e}{16}
        \fmfv{d.sh=circle,d.f=shaded,d.si=10mm,
          label.angle=-90,label.dist=11mm,label=$V_m^\Lambda$}{v}
        \fmfv{d.sh=circle,d.f=full,d.si=0.5mm}{e1,e2,e8,e9,e10,e16}
        \fmf{plain}{v,e3}
        \fmf{plain}{v,e5}
        \fmf{plain}{v,e7}
        \fmf{plain}{v,e11}
        \fmf{plain}{v,e13}
        \fmf{plain}{v,e15}
      \end{fmfgraph*}}}
  =
  \parbox{23mm}{\unitlength=1mm\fmfframe(3,12)(2,12){
      \begin{fmfgraph*}(18,18)
        \fmfsurroundn{e}{16}
        \fmfv{d.sh=circle,d.f=shaded,d.si=10mm,
          label.angle=-90,label.dist=11mm,label=$V_{m+1}^\Lambda$}{v}
        \fmfv{d.sh=circle,d.f=full,d.si=0.5mm}{e1,e2,e8,e9,e10,e16}
        \fmfleft{i}\fmfright{o}\fmf{phantom}{i,v,o}\fmffreeze
        \fmf{plain}{v,e3}
        \fmf{plain}{v,e7}
        \fmf{plain}{v,e11}
        \fmf{plain}{v,e13}
        \fmf{plain}{v,e15}
        \fmf{slplaino,right=100,tension=0.4,label=$\dot C^\Lambda$}{v,v}
      \end{fmfgraph*}}}
  + \sum_k
  \parbox{55mm}{\unitlength=1mm\fmfframe(3,0)(2,0){
      \begin{fmfgraph*}(50,18)
        \fmfleftn{l}{7}
        \fmfrightn{r}{7}
        \fmfv{d.sh=circle,d.f=shaded,d.si=10mm,
          label.angle=-90,label.dist=11mm,label=$V_k^\Lambda$}{vl}
        \fmfv{d.sh=circle,d.f=shaded,d.si=10mm,
          label.angle=-70,label.dist=12mm,label=$V_{m-k+1}^\Lambda$}{vr}
        \fmffixed{(.55w,0)}{vl,vr}
        \fmfleft{i}\fmfright{o}\fmf{phantom}{i,vl,vr,o}\fmffreeze
        \fmf{plain}{vl,l1}
        \fmf{plain}{vl,l3}
        \fmfv{d.sh=circle,d.f=full,d.si=0.5mm}{l4}
        \fmfv{d.sh=circle,d.f=full,d.si=0.5mm}{l5}
        \fmfv{d.sh=circle,d.f=full,d.si=0.5mm}{l6}
        \fmf{plain}{vl,l7}
        \fmf{plain}{vr,r1}
        \fmfv{d.sh=circle,d.f=full,d.si=0.5mm}{r2}
        \fmfv{d.sh=circle,d.f=full,d.si=0.5mm}{r3}
        \fmfv{d.sh=circle,d.f=full,d.si=0.5mm}{r4}
        \fmf{plain}{vr,r5}
        \fmf{plain}{vr,r7}
        \fmf{slplain,label.side=left,label=$\dot C^\Lambda$}{vl,vr}
      \end{fmfgraph*}}}  
\end{align*}


\subsection{Wick-ordered Green function flow}
\label{sec:rg:flow:wick}

The connected amputated Green functions are the expansion coefficients
of the generating functional $\V$ in terms of \emph{monomials} of the
source fields $\chi$, $\bar\chi$, whereas the \emph{Wick-ordered}
Green functions are the expansion coefficients in terms of
Wick-ordered \emph{polynomials} of the source fields,
$e^{\Laplace{D^\Lambda}} (\bar\chi \chi)^m$.  The construction is
completely analogous to the use of Hermite polynomials
$e^{\partial_x^2} x^m$ as compared to the monomials $x^m$: any
analytical function $f(x)$ can be expanded uniquely in terms of $x^m$
(Taylor expansion) or in terms of Hermite polynomials, which provide
another complete orthogonal basis.  In appendix~\ref{sec:app:heat} I
solve the heat equation in real space using Hermite polynomials; while
this is different from the usual textbook solution, it provides a
low-dimensional and intuitive example of the functional formalism.
The Wick-ordered generating functional is defined as
\cite{Wie88,Salmhofer:9706188,Salmhofer:1999,SH01}
\begin{align}
  \label{eq:Wdef}
  \W^\Lambda[\chi,\bar\chi]
  & = e^{\Laplace{D^\Lambda}}\, \V^\Lambda[\chi,\bar\chi].
\end{align}
Diagrammatically, the effect of Wick ordering is that the coefficients
$W_m^\Lambda$ of $\W^\Lambda$ contain all possible $D^\Lambda$ loops
(tadpoles) on the connected amputated Green functions $V_m^\Lambda$:
\begin{align*}
  \parbox{15mm}{\unitlength=1mm\fmfframe(0,0)(0,0){
      \begin{fmfgraph*}(15,15)
        \fmfsurroundn{e}{16}
        \fmfv{d.sh=circle,d.f=hatched,d.si=9mm,
          label.angle=-90,label.dist=11mm,label=$W_m^\Lambda$}{v}
        \fmfv{d.sh=circle,d.f=full,d.si=0.5mm}{e12,e13,e14}
        \fmf{plain}{v,e3}
        \fmf{plain}{v,e7}
        \fmf{plain}{v,e11}
        \fmf{plain}{v,e15}
      \end{fmfgraph*}}}
  & =
  \parbox{15mm}{\unitlength=1mm\fmfframe(0,0)(0,0){
      \begin{fmfgraph*}(15,15)
        \fmfsurroundn{e}{16}
        \fmfv{d.sh=circle,d.f=shaded,d.si=9mm,
          label.angle=-90,label.dist=11mm,label=$V_m^\Lambda$}{v}
        \fmfv{d.sh=circle,d.f=full,d.si=0.5mm}{e12,e13,e14}
        \fmf{plain}{v,e3}
        \fmf{plain}{v,e7}
        \fmf{plain}{v,e11}
        \fmf{plain}{v,e15}
      \end{fmfgraph*}}}
  +
  \parbox{19mm}{\unitlength=1mm\fmfframe(2,8)(2,8){
      \begin{fmfgraph*}(15,15)
        \fmfsurroundn{e}{16}
        \fmfv{d.sh=circle,d.f=shaded,d.si=9mm,
          label.angle=-90,label.dist=11mm,label=$V_{m+1}^\Lambda$}{v}
        \fmfv{d.sh=circle,d.f=full,d.si=0.5mm}{e12,e13,e14}
        \fmfleft{i}\fmfright{o}\fmf{phantom}{i,v,o}\fmffreeze
        \fmf{plain}{v,e1}
        \fmf{plain}{v,e9}
        \fmf{plain}{v,e11}
        \fmf{plain}{v,e15}
        \fmf{plain,right=1,tension=0.5,label=$D^\Lambda$}{v,v}
      \end{fmfgraph*}}}
  +
  \parbox{19mm}{\unitlength=1mm\fmfframe(2,0)(2,0){
      \begin{fmfgraph*}(15,15)
        \fmfsurroundn{e}{16}
        \fmfv{d.sh=circle,d.f=shaded,d.si=9mm,
          label.angle=-90,label.dist=11mm,label=$V_{m+2}^\Lambda$}{v}
        \fmfv{d.sh=circle,d.f=full,d.si=0.5mm}{e12,e13,e14}
        \fmfleft{i}\fmfright{o}\fmf{phantom}{i,v,o}\fmffreeze
        \fmf{plain}{v,e1}
        \fmf{plain}{v,e9}
        \fmf{plain}{v,e11}
        \fmf{plain}{v,e15}
        \fmf{phantom}{v,e5}
        \fmf{plain,right=1,tension=0.45,l.d=1mm,label=$D^\Lambda$}{v,v}
        \fmf{plain,right=100,tension=0.45,l.d=1mm,label=$D^\Lambda$}{v,v}
      \end{fmfgraph*}}}
  + \dotsb
\end{align*}
The flow equation for $\W^\Lambda$ can be derived from the Polchinski
flow equation \eqref{eq:vflow} with the replacement $\Laplace{\dot
  C^\Lambda} = -\Laplace{\dot D^\Lambda}$ (since
$C^\Lambda+D^\Lambda=C$ is independent of $\Lambda$),
\begin{align*}
  \dl \W^\Lambda & = (\dl e^{\Laplace{D^\Lambda}}) \, \V^\Lambda +
  e^{\Laplace{D^\Lambda}} \, \dl \V^\Lambda \\
  & = \Laplace{\dot D^\Lambda} e^{\Laplace{D^\Lambda}} \V^\Lambda
  + e^{\Laplace{D^\Lambda}} (-\Laplace{\dot D^\Lambda} \V^\Lambda) +
  e^{\Laplace{D^\Lambda}} (\tfrac{1}{2} \Laplace{\dot D^\Lambda}^{12}
  \V_1^\Lambda \V_2^\Lambda) \\
  & = \exp \, \{\Laplace{D^\Lambda}^{11+12+22}\} \; \tfrac{1}{2}
  \Laplace{\dot D^\Lambda}^{12} \, \V_1^\Lambda \, \V_2^\Lambda \\
  & = \tfrac{1}{2}
  (\Laplace{\dot D^\Lambda}^{12} e^{\Laplace{D^\Lambda}^{12}}) \,
  (e^{\Laplace{D^\Lambda}^{11}} \V_1^\Lambda) \,
  (e^{\Laplace{D^\Lambda}^{22}} \V_2^\Lambda).
\end{align*}
The first two terms in the second line cancel each other, and
$\Laplace{D^\Lambda}^{11+12+22} = \Laplace{D^\Lambda}^{11} +
\Laplace{D^\Lambda}^{12} + \Laplace{D^\Lambda}^{22}$ contains
derivatives with respect to the fields labeled $1$ and $2$,
respectively.  Finally, we obtain the \emph{Wick-ordered flow
  equation}
\begin{align}
  \label{eq:wickflow}
  \boxed{
    \dl\W^\Lambda = \tfrac{1}{2} \dl (e^{\Laplace{D^\Lambda}^{12}}) \,
    \W_1^\Lambda \, \W_2^\Lambda}\,.
\end{align}
The initial condition is
\begin{align*}
  \W^{\Lambda_0}[\chi,\bar\chi]
  = e^{\Laplace{C}}\, V_0[\chi,\bar\chi]
  = \parbox{6mm}{\unitlength=1mm\fmfframe(-2,0)(-2,0){\begin{fmfgraph}(10,10)
        \fmfleft{i}
        \fmfright{o}
        \fmfdot{v}
        \fmf{phantom}{i,v,o}
        \fmf{plain}{v,v}
        \fmf{plain,left=90}{v,v}
      \end{fmfgraph}}}
  + \parbox{14mm}{\unitlength=1mm\fmfframe(2,0)(2,0){\begin{fmfgraph}(10,10)
        \fmfleft{i}
        \fmfright{o}
        \fmfdot{v}
        \fmf{plain}{i,v,v,o}
      \end{fmfgraph}}}
  + \parbox{6mm}{\unitlength=1mm\begin{fmfgraph}(6,6)
      \fmfleft{i1,i2}
      \fmfright{i4,i3}
      \fmfdot{v}
      \fmf{plain}{i1,v}
      \fmf{plain}{i2,v}
      \fmf{plain}{i3,v}
      \fmf{plain}{i4,v}
    \end{fmfgraph}}
  \;\; \text{(for a two-particle interaction)}.
\end{align*}
The Wick-ordered functional $\W^\Lambda$ differs from $\V^\Lambda$
only for $\Lambda>0$: as $\Lambda\to 0$, $\W^\Lambda\to\W=\V$, \ie,
$\W^{\Lambda_0}$ has a different starting point (initial condition)
from $\V^{\Lambda_0}$ but flows via a different route to the same
final functional (cf.\ figure~\ref{fig:overview}).  The diagrammatic
representation of the flow equation for the components $W_m^\Lambda$
of $\W^\Lambda$ is
\begin{align*}
  \frac{\partial}{\partial\Lambda}
  \parbox{23mm}{\unitlength=1mm\fmfframe(3,0)(2,0){
      \begin{fmfgraph*}(18,18)
        \fmfsurroundn{e}{16}
        \fmfv{d.sh=circle,d.f=hatched,d.si=10mm,
          label.angle=-90,label.dist=11mm,label=$W_m^\Lambda$}{v}
        \fmfv{d.sh=circle,d.f=full,d.si=0.5mm}{e1,e2,e8,e9,e10,e16}
        \fmf{plain}{v,e3}
        \fmf{plain}{v,e5}
        \fmf{plain}{v,e7}
        \fmf{plain}{v,e11}
        \fmf{plain}{v,e13}
        \fmf{plain}{v,e15}
      \end{fmfgraph*}}}  
  = \sum_{k,j} \sum_{\text{permutations}}
  \parbox{55mm}{\unitlength=1mm\fmfframe(3,8)(2,8){
      \begin{fmfgraph*}(50,18)
        \fmfleftn{l}{7}
        \fmfrightn{r}{7}
        \fmfv{d.sh=circle,d.f=hatched,d.si=10mm,
          label.angle=-90,label.dist=11mm,label=$W_k^\Lambda$}{vl}
        \fmfv{d.sh=circle,d.f=hatched,d.si=10mm,
          label.angle=-70,label.dist=12mm,label=$W_{m-k+j}^\Lambda$}{vr}
        \fmffixed{(.55w,0)}{vl,vr}
        \fmfleft{i}\fmfright{o}\fmf{phantom}{i,vl,vr,o}\fmffreeze
        \fmfv{d.sh=circle,d.f=full,d.si=0.5mm}{l4,l5,l6,r2,r3,r4}
        \fmf{plain}{vl,l1}
        \fmf{plain}{vl,l3}
        \fmf{plain}{vl,l7}
        \fmf{plain}{vr,r1}
        \fmf{plain}{vr,r5}
        \fmf{plain}{vr,r7}
        \fmf{slplain,left=0.85,label=$\dot D^\Lambda$}{vl,vr}
        \fmf{plain,left=0.4}{vl,vr}
        \fmf{plain,left=-0.2}{vl,vr}
        \fmf{plain,left=-0.85,label.side=left,label.dist=2mm,
          label=$D^\Lambda$}{vl,vr}
        \fmfv{d.sh=circle,d.f=full,d.si=0.5mm}{i1,i2,i3}
        \fmfforce{(0.5w,0.48h)}{i1}
        \fmfforce{(0.5w,0.58h)}{i2}
        \fmfforce{(0.5w,0.68h)}{i3}
      \end{fmfgraph*}}}
\end{align*}
Note that all terms on the \RHS\ of the flow equation are bilinear in
the vertices and are either tree or higher loop diagrams.  In the
exact hierarchy without truncation, the flow equation of each vertex
has infinitely many terms on the \RHS\ with higher vertices connected
by many loops.  One internal line is the single-scale propagator $\dot
D^\Lambda$, while all others are \emph{soft}-mode propagators
$D^\Lambda$.  As the cutoff scale is lowered, only momenta in a small
neighborhood of the Fermi surface appear on the internal lines.  This
justifies a particularly efficient parametrization of the coupling
functions which depend on momenta anywhere in the Brillouin zone: at a
low cutoff scale, all internal momenta are close to the Fermi surface,
so it is sufficient to parametrize the couplings by their values with
all momenta projected onto the Fermi surface.  This parametrization is
employed in computations of the \twoD\ Hubbard model \cite{HM00,RM04}.


\section{Summary}
\label{sec:rg:sum}

In this chapter I have derived the functional renormalization-group
flow equations in the \onePI, Polchinski and Wick-ordered schemes.
While this is not new, I believe that the derivations are formally
simpler and more straightforward than in much of the literature.
Furthermore, the treatment of the sharp cutoff on the functional level
(cf.\ section \ref{sec:rg:flow:gamma}) has not yet been published to
my knowledge.

\begin{figure}[ht!]%
    \centerline{\input{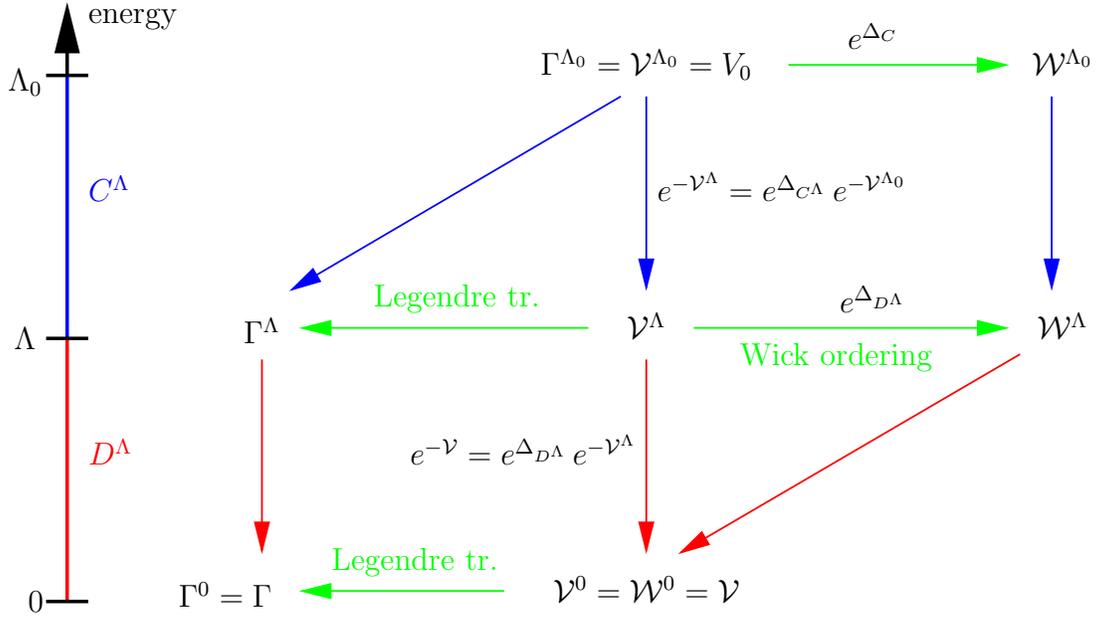}}
    \caption{\small\label{fig:overview}Overview over the relation between the
  different flow-equation schemes.}\end{figure}

Figure~\ref{fig:overview} provides an overview over the different
\fRG\ schemes: in the center is the \emph{Polchinski scheme} of
connected amputated Green functions $\V^\Lambda$ which have
essentially the same structure and flow equations as the connected
Green functions $\G^\Lambda$.  By Legendre transformation to
$\Gamma^\Lambda$ we have obtained the \emph{\onePI\ scheme} which
starts from the same initial condition but parametrizes the physical
properties in a different way particularly suited for symmetry
breaking.  It has the advantage that the internal lines are full
propagators taking into account all self-energy effects already known
at scale $\Lambda$, while it has the disadvantage that one has to
integrate the internal loop over the whole Brillouin zone.

On the other side, the Polchinski scheme leads to the
\emph{Wick-ordered} scheme $\W^\Lambda$ which starts from the
first-order Hartree-Fock solution as the initial value of the self
energy but converges towards the same connected amputated Green
functions $\V^\Lambda$ at the end of the flow.  Its main advantages
are that internal lines are restricted to a small neighborhood of the
Fermi surface, and that its simple power counting allows rigorous
proofs of renormalizability to all orders in perturbation theory.

The next difference arises with truncations.  While the full hierarchy
of flow equations in any scheme leads to the correct solution to all
orders in perturbation theory, the lowest orders in different basis
sets of correlation functions capture different aspects of the the
solution.  As an illustration, imagine a complicated real analytical
function $f(x)$, expand it in different sets of orthogonal polynomials
and retain only the first few coefficients: the approximate result
will look different for each basis set.  A third question is whether
the basis functions allow for efficient parametrizations in terms of a
small set of coefficients, for instance by projecting onto the Fermi
surface.

Hence there is not one single flow scheme superior for all
applications, but the method of choice depends on the particularities
of the model studied.



\chapter{Ward identities in the functional RG}
\label{sec:ward}

In constructing our microscopic model we demand that it satisfies
certain physical properties, such as charge conservation.  This
particular property is guaranteed by the requirement of $U(1)$
\emph{gauge invariance}\kern1pt: by the Noether theorem every continuous
symmetry has a conserved quantity associated with it, in this case the
electrical charge.  In a classical calculation, the conservation of
charge is expressed by the continuity equation
\begin{align*}
  \partial_t\, \rho(x) + \vec\nabla \cdot \vec{j}(x)
  \equiv \partial_\mu j^\mu(x)
  = 0
\end{align*}
where $\rho(x)$ and $\vec{j}(x)$ are the charge and current densities,
respectively.  In quantum theory the current $j^\mu(x)$ becomes an
operator $\hat j^\mu(x)$, and the continuity equation is evaluated
inside an expectation value together with insertions of other
operators.  If the time derivative acts not only on the current
operator but also on the time ordering in the expectation value, it
generates differences of Green functions \cite{IZ80}.  In momentum
space with momentum transfer $q$ to the external field, and in units
such that the electron charge $e=1$,
\begin{align}
  \label{eq:WIexample}
  \begin{split}
    q_\mu \vev{\hat j^\mu\, \psi_{p-q/2}\, \bar\psi_{p+q/2}}
    & = \vev{\psi_{p-q/2}\, \bar\psi_{p-q/2}}
    - \vev{\psi_{p+q/2}\, \bar\psi_{p+q/2}} \\
    & = G(p+q/2) - G(p-q/2) \\
    q_\mu
    \parbox{38mm}{\unitlength=1mm\fmfframe(4,0)(4,0){
        \begin{fmfgraph*}(30,20)
          \fmfsurroundn{e}{4}
          \fmfv{d.sh=circle,d.f=0,d.si=8mm}{v}
          \fmfforce{(15mm,10mm)}{v}
          \fmfv{d.sh=circle,d.f=full,d.size=0}{v3}
          \fmfforce{(15mm,23mm)}{v3}
          \fmffreeze
          \fmf{plain_arrow,label=$+$,label.side=right}{e3,v}
          \fmf{plain_arrow,label=$-$,label.side=right}{v,e1}
          \fmf{wiggly,label=$\mu$,label.side=right}{v,v3}
        \end{fmfgraph*}}}
    & =
    \parbox{32mm}{\unitlength=1mm\fmfframe(4,0)(4,0){
        \begin{fmfgraph*}(24,20)
          \fmfleft{i}
          \fmfright{o}
          \fmfv{d.sh=circle,d.f=0,d.si=8mm}{v}
          \fmf{plain_arrow,label=$+$,label.side=right}{i,v,o}
        \end{fmfgraph*}}}
    -
    \parbox{32mm}{\unitlength=1mm\fmfframe(4,0)(4,0){
        \begin{fmfgraph*}(24,20)
          \fmfleft{i}
          \fmfright{o}
          \fmfv{d.sh=circle,d.f=0,d.si=8mm}{v}
          \fmf{plain_arrow,label=$-$,label.side=right}{i,v,o}
        \end{fmfgraph*}}}.
  \end{split}
\end{align}
This is an example of a \emph{Ward identity} (\WI): these are
constraints on Green functions expressing the underlying symmetry
\cite{War50}.  For the one-particle Green functions equation
\eqref{eq:WIexample} can be easily derived by hand, but it will become
more tedious as more external legs are added, each corresponding to an
operator insertion in the expectation value.  Instead, we shall employ
the formalism introduced in Chapter~\ref{sec:rg} to derive a
functional Ward identity applicable to \emph{all} $m$-particle Green
functions simultaneously.  This will allow us to investigate general
properties of Ward identities to all orders in the fields and in the
interaction.

Let me mention one important point already now in order to avoid
confusion (the issue will be explained in more detail below): there
are two versions of the Ward identities in the literature, one
relating the difference of Green functions to the current as above
(\emph{response-function Ward identity}), the other relating it to the
two-particle interaction with a loop closed (\emph{self-consistent
  Ward identity}).  Both are equivalent in the exact solution and in
conserving approximations but generally differ in other approximations
such as truncated \fRG\ flows.  I will show that in truncated flows
the response-function Ward identities can still be satisfied in the
absence of a momentum cutoff, for instance in the temperature-flow
scheme, while self-consistency and thereby the self-consistent Ward
identities are in general violated.

We proceed as follows: after the functional derivation of Ward
identities in section \ref{sec:ward:gauge} we shall see how a momentum
cutoff breaks response-function Ward identities and how this can be
fixed in flow schemes without momentum cutoffs (section
\ref{sec:ward:cutoff}).  To answer the related question of
self-consistency (section \ref{sec:ward:sc}) we first review the
construction of self-consistent conserving approximations of Baym and
Kadanoff.  We then show that the self-consistent Ward identities are
generally violated by truncated flow equations.  The results are
summarized in section \ref{sec:ward:sum}.


\section{Gauge invariance and Ward identities}
\label{sec:ward:gauge}

We couple the fermionic action to an external field for two reasons:
it allows us to compute linear response by derivatives with respect to
the external field, and it is necessary to make the action satisfy the
desired symmetry.

In order to compute electrical transport properties, we need to study
the response of the system to the external electromagnetic potential.
This is done by including in the action a term coupling a current
operator $j^\mu(x)$ to the external field $A_\mu(x)$,
\begin{align}
  \label{eq:actionA}
  S[\psi,\bar\psi,A] := S[\psi,\bar\psi] - \inner{j^\mu}{A_\mu}
\end{align}
where the summation over $\mu$ is understood.  Then, the response
functions are derivatives of the Green functions with respect to
$A_\mu(x)$.  The current operator $j^\mu[\psi,\bar\psi,A]$ is a
composite operator of the fields $\psi$, $\bar\psi$ and generally
depends also on the external field $A$.  We shall now demand a
symmetry and construct the current such that this symmetry holds.


\subsection{Gauge transformation}
\label{sec:ward:gauge:trans}

The local $U(1)$ gauge transformation is \cite[chapter 12]{Zin02}
\begin{align}
  \label{eq:gaugetrans}
  \left\{
    \begin{aligned}
      & \psi'(x) := e^{-i\alpha(x)} \psi(x)
      & & \delta_\alpha \psi(x) = -i\alpha(x) \psi(x) \\
      & \bar\psi'(x) := e^{i\alpha(x)} \bar\psi(x)
      & & \delta_\alpha \bar\psi(x) = i\alpha(x) \bar\psi(x) \\
      & A_\mu'(x) := A_\mu(x) + \partial_\mu \alpha(x)
      & \qquad & \delta_\alpha A_\mu(x) = \partial_\mu \alpha(x),
    \end{aligned}
  \right.
\end{align}
where $\alpha(x)$ is a real function, and $\delta_\alpha$ an operator
which acts on fields and functionals by performing an infinitesimal
gauge transformation.  The condition of gauge invariance of our model
is
\begin{align*}
  \delta_\alpha \left( [d\psi\bar\psi]\; e^{S[\psi,\bar\psi,A]}
  \right) \overset{!}{=} 0
\end{align*}
\emph{without} the integral, \ie, the path integral measure times the
weight should stay invariant.  In the case of the $U(1)$ gauge
transformation \eqref{eq:gaugetrans} the measure remains invariant,
therefore the gauge invariance reduces to a condition on the action,
\begin{align*}
  \delta_\alpha S[\psi,\bar\psi,A] \overset{!}{=} 0.
\end{align*}
Consider first the gauge variation of the fermionic action,
$\delta_\alpha S[\psi,\bar\psi]$.  If $\alpha(x) = \mathit{const}$
this transformation probes \emph{global} charge conservation: this is
satisfied as long as each term in the action has the same number of
fermion annihilation and creation operators.  But if we admit an
$\alpha(x)$ varying in space-time, the action will in general change
proportionally to $\partial_\mu\alpha(x)$,
\begin{align}
  \label{eq:actionferm}
  \delta_\alpha S[\psi,\bar\psi]
  = \inner{j^\mu}{\partial_\mu \alpha} + \Ord(\alpha^2),
\end{align}
where we have called the coefficient of the $\partial_\mu\alpha$ term
$j^\mu$.  This is the motivation to introduce the gauge field $A_\mu$:
if it is coupled to $j^\mu$ by $\inner{j^\mu}{A_\mu}$, the gauge
variation of the $A_\mu$ term will cancel the gauge variation of the
fermionic action \eqref{eq:actionferm} to first order in $\alpha$.

This raises the question how to construct the current operator $j^\mu$
in the action \eqref{eq:actionA}.  In the continuum this is achieved
by the procedure of minimal coupling.  On the lattice, one has to use
lattice gauge theory which is technically quite different (for an
introduction, see for instance \cite[chapter 34]{Zin02}).  Note that
$(i)$ the general arguments about symmetries and Ward identities can
be seen already in the technically simpler continuum model, and $(ii)$
the lattice formalism will be used in this work only in the special
case of a \oneD\ lattice where the field-strength tensor vanishes.
Therefore, I derive the Ward identities pedagogically only for the
continuum case but give the final results also specifically for the
\oneD\ lattice (in section \ref{sec:lutt:condtech}) after the lattice
formulation has been introduced in Chapter~\ref{sec:lutt}.


\subsection{Minimal coupling in the continuum}
\label{sec:ward:gauge:coupl}

One way to construct a gauge-invariant continuum action is by minimally
coupling the gauge field to the fermions: in the quadratic part of
the action we replace the partial derivative $\partial_\mu$ by the
covariant derivative
\begin{align*}
  D_\mu & := \partial_\mu + iA_\mu(x) ,
\end{align*}
such that
\begin{align*}
  D_\mu' & = e^{-i\alpha} D_\mu e^{i\alpha},
  & \delta_\alpha D_\mu & = i\partial_\mu \alpha.
\end{align*}
Likewise, the inverse bare propagator $Q(\partial_\mu)$ is replaced by
$Q[A] \equiv Q(D_\mu)$ which is also gauge covariant,
\begin{align}
  \label{eq:Dcov}
  Q(D_\mu') & = e^{-i\alpha} Q(D_\mu) e^{i\alpha},
\end{align}
such that the kinetic term is gauge invariant,
\begin{align*}
  \inner{\bar\psi'}{Q(D_\mu')\psi'} & = \inner{\bar\psi}{Q(D_\mu)\psi},
  & \delta_\alpha \inner{\bar\psi}{Q(D_\mu)\psi} & = 0.
\end{align*}
Then, the current defined as the $A$ dependence of the kinetic term,
\begin{align*}
  \inner{\bar\psi}{Q(D_\mu)\psi} =:
  \inner{\bar\psi}{Q(\partial_\mu)\psi} - \inner{j_0^\mu}{A_\mu}
\end{align*}
makes the quadratic part of the action gauge invariant.  Expanding
both sides of equation \eqref{eq:Dcov} to first order in $\alpha$,
\begin{align}
  \label{eq:proptrans}
  \inner{\delta_\alpha A_\mu}{\frac{\delta Q[A]}{\delta iA_\mu}}
  & = \inner{\alpha}{\partial_\mu\frac{\delta Q[A]}{\delta iA_\mu}}
  = \bigl[\alpha,Q[A]\bigr]
\end{align}
where the commutator $\alpha(x) Q(x,y) - Q(x,y) \alpha(y) =
[\alpha(x)-\alpha(y)] Q(x,y)$, and we have integrated by
parts.\footnote{In this chapter we assume boundary conditions such
  that the integration by parts yields no boundary term.}

Consider now the interaction term.  The density-density interaction is
explicitly gauge invariant,
\begin{align*}
  \delta_\alpha V_0 & [\psi,\bar\psi]
  = \delta_\alpha \int dx\, dy\, V(x-y) \bar\psi(x) \bar\psi(y) \psi(y)
  \psi(x) \\
  & = i \int dx\, dy\, V(x-y) \bar\psi(x) \bar\psi(y) \psi(y)
  \psi(x) \, [\alpha(x) + \alpha(y) - \alpha(y) - \alpha(x)] \\
  & = 0.
\end{align*}
On the other hand, for a bare interaction which itself is not gauge
invariant ($\delta_\alpha V_0 \neq 0$) some further counter term
$\inner{j_V^\mu} {A_\mu} := -\delta_\alpha V_0$ is needed to make the
action gauge invariant.  Combining both the kinetic and the
interaction parts of the current operator,
\begin{align*}
  j^\mu(x) := j_0^\mu(x) + j_V^\mu(x),
\end{align*}
the action \eqref{eq:actionA} will be gauge invariant.

\subsubsection*{Example for quadratic dispersion}

Let us give a specific example: for nonrelativistic fermions of mass
$m$ in the continuum, the dispersion relation is linear in time and
quadratic in space, $Q(\partial_\mu) = -\partial_t +\vec\nabla^2/2m$.
We assume a gauge-invariant interaction $V_0$.  Minimal coupling to
the external electromagnetic potential $A_\mu(x) = (\varphi,-\vec{A})$
is given in real space by $\partial_t \mapsto \partial_t + \varphi(x)$,
$-i\vec{\nabla} \mapsto -i\vec\nabla - \vec A(x)$, hence the current
which makes the action gauge invariant is constructed as
\begin{align}
  \label{eq:nonreldisp}
  S[\psi,\bar\psi,A] & = \int dx\, \bar\psi(x)
  \left[ -\partial_t - \varphi -
    \tfrac{1}{2m} (-i\vec\nabla - \vec A)^2 \right] \psi(x)
  -V_0[\psi,\bar\psi] \\
  & \overset != \inner{\bar\psi}{Q\psi} - \inner{j^\mu}{A_\mu}
  - V_0[\psi,\bar\psi] \notag \\
  & \Rightarrow \; j^0 = n, \quad
  \vec j = \underbrace{\Bigl.\tfrac{1}{2mi}
    \left[ \bar\psi(\vec\nabla \psi) -
      (\vec\nabla \bar\psi)\psi\right]\Bigr.}
  _{\text{paramagnetic}}
  - \underbrace{\Bigl.\tfrac{1}{2m} n \vec A\Bigr.}
  _{\text{diamagn.}} \notag
\end{align}
with the density operator $n(x):=\bar\psi(x) \psi(x)$.  The current
$\vec j$ has a part independent of $A$ which is called the
paramagnetic current and a part proportional to $A$ which is called
the diamagnetic current.


\subsection{Functional derivation of Ward identities}
\label{sec:ward:gauge:func}

The current operator $j^\mu[\psi,\bar\psi,A]$ constructed in this way
is coupled to an external field $A_\mu(x)$.  Thereby, all generating
functionals depend on $A$ as a parameter,
\begin{align*}
  \Z[\eta,\bar\eta,A]
  & := \int \dmu{Q}{\psi}\,
  e^{-\inner{j^\mu}{\,A_\mu} - V_0[\psi,\bar\psi]
    - \inner{\bar\psi}{\eta} - \inner{\bar\eta}{\psi}}
\end{align*}
with measure \eqref{eq:measure}.  The source fields $\eta$, $\bar\eta$
transform under gauge transformations as
\begin{align}
  \label{eq:gaugetranseta}
  \delta_\alpha \eta(x) & := -i\alpha(x) \eta(x), &
  \delta_\alpha \bar\eta(x) & := i\alpha(x) \bar\eta(x).
\end{align}
Then, the gauge variation of $\Z$ to first order in $\alpha$,
\begin{align}
  \delta_\alpha \Z[\eta,\bar\eta,A]
  & = \left\{
    \inner{\delta_\alpha \eta} {\ddd{\eta}} +
    \inner{\delta_\alpha \bar\eta} {\ddd{\bar\eta}} +
    \inner{\delta_\alpha A_\mu} {\ddd{A_\mu}}
  \right\} \Z[\eta,\bar\eta,A] \notag \\
  \label{eq:zward}
  & = \left\{
    \inner{-i\alpha \eta} {\ddd{\eta}} +
    \inner{i\alpha \bar\eta} {\ddd{\bar\eta}} +
    \inner{\partial_\mu\alpha} {\ddd{A_\mu}}
  \right\} \Z[\eta,\bar\eta,A] \notag \\
  & = 0,
\end{align}
vanishes because $\Z$ is gauge invariant by construction.
Substituting $\Z=e^{-\G}$ we obtain the functional \emph{Ward identity
  for the connected Green functions},
\begin{align}
  \label{eq:gward}
  \boxed{
    \inner{\alpha}{\partial_\mu\frac{\delta\G}{\delta iA_\mu}}
    = \inner{\frac{\delta\G}{\delta\eta}}{\alpha\eta}
    + \inner{\bar\eta}{\alpha\frac{\delta\G}{\delta\bar\eta}}}\,.
\end{align}
By the Legendre transformation \eqref{eq:gammadef} in the presence of
the external field,
\begin{align}
  \label{eq:legA}
  \Gamma[\phi,\bar\phi,A] + \inner{\bar\phi}{Q[A]\phi}
  := \G[\eta,\bar\eta,A] +\inner{\bar\phi}{\eta} -\inner{\bar\eta}{\phi},
\end{align}
we rewrite equation \eqref{eq:gward} in terms of the \onePI\ 
functional $\Gamma[\phi,\bar\phi,A]$,
\begin{align*}
  \inner{\alpha}
  {\partial_\mu\frac{\delta}{\delta iA_\mu}
    \{\Gamma+\inner{\bar\phi}{Q[A]\phi}\}}
  & =
  \inner{\bar\phi}{\alpha\Bigl\{\frac{\delta\Gamma}{\delta\bar\phi}+
        Q[A]\phi\Bigr\}}
  +\inner{\Bigl\{\frac{\delta\Gamma}{\delta\phi}-
        Q^t[A]\bar\phi\Bigr\}}{\alpha\phi} \\
  & =
  \inner{\frac{\delta\Gamma}{\delta\phi}}{\alpha\phi}
  +\inner{\bar\phi}{\alpha\frac{\delta\Gamma}{\delta\bar\phi}}
  +\inner{\bar\phi}{\bigl[\alpha,Q[A]\bigr]\phi}.
\end{align*}
By equation \eqref{eq:proptrans} we obtain the \emph{\onePI\ Ward
  identity}
\begin{align}
  \label{eq:gammaward}
  \boxed{
    \inner{\alpha}{\partial_\mu\frac{\delta\Gamma}{\delta iA_\mu}}
    =\inner{\frac{\delta\Gamma}{\delta\phi}}{\alpha\phi}
    +\inner{\bar\phi}{\alpha\frac{\delta\Gamma}{\delta\bar\phi}}}\,.
\end{align}
By amputating bare propagators $C$
(without external field) from the external legs of equation
\eqref{eq:gward} using \eqref{eq:VGrel} we obtain
\begin{align*}
  \inner{\alpha}{\partial_\mu\frac{\delta\V}{\delta iA_\mu}}
  &=\inner{\left[C^t\frac{\delta\V}{\delta\chi}-\bar\chi\right]}{\alpha Q\chi}
 +\inner{Q^t\bar\chi}{\alpha\left[C\frac{\delta\V}{\delta\bar\chi}+\chi\right]}
\end{align*}
and by rearranging terms the \emph{Polchinski Ward identity}
\begin{align}
  \label{eq:vward}
  \boxed{
    \inner{\alpha}{\partial_\mu\frac{\delta\V}{\delta iA_\mu}}
    =\inner{\frac{\delta\V}{\delta\chi}}{C\alpha Q\chi}
    +\inner{\bar\chi}{Q\alpha C\frac{\delta\V}{\delta\bar\chi}}
    -\inner{\bar\chi}{[\alpha,Q]\chi}}\,.
\end{align}
On the \RHS\ of equation \eqref{eq:vward} the non-amputated external
legs are shifted.


\subsection{Momentum- and real-space formulation}
\label{sec:ward:gauge:momreal}

All of the above Ward identities still contain an inner product with
the arbitrary gauge-transformation parameter $\alpha_{-q}$.  In
Fourier space the transformation \eqref{eq:gaugetranseta} reads
\begin{align}
  \label{eq:sourcegt}
  \begin{split}
    \eta_k' & = \eta_k - i\sum_q \alpha_{-q} \eta_{k+q} + \Ord(\alpha^2) \\
    \bar\eta_k' & = \bar\eta_k + i\sum_q \alpha_{-q} \bar\eta_{k-q} +
    \Ord(\alpha^2)
  \end{split}
\end{align}
where the mode $\alpha_{-q}$ subtracts momentum $q$ from the fermion.
The coefficient of $\alpha_{-q}$ in the Ward identity for the
connected Green functions, equation \eqref{eq:gward}, is for example
\begin{align*}
  q_\mu \frac{\delta\G}{\delta A_\mu(-q)}
  & = \sum_k \left( 
    \frac{\delta\G}{\delta\eta_{k_-}} \eta_{k_+}
    +\bar\eta_{k_-} \frac{\delta\G}{\delta\bar\eta_{k_+}}
  \right)
\end{align*}
where the indices $k_\pm$ denote $k\pm q/2$.  The Ward identities for
the other functionals has an analogous form.  Taking only the
one-particle component of this functional Ward identity by applying
$\frac{\delta^2}{\delta\eta_{k_+} \delta\bar\eta_{k_-}}\vert
_{\eta=\bar\eta=0}$ we obtain the Ward identity given as an example in
equation \eqref{eq:WIexample}.  Generally, the Ward identities can be
represented diagrammatically as
\begin{align}
  \label{eq:gwarddiag}
  \parbox{28mm}{\unitlength=1mm\fmfframe(8,12)(2,12){
      \begin{fmfgraph*}(18,18)
        \fmfsurroundn{e}{16}
        \fmfv{d.sh=circle,d.f=0,d.si=7mm,
          label.angle=-90,label.dist=17mm,
          label=$q_\mu\tfrac{\delta G_m^\Lambda}{\delta A_\mu}$}{v}
        \fmfv{d.sh=circle,d.f=full,d.si=0.5mm}{e4,e6,e12,e14}
        \fmfleft{i}\fmfright{o}\fmf{phantom}{i,v,o}\fmffreeze
        \fmf{plain_arrow}{v,e3}\fmflabel{$k_m'$}{e3}
        \fmf{plain_arrow}{v,e5}\fmflabel{$k_i'$}{e5}
        \fmf{plain_arrow}{v,e7}\fmflabel{$k_1'$}{e7}
        \fmf{plain_arrow}{e11,v}\fmflabel{$k_1$}{e11}
        \fmf{plain_arrow}{e13,v}\fmflabel{$k_i$}{e13}
        \fmf{plain_arrow}{e15,v}\fmflabel{$k_m$}{e15}
        \fmf{wiggly}{v,e9}\fmflabel{$q_\mu$}{e9}
      \end{fmfgraph*}}}
  = \sum_i \left(
    \parbox{28mm}{\unitlength=1mm\fmfframe(5,12)(5,12){
        \begin{fmfgraph*}(18,18)
          \fmfsurroundn{e}{16}
          \fmfv{d.sh=circle,d.f=0,d.si=7mm,
            label.angle=-90,label.dist=18mm,label=$G_m^\Lambda$}{v}
          \fmfv{d.sh=circle,d.f=full,d.si=0.5mm}{e4,e6,e12,e14}
          \fmfleft{i}\fmfright{o}\fmf{phantom}{i,v,o}\fmffreeze
          \fmf{plain_arrow}{v,e3}\fmflabel{$k_m'$}{e3}
          \fmf{plain_arrow}{v,e5}\fmflabel{$k_i'+q$}{e5}
          \fmf{plain_arrow}{v,e7}\fmflabel{$k_1'$}{e7}
          \fmf{plain_arrow}{e11,v}\fmflabel{$k_1$}{e11}
          \fmf{plain_arrow}{e13,v}\fmflabel{$k_i$}{e13}
          \fmf{plain_arrow}{e15,v}\fmflabel{$k_m$}{e15}
        \end{fmfgraph*}}}
    -
    \parbox{28mm}{\unitlength=1mm\fmfframe(5,12)(5,12){
        \begin{fmfgraph*}(18,18)
          \fmfsurroundn{e}{16}
          \fmfv{d.sh=circle,d.f=0,d.si=7mm,
            label.angle=-90,label.dist=18mm,label=$G_m^\Lambda$}{v}
          \fmfv{d.sh=circle,d.f=full,d.si=0.5mm}{e4,e6,e12,e14}
          \fmfleft{i}\fmfright{o}\fmf{phantom}{i,v,o}\fmffreeze
          \fmf{plain_arrow}{v,e3}\fmflabel{$k_m'$}{e3}
          \fmf{plain_arrow}{v,e5}\fmflabel{$k_i'$}{e5}
          \fmf{plain_arrow}{v,e7}\fmflabel{$k_1'$}{e7}
          \fmf{plain_arrow}{e11,v}\fmflabel{$k_1$}{e11}
          \fmf{plain_arrow}{e13,v}\fmflabel{$k_i-q$}{e13}
          \fmf{plain_arrow}{e15,v}\fmflabel{$k_m$}{e15}
        \end{fmfgraph*}}}
  \right)
  \equiv
  \parbox{24mm}{\unitlength=1mm\fmfframe(3,8)(3,8){
      \begin{fmfgraph*}(18,18)
        \fmfsurroundn{e}{16}
        \fmfv{d.sh=circle,d.f=0,d.si=7mm,
          label.angle=-90,label.dist=13mm,label=$\Delta G_m^\Lambda$}{v}
        \fmfv{d.sh=circle,d.f=full,d.si=0.5mm}{e4,e6,e12,e14}
        \fmfleft{i}\fmfright{o}\fmf{phantom}{i,v,o}\fmffreeze
        \fmf{dots}{v,e3}
        \fmf{dots}{v,e5}
        \fmf{dots}{v,e7}
        \fmf{dots}{v,e11}
        \fmf{dots}{v,e13}
        \fmf{dots}{v,e15}
      \end{fmfgraph*}}}
\end{align}
where the wiggly line on the \LHS\ denotes the response function of
which the divergence is taken.  As momentum $q$ is transferred to the
external field, momentum conservation implies $k_1 + \dotsb + k_m = q
+ k_1' + \dotsb + k_m'$.  The Green functions in the large parentheses
are without external field, so any one of the external legs has to be
momentum-shifted to satisfy momentum conservation.  The rightmost
diagram with the dotted external legs shall abbreviate all the
combinations of momentum-shifted external lines in the middle.

Generally, for the case of \emph{global} gauge invariance there is
only the $q=0$ mode of $\alpha_{-q}$ and the Ward identity
\begin{align}
  \label{ggtglobal}
  \inner{\eta}{\frac{\delta\G}{\delta\eta}}
  = \inner{\bar\eta}{\frac{\delta\G}{\delta\bar\eta}}
\end{align}
implies only that there are as many creation as annihilation operators
in every monomial in the action, independent of their momenta or
positions.

In real space, the coefficient of $\alpha(x)$ in equation
\eqref{eq:gward} fulfills the Ward identity
\begin{align*}
  \partial_\mu \frac{\delta \G}{\delta iA_\mu(x)}
  = \frac{\delta \G}{\delta \eta(x)} \eta(x)
  + \bar\eta(x) \frac{\delta \G}{\delta \bar\eta(x)}.
\end{align*}


\section{Cutoff Ward identities}
\label{sec:ward:cutoff}

Generally, a momentum cutoff breaks local gauge invariance.  The cutoff
divides the fields $\psi$ into high and low modes.  However, a local
gauge transformation $\alpha(x)$ which is multiplicative in real
space, $\delta_\alpha \psi(x) \sim \alpha(x) \psi(x)$ from equation
\eqref{eq:gaugetrans}, becomes a convolution in momentum space,
$\delta_\alpha \psi_k \sim \sum_q \alpha_{-q} \psi_{k+q}$ from
equation \eqref{eq:sourcegt}.  This shift of the momenta spoils the
division of modes and poses a problem when treating gauge theories
using flow equations.  These problems would be solved if we could
somehow define a gauge-invariant cutoff propagator $Q^\Lambda[A]$.
However, in gauge theories, where one integrates over fluctuations of
the gauge field $A$ in the path integral, this is difficult except for
pure gauge theories \cite{DM96}.  Note that these problems are absent
for global gauge invariance (no momentum shift) and for the
temperature and interaction flow schemes which have no momentum
cutoff.

There are two ways in the high-energy physics literature to deal with
this problem: $(i)$ One can give up gauge invariance at intermediate
cutoff scales but try to ensure that the full Ward identities are
recovered as the cutoff is finally removed ($\Lambda\to 0$); this
approach of modified Ward identities is explained in section
\ref{sec:ward:cutoff:mWI}.  $(ii)$ One can manifestly satisfy Ward
identities in the presence of a cutoff with the help of an auxiliary
field, the external background gauge field $\bar A$ (section
\ref{sec:ward:cutoff:bg}).  Both approaches become cumbersome if one
has to truncate the flow-equation hierarchy.  However, if one uses a
manifestly gauge-invariant formulation (section
\ref{sec:ward:cutoff:ginv}) such as the temperature-flow scheme, we
show that the Ward identities can be satisfied exactly even in
truncated flows (section \ref{sec:ward:cutoff:trunc}).


\subsection{Modified Ward identities}
\label{sec:ward:cutoff:mWI}

The Ward identities \eqref{eq:gward} which hold without cutoff are
broken at intermediate cutoff scales ($\Lambda>0$) by modification
terms, leading to \emph{modified Ward identities} (\mWI).  These
modification terms vanish in the limit $\Lambda\to 0$.  Compatibility
of flow and \mWI\ ensures that the full hierarchy of flow equations
satisfies the \mWI\ at all scales $\Lambda$.  This is sufficient for
proving perturbative renormalizability of \abb{QED}; the \mWI\ for the
connected amputated Green functions were derived in \cite{KK91,KK96}
and used to prove bounds for the full hierarchy of Green functions
(see also \cite{BDM94}, and \cite{Ell94} for \mWI\ in the \onePI\ 
scheme).  For practical computations, however, the flow equations need
to be truncated, and the \mWI\ (and ultimately the original Ward
identities at $\Lambda=0$) are satisfied only to truncation order.
Alternatively, if there are only a few relevant components of the
flowing vertices, one can determine some of them not by the flow but
by the \mWI\ at every scale $\Lambda$, thus satisfying the \mWI\ 
exactly \cite{EHW96}.


\subsubsection*{Modified Ward identities for connected Green functions}

The definition of the cutoff connected Green functions
\eqref{eq:glambda} with the addition of the external field can be
split into a gauge-invariant part (without cutoff) and a part
containing the cutoff function,
\begin{align}
  \label{eq:glsplit}
  e^{-\G^\Lambda[\eta,\bar\eta,A]}
  = \int \frac{[d\psi\bar\psi]}{\det Q} \,
  e^{\inner{\bar\psi}{Q[A]\psi} -
    V_0[\psi,\bar\psi] - \inner{\bar\psi}{\eta} -
    \inner{\bar\eta}{\psi}} \times
  \frac{\det Q}{\det Q^\Lambda} \,
  e^{\inner{\bar\psi}{[Q^\Lambda-Q]\psi}}.
\end{align}
Performing the gauge transformation \eqref{eq:gaugetrans},
\eqref{eq:gaugetranseta} on all fields, the first part remains
invariant while the latter gives to first order in $\alpha$ a term
involving $[Q^\Lambda-Q]$ which breaks gauge invariance at
$\Lambda>0$,
\begin{align}
  \delta_\alpha e^{-\G^\Lambda[\eta',\bar\eta',A']}
  & = \int \frac{[d\psi\bar\psi]}{\det Q^\Lambda} \,
  e^{\inner{\bar\psi}{Q[A]\psi} -
    V_0[\psi,\bar\psi] - \inner{\bar\psi}{\eta} -
    \inner{\bar\eta}{\psi}} \\
  & \quad \times 
  \inner{\bar\psi}{i[\alpha,Q^\Lambda-Q]\psi} \,
  e^{\inner{\bar\psi}{[Q^\Lambda-Q]\psi}} \notag \\
  \label{eq:gltrans1}
  & = -\inner{\ddd\eta}{iX^\Lambda\ddd{\bar\eta}}
  e^{-\G^\Lambda[\eta,\bar\eta,A]} \,.
\end{align}
The $\eta$, $\bar\eta$ derivatives connect two legs of
$e^{-\G^\Lambda}$ with a propagator
\begin{align*}
  X^\Lambda := [\alpha,Q^\Lambda-Q]
\end{align*}
which vanishes as $\Lambda\to 0$ or for homogeneous $\alpha(x) \equiv
\alpha$.  On the other hand, expanding
$\G^\Lambda[\eta',\bar\eta',A']$ in the arguments to first order in
$\alpha$ as in equation \eqref{eq:zward} yields
\begin{align}
  \label{eq:gltrans2}
  \delta_\alpha \G^\Lambda[\eta',\bar\eta',A']
  & = \left\{ \inner{-i\alpha \eta} {\ddd{\eta}} +
    \inner{i\alpha \bar\eta} {\ddd{\bar\eta}} +
    \inner{\partial_\mu\alpha} {\ddd{A_\mu}}
  \right\} \G^\Lambda[\eta,\bar\eta,A].
\end{align}
We combine \eqref{eq:gltrans1} and \eqref{eq:gltrans2} into the
\emph{modified Ward identity}
\begin{align}
  \label{eq:gmward}
  \boxed{
    \inner{\alpha}{\partial_\mu\frac{\delta\G^\Lambda}{\delta iA_\mu}}
    - \Tr \left( X^\Lambda \frac{\delta^2 \G^\Lambda}{\delta\eta \,
      \delta\bar\eta} \right) +
    \inner{\frac{\delta\G^\Lambda}{\delta\eta}}{X^\Lambda
      \frac{\delta\G^\Lambda}{\delta\bar\eta}}
    = \inner{\frac{\delta\G^\Lambda}{\delta\eta}}{\alpha\eta}
      +\inner{\bar\eta}{\alpha\frac{\delta\G^\Lambda}{\delta\bar\eta}}}
\end{align}
which agrees with the usual Ward identity \eqref{eq:gward} in the
limit $\Lambda\to 0$ ($X^\Lambda\to 0$).  The modification terms have
the same structure as the \RHS\ of the flow equation \eqref{eq:gflow}
with the replacement $\dot Q^\Lambda \mapsto X^\Lambda$.


\subsubsection*{Modified Ward identities for 1PI vertex functions}

Using the Legendre transformation with cutoff \eqref{eq:leglambda} we
can rewrite equation \eqref{eq:gmward} as
\begin{align*}
  & \inner{\alpha}{\partial_\mu\frac{\delta\Gamma^\Lambda}{\delta iA_\mu}}
  - \Tr\bigl(X^\Lambda \bigl[Q^\Lambda+\Gamma^{(2)}\bigr]^{-1}\bigr)
  + \inner{\bar\phi}{X^\Lambda\phi} \\
  & = \inner{\frac{\delta\Gamma^\Lambda}{\delta\phi}}{\alpha\phi}
    + \inner{\bar\phi}{\alpha\frac{\delta\Gamma^\Lambda}{\delta\bar\phi}}
    + \inner{\bar\phi}{[\alpha,Q^\Lambda]\phi}
\end{align*}
and finally obtain the \emph{\onePI\ modified Ward identity}
\begin{align*}
  \boxed{
    \Bigl(\alpha,\partial_\mu\frac{\delta\Gamma^\Lambda}{\delta iA_\mu}\Bigr)
    - \Tr \bigl( X^\Lambda \bigl[Q^\Lambda+\Gamma^{(2)}\bigr]^{-1}
    \bigr)
    = \Bigl(\frac{\delta\Gamma^\Lambda}{\delta\phi},\alpha\phi\Bigr)
    +\Bigl(\bar\phi,\alpha\frac{\delta\Gamma^\Lambda}{\delta\bar\phi}\Bigr)
    +\inner{\bar\phi}{[\alpha,Q]\phi} }\,.
\end{align*}
The modification term is due to the inverse bare propagator $Q^\Lambda
- Q + Q[A]$ in equation \eqref{eq:glsplit}, where the cutoff acts only
on the $A$-independent part, while the $A$ dependence is treated as
part of the interaction (self energy), $Q^\Lambda -
\Sigma^\Lambda[A]$.  The modified Ward identities for the \onePI\ 
vertex functions $\Gamma^\Lambda$ may be written diagrammatically as
\begin{align*}
  \parbox{28mm}{\unitlength=1mm\fmfframe(8,8)(2,8){
      \begin{fmfgraph*}(18,18)
        \fmfsurroundn{e}{16}
        \fmfv{d.sh=circle,d.f=30,d.si=7mm,
          label.angle=-90,label.dist=11mm,
          label=$q_\mu\tfrac{\delta\gamma_m^\Lambda}{\delta A_\mu}$}{v}
        \fmfv{d.sh=circle,d.f=full,d.si=0.5mm}{e1,e2,e8,e10,e16}
        \fmfleft{i}\fmfright{o}\fmf{phantom}{i,v,o}\fmffreeze
        \fmf{plain}{v,e3}
        \fmf{plain}{v,e5}
        \fmf{plain}{v,e7}
        \fmf{plain}{v,e11}
        \fmf{plain}{v,e13}
        \fmf{plain}{v,e15}
        \fmf{wiggly}{v,e9}\fmflabel{$q_\mu$}{e9}
      \end{fmfgraph*}}}
  +
  \parbox{23mm}{\unitlength=1mm\fmfframe(3,12)(2,12){
      \begin{fmfgraph*}(18,18)
        \fmfsurroundn{e}{16}
        \fmfv{d.sh=circle,d.f=30,d.si=7mm,
          label.angle=-90,label.dist=11mm,label=$\gamma_{m+1}^\Lambda$}{v}
        \fmfv{d.sh=circle,d.f=full,d.si=0.5mm}{e1,e2,e8,e9,e10,e16}
        \fmfleft{i}\fmfright{o}\fmf{phantom}{i,v,o}\fmffreeze
        \fmf{plain}{v,e3}
        \fmf{plain}{v,e7}
        \fmf{plain}{v,e11}
        \fmf{plain}{v,e13}
        \fmf{plain}{v,e15}
        \fmf{dashesp,right=100,tension=0.4,
          label=$G^\Lambda X^\Lambda G^\Lambda$}{v,v}
      \end{fmfgraph*}}}
  +
  \parbox{55mm}{\unitlength=1mm\fmfframe(3,0)(2,0){
      \begin{fmfgraph*}(50,18)
        \fmfleftn{l}{7}
        \fmfrightn{r}{7}
        \fmfv{d.sh=circle,d.f=30,d.si=7mm,
          label.angle=-90,label.dist=11mm,label=$\gamma_{m'}^\Lambda$}{vl}
        \fmfv{d.sh=circle,d.f=30,d.si=7mm,
          label.angle=-70,label.dist=12mm,label=$\gamma_{m''}^\Lambda$}{vr}
        \fmffixed{(.55w,0)}{vl,vr}
        \fmfleft{i}\fmfright{o}\fmf{phantom}{i,vl,vr,o}\fmffreeze
        \fmf{plain}{vl,l1}
        \fmf{plain}{vl,l3}
        \fmfv{d.sh=circle,d.f=full,d.si=0.5mm}{l4}
        \fmfv{d.sh=circle,d.f=full,d.si=0.5mm}{l5}
        \fmfv{d.sh=circle,d.f=full,d.si=0.5mm}{l6}
        \fmf{plain}{vl,l7}
        \fmf{plain}{vr,r1}
        \fmfv{d.sh=circle,d.f=full,d.si=0.5mm}{r2}
        \fmfv{d.sh=circle,d.f=full,d.si=0.5mm}{r3}
        \fmfv{d.sh=circle,d.f=full,d.si=0.5mm}{r4}
        \fmf{plain}{vr,r5}
        \fmf{plain}{vr,r7}
        \fmf{dashesp,left=0.5,
          label=$G^\Lambda X^\Lambda G^\Lambda$}{vl,vr}
        \fmf{wiggly,right=0.5,label.side=left,label=expand}{vl,vr}
      \end{fmfgraph*}}}
  \\[-4ex]
  =
  \parbox{24mm}{\unitlength=1mm\fmfframe(3,8)(3,8){
      \begin{fmfgraph*}(18,18)
        \fmfsurroundn{e}{16}
        \fmfv{d.sh=circle,d.f=30,d.si=7mm,
          label.angle=-90,label.dist=11mm,label=$\Delta\gamma_m^\Lambda$}{v}
        \fmfv{d.sh=circle,d.f=full,d.si=0.5mm}{e1,e2,e8,e9,e10,e16}
        \fmfleft{i}\fmfright{o}\fmf{phantom}{i,v,o}\fmffreeze
        \fmf{dots}{v,e3}
        \fmf{dots}{v,e5}
        \fmf{dots}{v,e7}
        \fmf{dots}{v,e11}
        \fmf{dots}{v,e13}
        \fmf{dots}{v,e15}
      \end{fmfgraph*}}}
  +
  \parbox{16mm}{\unitlength=1mm\fmfframe(3,3)(3,3){
      \begin{fmfgraph*}(10,10)
        \fmfleft{i}
        \fmfright{o}
        \fmf{dots,label.side=left,label=$[\alpha\comma Q]$}{i,o}
      \end{fmfgraph*}}}
\end{align*}
where the dashed lines denote the propagator $G^\Lambda X^\Lambda
G^\Lambda$ and the dotted lines denote a momentum shift on any
external leg as in equation \eqref{eq:gwarddiag}.  The wiggly line in
the third diagram means that the expansion of the inverse second
derivative yields further one-loop diagrams with an appropriate number
of vertices.

The complete \LHS, \ie, the response function together with the
modification terms, will be interpreted in section
\ref{sec:ward:cutoff:ginv} as the divergence of the effective response
on scale $\Lambda$.  In terms of this new response function the Ward
identities will be satisfied \emph{without} modification even after
truncations (section \ref{sec:ward:cutoff:trunc}).


\subsubsection*{Modified Ward identities for connected amputated
  Green functions}

By amputating $C^\Lambda$ from the external legs of $\G^\Lambda$ in
equation \eqref{eq:gmward} we obtain the \emph{Polchinski modified
  Ward identity}
\begin{multline*}
  \inner{\alpha}{\partial_\mu\frac{\delta\V^\Lambda}{\delta iA_\mu}}
  - \Tr \left(C^\Lambda X^\Lambda C^\Lambda 
  \frac{\delta^2 \V^\Lambda}{\delta\chi\, \delta\bar\chi}\right)
  + \inner{\frac{\delta\V^\Lambda}{\delta\chi}}
  {C^\Lambda X^\Lambda C^\Lambda
    \frac{\delta\V^\Lambda}{\delta\bar\chi}} \\
  = \inner{\frac{\delta\V^\Lambda}{\delta\chi}}
  {[C^\Lambda\alpha Q + D^\Lambda Q \alpha]\chi}
  + \inner{\bar\chi}{[Q\alpha C^\Lambda + \alpha QD^\Lambda]
    \frac{\delta\V^\Lambda}{\delta\bar\chi}}
  - \inner{\bar\chi}{[\alpha,Q]\chi}.
\end{multline*}


\subsection{Background-field method}
\label{sec:ward:cutoff:bg}

The background-field method for \abb{QED} was developed in \cite{RW94b}
and summarized in \cite{FW96}: the primary goal is manifest gauge
invariance of the \onePI\ vertex functions at every scale $\Lambda$
such that one only has to consider the relevant gauge-invariant
couplings and not the much more numerous counter terms breaking gauge
invariance.  This comes, however, at the price of introducing an
auxiliary external background gauge field $\bar A$ in addition to the
internal fluctuating gauge field $A$.  The gauge transformation of
$\bar A$ is designed to cancel the modification terms in the \mWI,
such that $\Gamma^\Lambda[\phi,\bar\phi,A,\bar A]$ is gauge invariant
under simultaneous gauge transformations of all fields.

Still, gauge invariance in all fields is not sufficient to guarantee
gauge invariance in the physically relevant $\psi$, $\bar\psi$, $A$
fields: this has to be required separately by the background-field
identity constraining the $\bar A$ dependence of the vertex functions.
Moreover, the gauge-fixing term acquires a complicated scale
dependence.  If there is no fluctuating $A$ field, however, there is
no gauge-fixing term, and in special cases a simple gauge-invariant
construction is possible.


\subsection{Manifest gauge invariance}
\label{sec:ward:cutoff:ginv}

The discussion of the background-field method (cf.\ section
\ref{sec:ward:cutoff:bg}) raises the question whether we can construct
a gauge-invariant cutoff kinetic term $Q^\Lambda[A]$ in the presence
of an external field $A$.  For a Lorentz-invariant model, $-iD_\mu$ is
a hermitean operator and so is $Q[A] = Q(D_\mu)$, with real energy
eigenvalues $\epsilon(A)$ depending on the particular configuration of
$A$.  This allows to define a cutoff $\chi^\Lambda(\abs{\epsilon(A)})$
in terms of this energy, instead of frequency or momentum, by
\begin{align}
  \label{eq:gicutprop}
  Q^\Lambda[A] := \frac{Q[A]}
  {\chi^\Lambda\bigl(\sqrt{Q^\dagger[A] Q[A]}\bigr)}
\end{align}
for normal operators $Q[A]$.  For finite matrices and discrete
spectra, $Q[A]$ can be diagonalized and the cutoff applied separately
for each eigenvalue $\epsilon_k$, $Q^\Lambda(\epsilon_k) :=
Q(\epsilon_k) / \chi^\Lambda(\epsilon_k)$.

Because $Q^\Lambda[A]$ is gauge invariant by construction, equation
\eqref{eq:proptrans} now holds also with cutoff,
\begin{align}
  \label{eq:proptranslambda}
  \inner{\alpha}{\partial_\mu\frac{\delta Q^\Lambda[A]}{\delta iA_\mu}}
  = \bigl[\alpha,Q^\Lambda[A]\bigr].
\end{align}
In the background-field method the background field $\bar A$ is
coupled via the term $(Q^\Lambda[\bar A] - Q[\bar A])$ in the action.
Using equation \eqref{eq:proptranslambda}, the gauge transformation of
this term,
\begin{align*}
  \inner{\delta_\alpha \bar A_\mu}{\frac{\delta}{\delta \bar A_\mu}
    \{Q^\Lambda[\bar A] - Q[\bar A]\}}\at_{\bar A=0}
  = -iX^\Lambda,
\end{align*}
indeed cancels the modification term $X^\Lambda$ in equation
\eqref{eq:gltrans1}.  Hence, the $\bar A$-dependent cutoff propagator
restores the Ward identities without modification.

However, in this work we concentrate on non-relativistic applications
with a dispersion relation as in equation \eqref{eq:nonreldisp}, which
contains a \emph{real} time derivative or imaginary frequency.  Then
the situation is different: $Q[A]$ is not a hermitean operator any
more, and it is not even a normal operator if the commutator
\begin{align*}
  \bigl[Q^\dagger[A],Q[A]\bigr] & =
  -2 \partial_t \Bigl( \varphi + \frac{1}{2m} (i\vec\nabla\cdot
    \vec A + i\vec A\cdot\vec\nabla + \vec A^2) \Bigr)
\end{align*}
does not vanish, \ie, if the electromagnetic potential $A$ depends on
time.  Then $Q[A]$ and $Q^\dagger[A]$ are not diagonal in the same
basis, and the above definition \eqref{eq:gicutprop} of $Q^\Lambda[A]$
is not applicable.  Unfortunately, this condition excludes the
important case of a finite frequency transfer $\omega$ to the external
field, even in the limit $\omega\to 0$, which is essential for
transport.  Thus, energy-momentum cutoffs appear not to be useful in
constructing a gauge-invariant $Q^\Lambda[A]$ for non-relativistic
models at finite temperature.  If $Q$ couples only to the homogeneous
$q=0$ mode of $A$, the momentum transfer is zero and $Q[A]$ remains
diagonal in momentum space, such that one can use a momentum cutoff,
$Q^\Lambda[A](K) := Q[A](K)/\chi^\Lambda(K)$, for all frequency shifts
$\omega$.

Consider, therefore, alternative flow schemes without an
energy-momentum cutoff.  If the model is regularized by finite
temperature, the temperature- and interaction-flow schemes (cf.\ 
section \ref{sec:rg:flow:regu}) allow a trivial definition of
$Q^\Lambda[A]$.  For example, in the interaction-flow scheme with
$g=0\dotsc 1$ we can define $Q^g[A] := Q[A]/g$ for any momentum
transfer to the external field.  It is essential that the current
coupling to the external field is rescaled by the temperature or
interaction strength just like the $A$-independent quadratic part.

In those cases where a gauge-invariant construction of $Q^\Lambda[A]$
is possible we can define a new \emph{gauge-invariant generating
  functional}\footnote{Equations \eqref{eq:gginv} and
  \eqref{eq:leglambdaA} are equivalent to equation (4.10) in
  \cite{RW94b} without fluctuations of the gauge field, $a\equiv 0$.}
as
\begin{align}
  \label{eq:gginv}
  e^{-\G^{\text{gi},\Lambda}[\eta,\bar\eta,A]}
  & := \int \frac{[d\psi\bar\psi]}{\det Q^\Lambda} \,
  e^{\inner{\bar\psi}{Q^\Lambda[A]\psi} - V_0[\psi,\bar\psi]
    -\inner{j_V^\mu}{\,A_\mu}
    -\inner{\bar\psi}{\eta} - \inner{\bar\eta}{\psi}}
\end{align}
which differs from the previous definition \eqref{eq:glsplit} in that
the cutoff acts also on the $A$-de\-pen\-dent quadratic part.  This yields
the Ward identity
\begin{align}
  \label{eq:gwardmani}
  \boxed{
    \inner{\alpha}{\partial_\mu\frac{\delta\G^{\text{gi},\Lambda}}
      {\delta iA_\mu}}
    = \inner{\frac{\delta\G^{\text{gi},\Lambda}}{\delta\eta}}{\alpha\eta}
      +\inner{\bar\eta}{\alpha\frac{\delta\G^{\text{gi},\Lambda}}
        {\delta\bar\eta}}}
\end{align}
on any scale $\Lambda$.  Instead of the Legendre transformation
\eqref{eq:leglambda} we can now define
\begin{align}
  \label{eq:leglambdaA}
  \Gamma^{\text{gi},\Lambda}[\phi,\bar\phi,A] +
  \inner{\bar\phi}{Q^\Lambda[A]\phi}
  = \G^{\text{gi},\Lambda}[\eta,\bar\eta,A]
  +\inner{\bar\phi}{\eta} -\inner{\bar\eta}{\phi}.
\end{align}
Then $\Gamma^{\text{gi},\Lambda}$ evolves under the flow equation
\begin{align}
  \label{eq:gammaginvflow}
  \boxed{
    \dl \Gamma^{\text{gi},\Lambda} = \Tr \dot Q^\Lambda C^\Lambda 
    - \Tr \dot Q^\Lambda[A] 
    \Bigl(
    Q^\Lambda[A]+\frac{\delta^2\Gamma^{\text{gi},\Lambda}}{\delta\phi
      \, \delta\bar\phi} \Bigr)^{-1}}
\end{align}
and satisfies the Ward identity
\begin{align}
  \label{eq:gammawardmani}
  \boxed{
    \inner{\alpha}{\partial_\mu\frac{\delta\Gamma^{\text{gi},\Lambda}}
      {\delta iA_\mu}}
    = \inner{\frac{\delta\Gamma^{\text{gi},\Lambda}}{\delta\phi}}{\alpha\phi}
    + \inner{\bar\phi}
    {\alpha\frac{\delta\Gamma^{\text{gi},\Lambda}}{\delta\bar\phi}}}\,.
\end{align}
This new functional converges to the original one,
$\Gamma^{\text{gi},\Lambda} \to \Gamma$ in the limit $\Lambda\to 0$,
and $\Gamma^{\text{gi},\Lambda} = \Gamma^\Lambda$ for $A=0$.  However,
it has the advantage that its vertex functions are manifestly gauge
invariant during the whole flow, \ie, the Ward identities are not
modified.  Even for a truncated flow-equation hierarchy this remains
true as I will show in section \ref{sec:ward:cutoff:trunc}: while the
flowing response vertex at $\Lambda=0$ approximates the exact response
vertex only to truncation order, it still satisfies the Ward
identities exactly to all orders.


\subsubsection*{Gauge-invariant response vertex}

Let us give an explicit example of the current-response operator with
cutoff.  We diagonalize $Q[A]$, apply the cutoff on the eigenvalues
and take the $A$ derivative to obtain the current.  To leading order
in $A$, all modes $A(q)$ couple independently, and it is sufficient to
consider a single $q$ mode.  We therefore capture the generic
situation by considering a two-state system with momenta $k\pm q/2
\equiv \pm$ and an inverse propagator
\begin{align*}
  Q[A] = 
  \begin{pmatrix}
    Q_+ & -JA \\
    -(JA)^* & Q_-
  \end{pmatrix}
  + \Ord(A^2),
\end{align*}
which is diagonal for $A=0$, and where the coupling to $A$ transfers
momentum $q$ between the two states, $JA = J^\mu(q;+;-) A_\mu(q)$,
$(JA)^* = J^\mu(-q;-;+) A_\mu(-q)$, and the current
\begin{align*}
  J^\mu = 
  \begin{pmatrix}
    0 & J^\mu(q;+;-) \\
    J^\mu(-q;-;+) & 0
  \end{pmatrix}.
\end{align*}
Generally, $Q[A]$ does not need to be a hermitean matrix but only a
normal matrix; then the general formula \eqref{eq:gicutprop} has to be
used.  For a pedagogical derivation, however, we assume that $Q[A]$ is
hermitean, as for instance with a purely quadratic dispersion
$Q(k_\mu) = \abs{k}^2/2m$.  Then $Q[A]$ is diagonalized,
\begin{align*}
  Q[A] = 
  \begin{pmatrix}
    1 & \frac{JA}{Q_+-Q_-} \\
    -\frac{(JA)^*}{Q_+-Q_-} & 1
  \end{pmatrix}
  \begin{pmatrix}
    Q_+ & 0\phantom{\frac JJ} \\
    0\phantom{\frac JJ} & Q_-
  \end{pmatrix}
  \begin{pmatrix}
    1 & -\frac{JA}{Q_+-Q_-} \\
    \frac{(JA)^*}{Q_+-Q_-} & 1
  \end{pmatrix}
  + \Ord(A^2),
\end{align*}
where notably, the eigenvalues depend only quadratically on $A$.
Replacing the diagonal $Q$ by $Q^\Lambda = \left( \begin{smallmatrix}
    Q_+^\Lambda & 0 \\ 0 & Q_-^\Lambda \end{smallmatrix} \right)$, we
obtain
\begin{align*}
  Q^\Lambda[A] = 
  \begin{pmatrix}
    Q_+^\Lambda & -\frac{Q_+^\Lambda-Q_-^\Lambda}{Q_+-Q_-} JA \\
    -\frac{Q_+^\Lambda-Q_-^\Lambda}{Q_+-Q_-} (JA)^* & Q_-^\Lambda
  \end{pmatrix}
  + \Ord(A^2),
\end{align*}
and the cutoff current operator reads
\begin{align*}
  J^{\mu,\Lambda} = \frac{Q_+^\Lambda-Q_-^\Lambda}{Q_+-Q_-} J^\mu \,,
\end{align*}
\ie, the original current rescaled by a number.  As a check, the
current operator thus constructed has the right limit,
$J^{\mu,\Lambda} \to J^\mu$ as $\Lambda \to 0$, and satisfies $q^\mu
J^{\mu,\Lambda} = Q_+^\Lambda-Q_-^\Lambda$ as a consequence of the
gauge invariance of $Q^\Lambda[A]$, equation \eqref{eq:proptranslambda}.


\subsection{Ward identities in truncated flows}
\label{sec:ward:cutoff:trunc}

If the model and flow scheme permit the construction of a
gauge-invariant cutoff bare action as in section
\ref{sec:ward:cutoff:ginv}, \ie, if $Q^\Lambda[A]$ and the bare
interaction are manifestly gauge invariant at any scale, then the full
flow-equation hierarchy satisfies the \emph{unmodified} Ward
identities on all scales.  We shall see that in this case even
truncated flows can satisfy the Ward identities exactly.

The most commonly used truncation of the flow-equation hierarchy
\emph{without} external field is to set the flow of higher Green
functions to zero,
\begin{align}
  \label{eq:truncA0}
  \dl G_m^\Lambda(A=0) := 0 \quad \forall m\geq m_0,
\end{align}
for some $m_0>0$ usually determined by practical considerations and
justified perturbatively in the renormalized interaction.  I will now
show that the Ward identities are satisfied if we demand the same
truncation \eqref{eq:truncA0} also \emph{with} external field,
\begin{align}
  \label{eq:trunc}
  \boxed{
    \dl G_m^\Lambda(A) := 0 \quad \forall m\geq m_0 \text{ and }
    \forall A}
\end{align}
and take derivatives with respect to $A_\mu$ in order to obtain the
truncated flow equations of the response functions
$G_{m;1}^{\mu,\Lambda}(A) := \delta_{A_\mu} G_m^\Lambda(A)$.  In
particular, also the response functions with $m\geq m_0$ do not flow,
\begin{align}
  \label{eq:truncA}
  \dl G_{m;1}^{\mu,\Lambda}(A) := 0 \quad \forall m\geq m_0.  
\end{align}
The response-function Ward identities \eqref{eq:gwardmani} are of the
form
\begin{align}
  \label{eq:wiresp}
  \partial_\mu G_{m;1}^{\mu,\Lambda}(A)
  = i \, \mathsf{S} \, G_m^\Lambda(A) \quad \forall m
\end{align}
homogeneous in the number of fermion lines, where $\mathsf{S}$
represents the momentum shift on the external legs.  By the
truncation, neither side of equation \eqref{eq:wiresp} flows for
$m\geq m_0$, so the higher Green functions remain at their initial
condition---given by the bare action---which by construction is gauge
invariant and satisfies the Ward identities.

For $m<m_0$, the \RHS\ of the truncated flow equation is built up
completely from bare propagators $Q^\Lambda[A]$ and single-scale
propagators $\dot Q^\Lambda[A]$ which by construction are manifestly
gauge covariant on any scale $\Lambda$, as well as from Green
functions $G_m^\Lambda(A)$ which we assume to be gauge covariant on a
particular scale $\Lambda$.  Under a gauge transformation, all Green
functions and propagators acquire phase factors which cancel on all
internal lines, leaving only the phase factors on the external legs
which, in turn, imply gauge covariance of the \RHS.  Then, by
infinitesimal induction the Green functions
$G_m^{\Lambda-d\Lambda}(A)$ at an infinitesimally lower scale
$\Lambda-d\Lambda$ will also be gauge covariant, and hence the Ward
identities are satisfied during the complete \fRG\ flow.

This argument is valid in any \fRG\ scheme where the Ward identities
are homogenous in the number of fermion lines, which holds in the
Polchinski, \onePI\ and Wick-ordered schemes presented in this work.


\section{The role of self-consistency}
\label{sec:ward:sc}

The generating functional $\G^{\text{gi},\Lambda}[\eta,\bar\eta,A]$
defined in equation \eqref{eq:gginv} has a redundant parametrization:
there are two different ways to obtain the response functions, either
by taking a derivative with respect to $A_\mu$ or by inserting
$\inner{\bar\psi}{\frac{\delta Q^\Lambda[A]}{\delta A_\mu}\psi}$ into
the path integral.  Assuming from now on a gauge-invariant interaction
$V_0$ and dropping the label ``$\text{gi}$'', we have
\begin{align*}
  \ddd{A_\mu} e^{-\G^{\Lambda}[\eta,\bar\eta,A]}
  & = \int \frac{[d\psi\bar\psi]}{\det Q^\Lambda} \,
  \inner{\bar\psi}{\frac{\delta Q^\Lambda[A]}{\delta A_\mu}\psi} \;
  e^{\inner{\bar\psi}{Q^\Lambda[A]\psi} - V_0[\psi,\bar\psi]
    -\inner{\bar\psi}{\eta} - \inner{\bar\eta}{\psi}} \\
  & = -\inner{\frac\delta{\delta\eta}}
  {\frac{\delta Q^\Lambda[A]}{\delta A_\mu} \, \frac\delta{\delta\bar\eta}} \,
  e^{-\G^{\Lambda}[\eta,\bar\eta,A]} \,.
\end{align*}
The second derivative yields terms linear and quadratic in
$\G^{\Lambda}$,
\begin{align}
  \label{eq:selfcons}
  \boxed{
    \frac{\delta\G^{\Lambda}}{\delta A_\mu}
    = -\Tr\Bigl(\frac{\delta Q^\Lambda[A]}{\delta A_\mu} \,
    \frac{\delta^2 \G^{\Lambda}}{\delta\eta \, \delta\bar\eta}\Bigr) +
    \inner{\frac{\delta\G^{\Lambda}}{\delta\eta}}
    {\frac{\delta Q^\Lambda[A]}{\delta A_\mu} \,
      \frac{\delta\G^{\Lambda}}{\delta\bar\eta}}} \,.
\end{align}
This \emph{self-consistency equation} expresses the response functions
in terms of higher Green functions with a loop closed, and tree terms.
Equation \eqref{eq:scdiag} illustrates the relation between one- and
two-particle connected Green function diagrammatically,
\begin{align}
  \label{eq:scdiag}
  \parbox{28mm}{\unitlength=1mm\fmfframe(8,2)(2,2){
      \begin{fmfgraph*}(18,18)
        \fmfsurroundn{e}{16}
        \fmfv{d.sh=circle,d.f=0,d.si=7mm,
          label.angle=-90,label.dist=8mm,
          label=$q_\mu G_{1;1}^{\mu\comma\Lambda}$}{v}
        \fmfleft{i}\fmfright{o}\fmf{phantom}{i,v,o}\fmffreeze
        \fmf{plain}{v,e3}
        \fmf{plain}{v,e15}
        \fmf{wiggly}{v,e9}\fmflabel{$q_\mu$}{e9}
      \end{fmfgraph*}}}
  =
  \parbox{23mm}{\unitlength=1mm\fmfframe(3,2)(2,2){
      \begin{fmfgraph*}(18,18)
        \fmfsurroundn{e}{16}
        \fmfv{d.sh=circle,d.f=0,d.si=7mm,
          label.angle=-90,label.dist=8mm,label=$G_2^\Lambda$}{v}
        \fmfleft{i}\fmfright{o}\fmf{phantom}{i,v,o}\fmffreeze
        \fmf{plain}{v,e11}
        \fmf{plain}{v,e15}
        \fmf{dashesp,right=100,tension=0.5}{v,v}
      \end{fmfgraph*}}}
  +
  \parbox{55mm}{\unitlength=1mm\fmfframe(3,0)(2,0){
      \begin{fmfgraph*}(50,18)
        \fmfleftn{l}{7}
        \fmfrightn{r}{7}
        \fmfv{d.sh=circle,d.f=0,d.si=7mm,
          label.angle=-90,label.dist=8mm,label=$G_1^\Lambda$}{vl}
        \fmfv{d.sh=circle,d.f=0,d.si=7mm,
          label.angle=-70,label.dist=8mm,label=$G_1^\Lambda$}{vr}
        \fmffixed{(.55w,0)}{vl,vr}
        \fmfleft{i}\fmfright{o}\fmf{phantom}{i,vl,vr,o}\fmffreeze
        \fmf{plain}{vl,l4}
        \fmf{plain}{vr,r4}
        \fmf{dashesp}{vl,vr}
      \end{fmfgraph*}}}  
\end{align}
where the dashed lines are $\delta Q^\Lambda[A]/\delta A_\mu$.  This
relation is generally broken by truncated flows, as we shall see in
section \ref{sec:ward:sc:higher}.  Using the gauge transformation of
the quadratic part \eqref{eq:proptranslambda} we can write the
divergence of equation \eqref{eq:selfcons} as
\begin{align}
  \label{eq:scG}
  \inner{\alpha}{\partial_\mu\frac{\delta\G^{\Lambda}}
    {\delta iA_\mu}}
  = -\Tr \Bigl( \bigl[\alpha,Q^\Lambda[A]\bigr]
  \frac{\delta^2\G^{\Lambda}}{\delta\eta \, \delta\bar\eta} \Bigr) +
  \inner{\frac{\delta\G^{\Lambda}}{\delta\eta}}
  {\bigl[\alpha,Q^\Lambda[A]\bigr]
    \frac{\delta\G^{\Lambda}}{\delta\bar\eta}},
\end{align}
and replace the response function in the Ward identity
\eqref{eq:gwardmani} to obtain the \emph{self-consistent Ward
  identity}
\begin{align}
  \label{eq:gscward}
  \boxed{
    - \Tr \Bigl( \bigl[\alpha,Q^\Lambda[A]\bigr]
    \frac{\delta^2 \G^{\Lambda}}{\delta\eta \, \delta\bar\eta} \Bigr) +
    \inner{\frac{\delta\G^{\Lambda}}{\delta\eta}}
    {\bigl[\alpha,Q^\Lambda[A]\bigr]
      \frac{\delta\G^{\Lambda}}{\delta\bar\eta}}
    = \inner{\frac{\delta\G^{\Lambda}}{\delta\eta}}{\alpha\eta}
    + \inner{\bar\eta}{\alpha\frac{\delta\G^{\Lambda}}
      {\delta\bar\eta}}}\,.
\end{align}
Note that we would arrive at the same self-consistent Ward identities
to leading order in $A$ if we would start with the generating
functional \eqref{eq:glsplit} and the modified Ward identities
\eqref{eq:gmward}.  By Legendre transformation \eqref{eq:leglambdaA}
we obtain the self-consistent Ward identity for the \onePI\ vertex
functions,
\begin{align}
  \label{eq:gammascward}
  \boxed{
    - \Tr \Bigl( \bigl[\alpha,Q^\Lambda[A]\bigr] \,
    \bigl( Q^\Lambda[A]+\delta^2 \Gamma^{\Lambda}\bigr)^{-1} \Bigr)
    = \inner{\frac{\delta\Gamma^{\Lambda}}{\delta\phi}}{\alpha\phi}
    + \inner{\bar\phi}
    {\alpha\frac{\delta\Gamma^{\Lambda}}{\delta\bar\phi}}}\,.
\end{align}
The topological structure of the $A$ derivative on the \LHS\ is the
same as that of the $\Lambda$ derivative in equations
\eqref{eq:gflow}, \eqref{eq:gammaflow}.  The self-consistent Ward
identities for the \onePI\ vertex functions are represented
diagrammatically as
\begin{align*}
  \parbox{23mm}{\unitlength=1mm\fmfframe(3,2)(2,2){
      \begin{fmfgraph*}(18,18)
        \fmfsurroundn{e}{16}
        \fmfv{d.sh=circle,d.f=30,d.si=7mm,
          label.angle=-90,label.dist=11mm,label=$\gamma_{m+1}^\Lambda$}{v}
        \fmfv{d.sh=circle,d.f=full,d.si=0.5mm}{e1,e2,e8,e9,e10,e16}
        \fmfleft{i}\fmfright{o}\fmf{phantom}{i,v,o}\fmffreeze
        \fmf{plain}{v,e3}
        \fmf{plain}{v,e7}
        \fmf{plain}{v,e11}
        \fmf{plain}{v,e13}
        \fmf{plain}{v,e15}
        \fmf{dashesp,right=100,tension=0.4}{v,v}
      \end{fmfgraph*}}}
  +
  \parbox{55mm}{\unitlength=1mm\fmfframe(3,0)(2,0){
      \begin{fmfgraph*}(50,18)
        \fmfleftn{l}{7}
        \fmfrightn{r}{7}
        \fmfv{d.sh=circle,d.f=30,d.si=7mm,
          label.angle=-90,label.dist=11mm,label=$\gamma_{m'}^\Lambda$}{vl}
        \fmfv{d.sh=circle,d.f=30,d.si=7mm,
          label.angle=-70,label.dist=12mm,label=$\gamma_{m''}^\Lambda$}{vr}
        \fmffixed{(.55w,0)}{vl,vr}
        \fmfleft{i}\fmfright{o}\fmf{phantom}{i,vl,vr,o}\fmffreeze
        \fmf{plain}{vl,l1}
        \fmf{plain}{vl,l3}
        \fmfv{d.sh=circle,d.f=full,d.si=0.5mm}{l4}
        \fmfv{d.sh=circle,d.f=full,d.si=0.5mm}{l5}
        \fmfv{d.sh=circle,d.f=full,d.si=0.5mm}{l6}
        \fmf{plain}{vl,l7}
        \fmf{plain}{vr,r1}
        \fmfv{d.sh=circle,d.f=full,d.si=0.5mm}{r2}
        \fmfv{d.sh=circle,d.f=full,d.si=0.5mm}{r3}
        \fmfv{d.sh=circle,d.f=full,d.si=0.5mm}{r4}
        \fmf{plain}{vr,r5}
        \fmf{plain}{vr,r7}
        \fmf{dashesp,left=0.5}{vl,vr}
        \fmf{wiggly,right=0.5,label.side=left,label=expand}{vl,vr}
      \end{fmfgraph*}}}
  =
  \parbox{24mm}{\unitlength=1mm\fmfframe(3,8)(3,8){
      \begin{fmfgraph*}(18,18)
        \fmfsurroundn{e}{16}
        \fmfv{d.sh=circle,d.f=30,d.si=7mm,
          label.angle=-90,label.dist=11mm,label=$\Delta\gamma_m^\Lambda$}{v}
        \fmfv{d.sh=circle,d.f=full,d.si=0.5mm}{e1,e2,e8,e9,e10,e16}
        \fmfleft{i}\fmfright{o}\fmf{phantom}{i,v,o}\fmffreeze
        \fmf{dots}{v,e3}
        \fmf{dots}{v,e5}
        \fmf{dots}{v,e7}
        \fmf{dots}{v,e11}
        \fmf{dots}{v,e13}
        \fmf{dots}{v,e15}
      \end{fmfgraph*}}}
\end{align*}
where the dashed lines feature the propagator $G^\Lambda
\bigl[\alpha,Q^\Lambda[A]\bigr] G^\Lambda$.

By amputating bare propagators $C^\Lambda[A]$ from the external legs
in equation \eqref{eq:gscward} we obtain the \emph{self-consistent
  Polchinski Ward identity}
\begin{align}
  \label{eq:vscward}
  \boxed{
    \Tr \Bigl( \bigl[\alpha,C^\Lambda[A]\bigr]
    \frac{\delta^2 \V^\Lambda} {\delta\chi \, \delta\bar\chi} \Bigr)
    - \inner{\frac{\delta\V^\Lambda}{\delta\chi}}
    {\bigl[\alpha,C^\Lambda[A]\bigr]
      \frac{\delta\V^\Lambda}{\delta\bar\chi}}
    = \inner{\frac{\delta\V^\Lambda}{\delta\chi}}{\alpha\chi}
    + \inner{\bar\chi}{\alpha\frac{\delta\V^\Lambda}{\delta\bar\chi}}}\,.
\end{align}

While the previous response-function Ward identities were homogeneous
in the number of external electron legs, the self-consistent Ward
identities close a loop on higher Green or vertex functions and
thereby decrease the number of external legs.  This ``inhomogeneity''
leads to severe problems: when the flow is truncated the lower Green
functions are flowing but the higher ones are not, hence
self-consistency is violated, and the above self-consistent Ward
identities \eqref{eq:gscward}, \eqref{eq:gammascward}, and
\eqref{eq:vscward} are only satisfied to truncation order.  Despite
intensive search a general solution to the problem of self-consistency
remains elusive.

In the upcoming section \ref{sec:ward:sc:higher} I will illustrate how
the self-consistent Ward identities are broken by the standard
truncation that we have used in section \ref{sec:ward:cutoff:trunc}.
In section \ref{sec:ward:sc:cons} I review the conserving
approximations of Baym and Kadanoff which are self-consistent and
satisfy these Ward identities, but turn out to be in general
incompatible with truncated flow equations except for special cases
(section \ref{sec:ward:sc:import}).


\subsection{Self-consistent Ward identities in truncated flows}
\label{sec:ward:sc:higher}

Let us give an example of how a simple truncation breaks the
self-consistent Ward identities.  This example shows how the problem
arises generically.  Consider the Polchinski scheme with the truncated
flow equations \eqref{eq:vflow},
\begin{align*}
  \parbox{16mm}{\unitlength=1mm\fmfframe(2,0)(2,0){
      \begin{fmfgraph*}(12,12)
        \fmfv{d.sh=circle,d.f=full,d.si=0.8mm}{i1}
        \fmfforce{(6mm,11mm)}{i1}
        \fmfv{d.sh=circle,d.f=shaded,d.si=6mm}{v}
        \fmfleft{i}
        \fmfright{o}
        \fmf{plain}{i,v,o}
      \end{fmfgraph*}}}
  & =
  \parbox{10mm}{\unitlength=1mm\fmfframe(0,4)(0,0){
      \begin{fmfgraph*}(10,10)
        \fmfsurroundn{e}{8}
        \fmfv{d.sh=circle,d.f=shaded,d.si=6mm}{v}
        \fmfleft{i}\fmfright{o}\fmf{phantom}{i,v,o}\fmffreeze
        \fmf{plain}{v,e6}
        \fmf{plain}{v,e8}
        \fmf{slplaino,right=100,tension=0.4}{v,v}
      \end{fmfgraph*}}}
  +
  \parbox{28mm}{\unitlength=1mm\fmfframe(2,0)(2,0){
      \begin{fmfgraph*}(24,10)
        \fmfv{d.sh=circle,d.f=shaded,d.si=6mm}{vl,vr}
        \fmfforce{(6mm,5mm)}{vl}
        \fmfforce{(18mm,5mm)}{vr}
        \fmfleft{i}\fmfright{o}
        \fmf{plain}{i,vl}
        \fmf{slplainm}{vl,vr}
        \fmf{plain}{vr,o}
      \end{fmfgraph*}}}
  & 
  \parbox{16mm}{\unitlength=1mm\fmfframe(2,2)(2,2){
      \begin{fmfgraph*}(12,12)
        \fmfsurroundn{e}{8}
        \fmfv{d.sh=circle,d.f=shaded,d.si=6mm}{v}
        \fmfv{d.sh=circle,d.f=full,d.si=0.8mm}{i1}
        \fmfforce{(6mm,11mm)}{i1}
        \fmfleft{i}\fmfright{o}\fmf{phantom}{i,v,o}\fmffreeze
        \fmf{plain}{v,e2}
        \fmf{plain}{v,e4}
        \fmf{plain}{v,e6}
        \fmf{plain}{v,e8}
      \end{fmfgraph*}}}
  & =0 \,,
\end{align*}
where slashed lines denote $\dot C^\Lambda$.  The corresponding Ward
identities \eqref{eq:vscward} are
\begin{align}
  \label{eq:polwardtrunc1}
  \parbox{16mm}{\unitlength=1mm\fmfframe(2,0)(2,0){
      \begin{fmfgraph*}(12,12)
        \fmfv{d.sh=circle,d.f=shaded,d.si=6mm}{v}
        \fmfleft{i}\fmfright{o}
        \fmf{plain,label=$+$}{i,v}
        \fmf{plain,label=$+$}{v,o}
      \end{fmfgraph*}}}
  - \parbox{16mm}{\unitlength=1mm\fmfframe(2,0)(2,0){
      \begin{fmfgraph*}(12,12)
        \fmfv{d.sh=circle,d.f=shaded,d.si=6mm}{v}
        \fmfleft{i}\fmfright{o}
        \fmf{plain,label=$-$}{i,v}
        \fmf{plain,label=$-$}{v,o}
      \end{fmfgraph*}}}
  \equiv
  \parbox{16mm}{\unitlength=1mm\fmfframe(2,0)(2,0){
      \begin{fmfgraph*}(12,12)
        \fmfv{d.sh=circle,d.f=shaded,d.si=6mm}{v}
        \fmfleft{i}\fmfright{o}
        \fmf{plain,label=$\pm$}{i,v}
        \fmf{plain,label=$\pm$}{v,o}
      \end{fmfgraph*}}}
  & =
  \parbox{20mm}{\unitlength=1mm\fmfframe(5,4)(5,4){
      \begin{fmfgraph*}(10,10)
        \fmfsurroundn{e}{8}
        \fmfv{d.sh=circle,d.f=shaded,d.si=6mm}{v}
        \fmfleft{i}\fmfright{o}\fmf{phantom}{i,v,o}\fmffreeze
        \fmf{plain}{v,e6}\fmflabel{$+$}{e6}
        \fmf{plain}{v,e8}\fmflabel{$-$}{e8}
        \fmf{dashesp,right=100,tension=0.4}{v,v}
      \end{fmfgraph*}}}
  +
  \parbox{28mm}{\unitlength=1mm\fmfframe(2,0)(2,0){
      \begin{fmfgraph*}(24,10)
        \fmfv{d.sh=circle,d.f=shaded,d.si=6mm}{vl,vr}
        \fmfforce{(6mm,5mm)}{vl}
        \fmfforce{(18mm,5mm)}{vr}
        \fmfleft{i}\fmfright{o}
        \fmf{plain,label=$+$,label.side=right}{i,vl}
        \fmf{dashesp}{vl,vr}
        \fmf{plain,label=$-$}{vr,o}
      \end{fmfgraph*}}}
  \\
  \label{eq:polwardtrunc2}
  \parbox{16mm}{\unitlength=1mm\fmfframe(2,2)(2,2){
      \begin{fmfgraph*}(12,12)
        \fmfsurroundn{e}{8}
        \fmfv{d.sh=circle,d.f=shaded,d.si=6mm}{v}
        \fmfleft{i}\fmfright{o}\fmf{phantom}{i,v,o}\fmffreeze
        \fmf{dots}{v,e2}
        \fmf{dots}{v,e4}
        \fmf{dots}{v,e6}
        \fmf{dots}{v,e8}
      \end{fmfgraph*}}}
  & =0 \,,
\end{align}
where dots on the external legs denote a momentum shift on any one of
the legs as in equation \eqref{eq:gwarddiag}, and dashed lines denote
$[\alpha,C^\Lambda]$.  This truncation is simpler than the ones used
in practice; however, it is useful pedagogically because it
demonstrates the problems already at second order in the renormalized
interaction.  A truncation at a higher level would display the same
type of problem, only further down in the hierarchy where the
diagrammatics is more tedious.

If the Ward identity is to be compatible with the truncated flow, the
$\Lambda$ derivatives of the \LHS\ and \RHS\ of the Ward identity
\eqref{eq:polwardtrunc1} for the one-particle function should agree:
\begin{align*}
  \dl \text{\LHS(\WI)} & =
  \parbox{16mm}{\unitlength=1mm\fmfframe(2,0)(2,0){
      \begin{fmfgraph*}(12,12)
        \fmfv{d.sh=circle,d.f=full,d.si=0.8mm}{i1}
        \fmfforce{(6mm,11mm)}{i1}
        \fmfv{d.sh=circle,d.f=shaded,d.si=6mm}{v}
        \fmfleft{i}
        \fmfright{o}
        \fmf{plain,label=$\pm$}{i,v}
        \fmf{plain,label=$\pm$}{v,o}
      \end{fmfgraph*}}}
  =
  \parbox{14mm}{\unitlength=1mm\fmfframe(2,4)(2,4){
      \begin{fmfgraph*}(10,10)
        \fmfsurroundn{e}{8}
        \fmfv{d.sh=circle,d.f=shaded,d.si=6mm}{v}
        \fmfleft{i}\fmfright{o}\fmf{phantom}{i,v,o}\fmffreeze
        \fmf{plain}{e8,v,e6}\fmflabel{$\pm$}{e6}\fmflabel{$\pm$}{e8}
        \fmf{slplaino,right=100,tension=0.4}{v,v}
      \end{fmfgraph*}}}
  +
  \parbox{28mm}{\unitlength=1mm\fmfframe(2,0)(2,0){
      \begin{fmfgraph*}(24,10)
        \fmfv{d.sh=circle,d.f=shaded,d.si=6mm}{vl,vr}
        \fmfforce{(6mm,5mm)}{vl}
        \fmfforce{(18mm,5mm)}{vr}
        \fmfleft{i}\fmfright{o}
        \fmf{plain,label=$\pm$,label.side=right}{i,vl}
        \fmf{slplainm}{vl,vr}
        \fmf{plain,label=$\pm$}{vr,o}
      \end{fmfgraph*}}} \\
  \dl \text{\RHS(\WI)} & =
  \parbox{10mm}{\unitlength=1mm\fmfframe(0,4)(0,4){
      \begin{fmfgraph*}(10,10)
        \fmfsurroundn{e}{8}
        \fmfv{d.sh=circle,d.f=shaded,d.si=6mm}{v}
        \fmfleft{i}\fmfright{o}\fmf{phantom}{i,v,o}\fmffreeze
        \fmf{plain}{e8,v,e6}\fmflabel{$+$}{e6}\fmflabel{$-$}{e8}
        \fmf{sldasheso,right=100,tension=0.4}{v,v}
      \end{fmfgraph*}}}
  +
  \parbox{28mm}{\unitlength=1mm\fmfframe(2,4)(2,4){
      \begin{fmfgraph*}(24,10)
        \fmfv{d.sh=circle,d.f=shaded,d.si=6mm}{vl,vr}
        \fmfforce{(6mm,5mm)}{vl}
        \fmfforce{(18mm,5mm)}{vr}
        \fmfleft{i}\fmfright{o}
        \fmf{plain,label=$+$,label.side=right}{i,vl}
        \fmf{dashesp}{vl,vr}
        \fmf{plain,label=$-$}{vr,o}
        \fmffreeze
        \fmf{slplaino,right=1,tension=0.7}{vl,vl}
      \end{fmfgraph*}}}
  +
  \parbox{40mm}{\unitlength=1mm\fmfframe(2,0)(2,0){
      \begin{fmfgraph*}(36,10)
        \fmfv{d.sh=circle,d.f=shaded,d.si=6mm}{vl,vm,vr}
        \fmfforce{(6mm,5mm)}{vl}
        \fmfforce{(18mm,5mm)}{vm}
        \fmfforce{(30mm,5mm)}{vr}
        \fmfleft{i}\fmfright{o}
        \fmf{plain,label=$+$,label.side=right}{i,vl}
        \fmf{slplainm}{vl,vm}
        \fmf{dashesp}{vm,vr}
        \fmf{plain,label=$-$}{vr,o}
      \end{fmfgraph*}}}
  +
  \parbox{28mm}{\unitlength=1mm\fmfframe(2,0)(2,0){
      \begin{fmfgraph*}(24,10)
        \fmfv{d.sh=circle,d.f=shaded,d.si=6mm}{vl,vr}
        \fmfforce{(6mm,5mm)}{vl}
        \fmfforce{(18mm,5mm)}{vr}
        \fmfleft{i}\fmfright{o}
        \fmf{plain,label=$+$,label.side=right}{i,vl}
        \fmf{sldashesm}{vl,vr}
        \fmf{plain,label=$-$}{vr,o}
      \end{fmfgraph*}}}.
\end{align*}
The second line can be rewritten using both Ward identities above and
$\dl [\alpha,C^\Lambda] = [\alpha,\dot C^\Lambda]$,
\begin{align*}
  \dl \text{\RHS(\WI)} & =
  \parbox{18mm}{\unitlength=1mm\fmfframe(4,4)(4,4){
      \begin{fmfgraph*}(10,10)
        \fmfsurroundn{e}{8}
        \fmfv{d.sh=circle,d.f=shaded,d.si=6mm}{v}
        \fmfleft{i}\fmfright{o}\fmf{phantom}{i,v,o}\fmffreeze
        \fmf{plain}{e8,v,e6}\fmflabel{$\pm$}{e6}\fmflabel{$\pm$}{e8}
        \fmf{slplaino,right=100,tension=0.4}{v,v}
      \end{fmfgraph*}}}
  +
  \parbox{28mm}{\unitlength=1mm\fmfframe(2,0)(2,0){
      \begin{fmfgraph*}(24,10)
        \fmfv{d.sh=circle,d.f=shaded,d.si=6mm}{vl,vr}
        \fmfforce{(6mm,5mm)}{vl}
        \fmfforce{(18mm,5mm)}{vr}
        \fmfleft{i}\fmfright{o}
        \fmf{plain,label=$\pm$,label.side=right}{i,vl}
        \fmf{slplainm}{vl,vr}
        \fmf{plain,label=$\pm$}{vr,o}
      \end{fmfgraph*}}}
  +
  \boxed{
    \parbox{28mm}{\unitlength=1mm\fmfframe(2,6)(2,6){
        \begin{fmfgraph*}(24,10)
          \fmfv{d.sh=circle,d.f=shaded,d.si=6mm}{vl,vr}
          \fmfforce{(6mm,5mm)}{vl}
          \fmfforce{(18mm,5mm)}{vr}
          \fmfleft{i}\fmfright{o}
          \fmf{plain,label=$+$,label.side=right}{i,vl}
          \fmf{dashesp}{vl,vr}
          \fmf{plain,label=$-$}{vr,o}
          \fmffreeze
          \fmf{slplaino,right=1,tension=0.7}{vl,vl}
        \end{fmfgraph*}}}
    -
    \parbox{28mm}{\unitlength=1mm\fmfframe(2,6)(2,6){
        \begin{fmfgraph*}(24,10)
          \fmfv{d.sh=circle,d.f=shaded,d.si=6mm}{vl,vr}
          \fmfforce{(6mm,5mm)}{vl}
          \fmfforce{(18mm,5mm)}{vr}
          \fmfleft{i}\fmfright{o}
          \fmf{plain,label=$+$,label.side=right}{i,vl}
          \fmf{slplainm}{vl,vr}
          \fmf{plain,label=$-$}{vr,o}
          \fmffreeze
          \fmf{dashesp,right=1,tension=0.7}{vl,vl}
        \end{fmfgraph*}}}}\,.
\end{align*}
The \LHS\ and \RHS\ differ, thus we cannot complete a proof of the
Ward identities by induction: even if the Ward identities are
satisfied at some scale $\Lambda$, the flow violates them by the terms
highlighted in the box.  There is no reason why these terms should in
general vanish, or why the Ward identities should be miraculously
satisfied at the end of the flow even though they are violated during
the flow.  Similarly, in the \onePI\ scheme with a truncation such
that the two-particle vertex function does not flow, the violation is
of the form
\begin{align}
  \label{eq:1piflowtrunc1}
  \boxed{
    \parbox{32mm}{\unitlength=1mm\fmfframe(0,2)(2,2){
        \begin{fmfgraph*}(30,10)
          \fmfleftn{l}{3}
          \fmfrightn{r}{3}
          \fmfv{d.sh=circle,d.f=30,d.si=6mm}{vl,vr}
          \fmfforce{(8mm,5mm)}{vl}
          \fmfforce{(22mm,5mm)}{vr}
          \fmfleft{i}\fmfright{o}\fmf{phantom}{i,vl,vr,o}\fmffreeze
          \fmf{plain}{l1,vl,l3}
          \fmf{slplain,left=1}{vr,r2}
          \fmf{plain,left=1}{r2,vr}
          \fmf{dashesp,left=0.5}{vl,vr}
          \fmf{plain,right=0.5}{vl,vr}
        \end{fmfgraph*}}}
    -
    \parbox{32mm}{\unitlength=1mm\fmfframe(0,2)(2,2){
        \begin{fmfgraph*}(30,10)
          \fmfleftn{l}{3}
          \fmfrightn{r}{3}
          \fmfv{d.sh=circle,d.f=30,d.si=6mm}{vl,vr}
          \fmfv{d.sh=circle,d.f=full,d.si=1.6mm}{r2}
          \fmfforce{(8mm,5mm)}{vl}
          \fmfforce{(22mm,5mm)}{vr}
          \fmfleft{i}\fmfright{o}\fmf{phantom}{i,vl,vr,o}\fmffreeze
          \fmf{plain}{l1,vl,l3}
          \fmf{dashes,left=1}{vr,r2,vr}
          \fmf{slplainm,left=0.5}{vl,vr}
          \fmf{plain,right=0.5}{vl,vr}
        \end{fmfgraph*}}}}\,.
\end{align}
If one had not truncated the flow and Ward identity of the
two-particle Green function, they would have generated the missing
terms.  The difference to the formulation of the Ward identities in
terms of response functions can be seen in \eqref{eq:wiresp} and
\eqref{eq:polwardtrunc1}: while the one-particle response function
follows the same flow equation \eqref{eq:trunc} as the one-particle
Green function, only with an additional $A$ derivative, the
two-particle Green function follows by truncation a different flow
equation and is, therefore, not determined by the same approximation
(in the sense explained below) as the one-particle function.  Hence,
self-consistency and with it the self-consistent Ward identities are
violated.  This will be illustrated more clearly in the next section
\ref{sec:ward:sc:cons} on conserving approximations.

Note, however, that this violation may be not so bad numerically: the
violation terms highlighted in the boxes all have the structure of the
terms neglected by the truncation,
\begin{align}
  \label{eq:truncated}
  \parbox{28mm}{\unitlength=1mm\fmfframe(2,0)(2,0){
      \begin{fmfgraph*}(24,10)
        \fmfv{d.sh=circle,d.f=shaded,d.si=6mm}{vl,vr}
        \fmfforce{(6mm,5mm)}{vl}
        \fmfforce{(18mm,5mm)}{vr}
        \fmfleftn{l}{3}
        \fmfright{o}
        \fmf{plain}{l2,vl}
        \fmf{slplainm}{vl,vr}
        \fmf{plain}{vr,o}
        \fmffreeze
        \fmf{plain}{l1,vl}
        \fmf{plain}{l3,vl}
      \end{fmfgraph*}}}
  & \text{ (Polchinski),} &
  \parbox{32mm}{\unitlength=1mm\fmfframe(0,2)(2,2){
      \begin{fmfgraph*}(30,10)
        \fmfleftn{l}{3}
        \fmfrightn{r}{3}
        \fmfv{d.sh=circle,d.f=30,d.si=6mm}{vl,vr}
        \fmfforce{(8mm,5mm)}{vl}
        \fmfforce{(22mm,5mm)}{vr}
        \fmfleft{i}
        \fmfright{o}
        \fmf{phantom}{i,vl,vr,o}
        \fmffreeze
        \fmf{plain}{l1,vl,l3}
        \fmf{plain}{r1,vr,r3}
        \fmf{slplainm,left=0.5}{vl,vr}
        \fmf{plain,right=0.5}{vl,vr}
      \end{fmfgraph*}}}
  & \text{(\onePI)}
\end{align}
but with another dashed line closed, and vice versa with the slashed
and dashed lines interchanged.  Therefore, if the truncation is
justified because the truncated terms are small, then also the
violation will be small.


\subsection{Conserving approximations of Baym and Kadanoff}
\label{sec:ward:sc:cons}

Baym and Kadanoff \cite{BK61,Bay62} introduced a formalism to obtain
\emph{conserving approximations} which by construction satisfy the
self-consistent Ward identities.  We shall compare the \fRG\ to these
approximations and see whether, under certain conditions, the \fRG\ 
might also provide conserving approximations.

\subsubsection*{Proof of number conservation}

In order to make the reader familiar with the derivation of conserving
approximations, we shall discuss the proof of the number conservation
law due to \cite{Bay62}.  During the proof it will become clear what
the requirements are, and we will try to answer the question whether
truncated \fRG\ schemes might satisfy these requirements.

Baym and Kadanoff formulate their theory in terms of propagator lines
as basic objects, not the source fields $\eta$, $\bar\eta$ we have
used.  Therefore, they add an external field to the partition function,
\begin{align*}
  Z(U) := \int [d\psi\bar\psi]\, e^{S[\psi,\bar\psi] -
    \inner{\bar\psi}{U\psi}}
\end{align*}
with $U=U(1,1')$ \emph{bilocal} in space-time:\footnote{This should
  not easily be confused with the two-particle interaction $U^\Lambda$
  in Chapter~\ref{sec:lutt}.} $\inner{\bar\psi}{U\psi} = \int d1 \,
d1' \, \bar\psi(1) \, U(1,1') \, \psi(1')$.  The labels $1$, $1'$ are
abbreviations for $x_1$, $x_{1'}$ with all space-time components, and
the integral $\int d1 \equiv \int dr_1 \, \int_0^{-i\beta} dt_1$.
Thereby, all Green functions are functionals of the external field
$U$, such as the bare propagator $G_0(1-1';U)$ and the full propagator
$G(1,1';U)$.  The linear response of $G(1,1';U)$ to the external field
is the two-particle correlation function,
\begin{align*}
  L(12,1'2') := -\frac{\delta G(1,1';U)} {\delta U(2',2)}\at_{U=0} 
  = [G_2(12,1'2') - G(1-1') G(2-2')]_{U=0} \,.
\end{align*}
The Dyson equation
\begin{align}
  \label{eq:dyson1}
  G^{-1}(1,1';U) = G_0^{-1}(1-1') - U(1,1') - \Sigma(1,1';G(U))
\end{align}
shall be satisfied exactly even by an approximate $G(U)$ and self
energy $\Sigma$.

The first requirement is that $\Sigma$ be a functional of the full
propagator $G(U)$ and the \emph{bare} gauge-invariant density-density
interaction $V_0$.  We shall see below that this may be relaxed a
little to include any propagator that transforms in the same way as
$G(U)$ under change of $U$.

In order to show that the approximate $L$ satisfies the local number
conservation law, we choose an external disturbance $U$ that corresponds
to a gauge transformation,
\begin{align*}
  \inner{\bar\psi}{U\psi}
  = \int d1\, \left[ 
    \frac{\partial \alpha(1)}{\partial t_1} \rho(1) +
    \vec\nabla \alpha(1) \cdot
    \bigl\{\vec{j}(1)
    + \frac{1}{2m} \vec\nabla \alpha(1) \rho(1)\bigr\} 
  \right].
\end{align*}
The equation for the bare propagator becomes
\begin{align}
  \label{eq:gauge-g0}
  \Bigl\{ i\frac{\partial}{\partial t_1}
    - \frac{\partial \alpha(1)}{\partial t_1} 
    + \frac{1}{2m} \bigl[\vec\nabla_1
    + i\vec\nabla \alpha(1)\bigr]^2
  \Bigr\} G_0(1,1';\alpha) = \delta(1-1')
\end{align}
with the solution
\begin{align}
  \label{eq:g0trans}
  G_0(1,1';\alpha) = e^{-i\alpha(1)}\, G_0(1-1')\, e^{i\alpha(1')}
\end{align}
where $G_0(1-1')$ satisfies \eqref{eq:gauge-g0} with $\alpha \equiv
0$.  If we assume the boundary condition of the external disturbance
to be
\begin{align}
  \label{eq:bndcond}
  \alpha(r,\tau=0) = \alpha(r,\tau=-i\beta)
\end{align}
then also the solution $G_0(\alpha)$ satisfies this boundary
condition.  The full propagator is the solution of the Dyson equation
\eqref{eq:dyson1},
\begin{align}
  & \Bigl\{ i\frac{\partial}{\partial t_1}
    - \frac{\partial \alpha(1)}{\partial t_1} 
    + \frac{1}{2m} \bigl[\vec\nabla_1
    + i\vec\nabla \alpha(1)\bigr]^2
  \Bigr\} G(1,1';\alpha)
  - \int d\bar 1\, \Sigma(1,\bar 1;G(\alpha))\, G(\bar 1,1';\alpha)
  \notag \\
  \label{eq:gauge-g}
  & = \delta(1-1').
\end{align}
We will now show that if $\Sigma$ is a functional of $G(\alpha)$ (as
we have assumed), also $G(\alpha)$ will transform analogously to
equation \eqref{eq:g0trans}.  To this end, substitute
\begin{align}
  \label{eq:gsubst}
  G(1,1';\alpha) \mapsto e^{-i\alpha(1)}\, \bar G(1,1';\alpha)\,
  e^{i\alpha(1')} 
\end{align}
in equation \eqref{eq:gauge-g} and in the functional
$\Sigma(G(\alpha))$.  If and only if all propagators in $\Sigma$
depend on $U$ in the same way, at each vertex $V_0$ there will be four
phase factors.  Because particles are assumed to be
conserved at each vertex (by the gauge invariance $\delta_\alpha V_0 =
0$), these factors cancel exactly at each interaction vertex and
remain only at the external legs, such that $\Sigma$ transforms as
\begin{align*}
  \Sigma(1,1';G(\alpha)) = e^{-i\alpha(1)}\, \Sigma(1,1';\bar G)\,
  e^{i\alpha(1')}.
\end{align*}
Then, equation \eqref{eq:gauge-g} becomes
\begin{align*}
  & \Bigl\{ i\frac{\partial}{\partial t_1}
    - \frac{\partial \alpha(1)}{\partial t_1} 
    + \frac{1}{2m} \bigl[\vec\nabla_1 + i\vec\nabla \alpha(1)\bigr]^2
  \Bigr\} e^{-i\alpha(1)}\, \bar G(1,1';\alpha)\,
  e^{i\alpha(1')} \\
  & \quad - \int d\bar 1\, e^{-i\alpha(1)}\, \Sigma(1,\bar 1;\bar G)\,
  \bar G(\bar 1,1';\alpha)\, e^{i\alpha(1')} \\
  & = e^{-i\alpha(1)}\, \left[ 
    \Bigl\{ i\frac{\partial}{\partial t_1} + \frac{1}{2m} \vec\nabla_1^2
    \Bigr\} \bar G - \int \Sigma\, \bar G
  \right] e^{i\alpha(1')}
  = \delta(1-1').
\end{align*}
Because of the $\delta$ function on the \RHS\ the two phase factors
cancel, and we obtain
\begin{align*}
  \Bigl\{ i\frac{\partial}{\partial t_1} + \frac{1}{2m} \vec\nabla_1^2
  \Bigr\} \bar G - \int \Sigma\, \bar G
  = \delta(1-1'),
\end{align*}
which is the Dyson equation \eqref{eq:gauge-g} defining
$G(1-1';\alpha=0)$.  Because the solution to this equation is unique,
\begin{align*}
  \bar G(1,1';\alpha) = G(1-1';\alpha=0)\,,
\end{align*}
and using equation \eqref{eq:gsubst},
\begin{align}
  \label{eq:gtrans}
  G(1,1';\alpha) = e^{-i\alpha(1)}\, G(1-1';\alpha=0)\,
  e^{i\alpha(1')}\,,
\end{align}
which is the same transformation law as for the bare propagator,
equation \eqref{eq:g0trans}.  Now we expand both sides of equation
\eqref{eq:gtrans} to first order in $\alpha$:
\begin{align*}
  & \int d2\, \Bigl\{
    \frac{\partial\alpha(2)}{\partial t_2} L(12,1'2) +
    \vec\nabla\alpha(2) \cdot \left[ \frac{\vec\nabla_2 -
          \vec\nabla_2'}{2im} L(12,1'2') \right]_{2'=2}
  \Bigr\} \\
  & = i[\alpha(1)-\alpha(1')] G(1-1').
\end{align*}
Integrating by parts on the \LHS\ using the boundary condition
\eqref{eq:bndcond} and comparing the coefficient of $\alpha(2)$ yields
the number conservation law for $L$, which is equivalent to the
self-consistent Ward identity \eqref{eq:gscward}:
\begin{align*}
  \frac{\partial}{\partial t_2} L(12,1'2) +
  \vec\nabla_2 \cdot \left[ \frac{\vec\nabla_2 - \vec\nabla_2'}{2im} L(12,1'2')
  \right]_{2'=2} = -i[\delta(1-2)-\delta(1'-2)] G(1-1').
\end{align*}

The conservation law in the $1,1'$ variables of $L$ follows if we
demand in addition that $L$ be symmetric in $1,1' \leftrightarrow
2,2'$:
\begin{align*}
  \frac{\delta G(1,1')}{\delta U(2',2)} = 
  \frac{\delta G(2,2')}{\delta U(1',1)}.
\end{align*}
This requirement of vanishing ``curl'' implies, except for
pathological cases, that there exists a functional $W(U)$ such that
\begin{align*}
  G(1,1') = \frac{\delta W}{\delta U(1',1)}.
\end{align*}

\subsubsection*{The $\boldsymbol\Phi$ functional}

For completeness, let us mention that Baym and Kadanoff express the
$1,1'$ conservation of $L$ also as a condition on $\Sigma$.  $\Sigma$
is assumed there---in contrast to this work---to be a functional only
of $G(U)$ and $V_0$ but not of other propagators like $G_0(U)$, or $U$
directly.  Then they derive that a similar vanishing-``curl''
condition must be required of $\Sigma$,
\begin{align*}
  \frac{\delta \Sigma(1,1')}{\delta G(2',2)} = 
  \frac{\delta \Sigma(2,2')}{\delta G(1',1)},
\end{align*}
and hence, there exists a functional $\Phi[G(U),V_0]$, such that
\begin{align*}
  \Sigma(1,1') = \frac{\delta \Phi}{\delta G(1',1)},
\end{align*}
and $\delta\Sigma/\delta G$ is the \emph{effective particle-hole
  interaction}.  In equilibrium when $U(1,1') = \delta(t_1-t_1')\,
\bar U(r_1,r_1')$, $W$ and $\Phi$ are related by $W = \Phi -
\tr(\Sigma G) - \tr \ln (-G)$.  An approximation for $\Sigma[G]$ that
can be written as $\delta \Phi / \delta G$ is called
\emph{$\Phi$-derivable}.

\subsubsection*{Diagrammatic interpretation of conserving approximations}

Let us give a diagrammatic illustration of the Baym-Kadanoff
formalism.  In the exact theory $W(U)$ contains every vacuum diagram
of perturbation theory with bare vertices $V_0$ and propagators
$G_0(U)$.  A derivative with respect to $U$ acts on the $G_0(U)$ lines
in each diagram by plucking out one line.  The full propagator $G(U) =
\delta W/\delta U$ is, therefore, made up of all connected vacuum
diagrams with two external legs.  A second derivative with respect to
$U$ plucks out a second line from each diagram, generating $L=-\delta
G/\delta U$, which consists of all diagrams with four external legs.

Consider approximations defined by choosing a possibly infinite subset
of all Feynman diagrams for $W(U)$, $G(U)$, and $L$.  In principle,
one could choose completely different sets of diagrams for each
correlation function.  But an approximation is conserving if and only
if $L=-\delta G/\delta U$ is satisfied \emph{exactly}, \ie, the
diagrams contributing to $L$ are obtained by plucking out one line in
each diagram of $G$, or equivalently, the $G$ diagrams are obtained by
closing one loop on each diagram of $L$.  Such an approximation for
$L$ is conserving in the $2,2'$ variables; it can be made conserving
also in the $1,1'$ variables if we demand further that the diagrams of
$G$ are derived from those of $W$ by plucking out one line.

Obviously this leaves no room for truncations only on a certain level
of the hierarchy of Green functions: the flow equation for the
zero-particle component $W(U)$ must completely determine all higher
flow equations in order that the classes of diagrams contributing to
each Green function are compatible.


\subsection{How important is self-consistency?}
\label{sec:ward:sc:import}

It has become clear that at the core of the problem are not the
response-function Ward identities but self-consistency: the relation
$L=-\delta G/\delta U$ is not satisfied in the above example where the
one-particle function flows but the flow of the two-particle function
is truncated.  This incompatibility of the truncated flow with
conserving approximations raises the question under which
circumstances it is a problem not to have a conserving approximation.
The answer depends on the physical problem at hand: there are problems
which work surprisingly well in the truncated \fRG, while others fail
miserably.

A favorable example are the \oneD\ impurity problems presented in
Chapters~\ref{sec:lutt} and \ref{sec:results}.  Already for simple
truncations the \fRG\ results are very close to the exact asymptotic
solution known from Bethe ansatz.  This suggests that the terms
neglected by the truncation are small.  Following the discussion
surrounding equation \eqref{eq:truncated}, also self-consistency is
then fulfilled to a high degree of accuracy.  Moreover, in the model
of Chapter~\ref{sec:lutt} the self-consistent Hartree-Fock
approximation, a simple conserving approximation, is known to produce
the wrong physical phase (charge-density wave).  This is due to the
fact that an approximation, while being conserving, can miss important
contributions.  Generally, if the truncated flow does not diverge, one
can even determine some components of $G(U)$ and $L$
self-consistently: one can use any truncated flow to obtain an
approximate $W^\Lambda(U)$ in the presence of $U$ and take numerical
derivatives with respect to $U$ to obtain approximate values for
$G(U)$ and $L$ which are conserving by construction (cf.\ section
\ref{sec:lutt:density}).

An example of the opposite situation where the violation of the
self-consistent Ward identities is disastrous is the reduced \abb{BCS}
model \cite{SHML04}.  This model is solved exactly by the
self-consistent Hartree-Fock approximation.  In the presence of a tiny
symmetry-breaking term (gap) of magnitude $\epsilon$ in the action,
the interaction grows very large to $1/\epsilon$ but does not diverge.
In the truncated flow equation, this result is only reproduced if one
has exactly the correct value of the gap (a component of the
one-particle function) in the flow equation for the two-particle
interaction.  This would be guaranteed by the self-consistent Ward
identity relating the one- and two-particle functions; however, even
if the truncated flow violates the Ward identity only slightly the
interaction may diverge prematurely at $\Lambda>0$.  This problem can
be solved by a modification of the flow-equation hierarchy
\cite{Kat04}: if the single-scale propagator $S^\Lambda$ in the
\onePI\ scheme is replaced by the $\Lambda$ derivative of the full
propagator $-\dl G^\Lambda = G^\Lambda (\dl Q^\Lambda - \dl
\Sigma^\Lambda) G^\Lambda$ in the truncated flow equation for the
two-particle vertex, then the \fRG\ flow reproduces exactly the
self-consistent Hartree-Fock solution for models where the
two-particle interaction has a reduced momentum dependence.  This
modification leads to significant improvements also in the
single-impurity Anderson model \cite{HMPS04}.

There are promising approaches to self-consistency by writing the
\fRG\ flow equations in terms of both fermionic and bosonic degrees of
freedom.  For instance, $(i)$ \cite{Wet02} thereby obtains the flow of
\abb{2PI} vertex functions.  $(ii)$ The Luttinger model without
backscattering is treated in \cite{Schuetz04}; using the separate
conservation of the number of left and right movers the flow-equation
hierarchy can be closed and solved exactly.


\section{Summary}
\label{sec:ward:sum}

In this chapter I have shown how Ward identities, which express the
symmetry of the Hamiltonian in terms of Green or vertex functions, are
derived in the functional formalism.  There are two common
formulations of Ward identities.  In the field-theoretical and
high-energy physics literature, Ward identities are written in terms
of response functions.  This form of the Ward identities relates for
example the self energy to the current response but assumes no
particular relation between the self energy and the interaction.  A
momentum cutoff breaks these response-function Ward identities.  This
is a problem for the treatment of gauge theories, leading either to
modified Ward identities or the introduction of a background gauge
field $\bar A$.  Alternatively, one can use a manifestly
gauge-invariant flow scheme such as the temperature flow.  It is shown
that in this case even truncated flows satisfy the unmodified Ward
identities exactly on all scales.

In the condensed-matter literature, a different form of Ward identity
is more common which assumes self-consistency: the response functions
in the Ward identities can be expressed as higher Green functions with
a loop closed by a special propagator.  An example of approximations
which satisfy these self-consistent Ward identities without cutoff are
the conserving approximations of Baym and Kadanoff.  However, I have
shown that self-consistency is violated by common truncations, which
neglect the flow of Green functions beyond a certain level in the
flow-equation hierarchy.  For special models with exact mean-field
solutions, modified truncated flow equations are known which reproduce
these solutions.


\end{fmffile}


\chapter{Functional RG technique in one dimension}
\label{sec:lutt}

In this chapter I introduce the one-dimensional lattice model of a
Luttinger liquid (cf.\ Chapter~\ref{sec:results}) and explain in
detail how the \fRG\ is used to compute observables such as the
effective impurity potential, the density-response vertex and the
conductance.  This chapter is organized as follows: in the first
section \ref{sec:lutt:model} the microscopic lattice model is defined.
In section \ref{sec:lutt:flow} I show how the flow equations are set
up and solved, with an emphasis on the finite-temperature flow and a
new efficient algorithm in \oneD.  I proceed to explain how to compute
the conductance in the \fRG\ framework in section
\ref{sec:lutt:condtech}, giving an argument why vertex corrections
play no role in our approximation.


\section{Microscopic model}
\label{sec:lutt:model}

Consider a model of spinless fermions on a \oneD\ lattice with
nearest-neighbor interaction and various types of impurity potentials.
Following \cite{AEMMSS04}, the Hamiltonian has the form
\begin{align*}
  H = H_0 + H_\tint + H_\imp
\end{align*}
with the kinetic term given by nearest-neighbor hopping with an
amplitude $-t$ and chemical potential $\mu$ (I will henceforth choose
units such that $t=1$, and the lattice spacing $a=1$),
\begin{align*}
  H_0 = -t \sum_j \bigl( c_{j+1}^\dagger c_j^{\phantom\dagger} +
  c_j^\dagger c_{j+1}^{\phantom\dagger} \bigr)
  - \mu \sum_j n_j,
\end{align*}
where $n_j = c_j^\dagger c_j^{\phantom\dagger}$ is the local density
operator.  The nearest-neighbor interaction of strength $U_{j,j+1} =
U_{j+1,j}$ on the bond between sites $j$ and $j+1$ enters as
\begin{align}
  \label{eq:Hint}
  H_\tint = \sum_j U_{j,j+1} n_j n_{j+1}
\end{align}
while the static impurity potential $V_{jj'}$ is represented by a term
\begin{align*}
  H_\imp = \sum_{j,j'} V_{jj'}^{\phantom\dagger}
  \, c_j^\dagger c_{j'}^{\phantom\dagger}.
\end{align*}

\fig[width=0.8\linewidth]{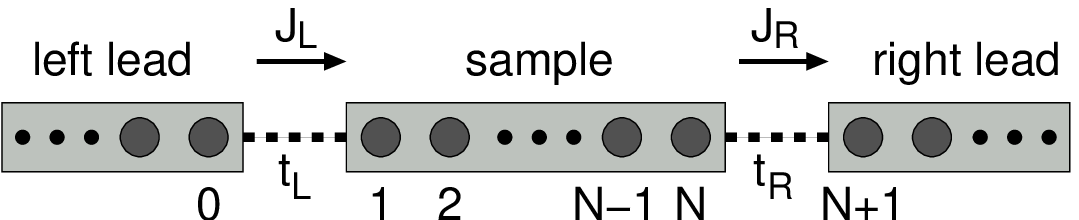}{fig:leads}{The microscopic setup of
  the system with leads.  The current operators $J_{L,R}$ will only be
  needed for the conductance calculations.}
For the conductance calculations, we will couple an interacting system
on lattice sites $1,\dotsc,N$ on both sides to semi-infinite,
non-interacting leads which are described completely by $H_0$ (figure
\ref{fig:leads}).  In order that electrons do not scatter off the
beginning of the interacting region, the interaction has to be
switched on smoothly.  Explicitly, we choose a spatial profile
\begin{align*}
  U_{j,j+1} & := U \frac{\arctan{[(j-j_s)/w]}-\arctan{[(1-j_s)/w}]}
  {\arctan{[(N/2-j_s)/w]}-\arctan{[(1-j_s)/w}]} && (j=1,\dotsc,N/2)
\end{align*}
for the left side of the system, and likewise for the right, where $U$
is the bulk interaction.  We have chosen the parameters $w=4$ and
$j_s=56$ such that the interaction falls to 10\% of its value over a
typical distance 22 lattice sites.\footnote{Although the difference
  $\abs{U_{j_s-1,j_s} - U_{j_s,j_s+1}} \approx 0.08 \, U$ is rather
  large, the corresponding backscattering component of the effective
  potential $\Sigma$ is typically below $10^{-4}$, and the conductance
  at $T=0$ deviates from the unitary limit by less than $10^{-8}$.}


\subsection{Projection method applied to the wire}
\label{sec:lutt:model:wireproj}

In order to treat this infinite system with leads numerically, we
express it (exactly) by an effective Hamiltonian on the $N$-site
interacting region via the projection technique.  Consider splitting
the Hilbert space of the Hamiltonian $H$ into disjoint subspaces with
projection operators $P+Q=\id$:
\begin{align*}
  H =
  \begin{pmatrix}
    H_{PP} & H_{PQ} \\
    H_{QP} & H_{QQ}
  \end{pmatrix}.
\end{align*}
The one-particle Green function is the resolvent
\begin{align*}
  G(z) := \frac{1}{z-H} =
  \begin{pmatrix}
    G_{PP}(z) & G_{PQ}(z) \\
    G_{QP}(z) & G_{QQ}(z)
  \end{pmatrix}
\end{align*}
with components \cite[equation (2.7.23)]{Pre86}
\begin{align}
  \label{eq:GPP}
  G_{PP}(z) & = P\frac{1}{z-H}P = \frac{1}{zP-H_{PP}
    - \underbrace{H_{PQ}\frac{1}{zQ-H_{QQ}}H_{QP}}_{=:\Sigma_{PP}(z)}} \\
  G_{PQ}(z) & = P\frac{1}{z-H}Q = -G_{PP}(z)H_{PQ}\frac{1}{zQ-H_{QQ}} \notag \\
  G_{QP}(z) & = Q\frac{1}{z-H}P = -\frac{1}{zQ-H_{QQ}}H_{QP}G_{PP}(z) \notag \\
  G_{QQ}(z) & = Q\frac{1}{z-H}Q = \frac{1}{zQ-H_{QQ}} +
    \frac{1}{zQ-H_{QQ}} H_{QP} G_{PP}(z) H_{PQ} \frac{1}{zQ-H_{QQ}} \notag .
\end{align}

For the wire, $P$ shall project onto the Hilbert space of the states
in the interacting sample with site indices $1,\dotsc,N$, while $Q$
shall denote the remaining non-interacting leads on sites $\dotsc,0$
and $N+1,\dotsc$.  Thus,
\begin{align*}
  H_{PP} & = -t \sum_{n=1}^{N-1}
  \bigl( \ket{n+1}\bra{n}+\ket{n}\bra{n+1} \bigr)
  - \mu \sum_{n=1}^{N} \ket{n}\bra{n} + H_\tint + H_\imp \\
  H_{PQ} & = -t_L \ket{1}\bra{0} - t_R \ket{N}\bra{N+1} \\
  H_{QP} & = -t_L \ket{0}\bra{1} - t_R \ket{N+1}\bra{N} \\
  H_{QQ} & = -t \sum_{\substack{n<0,\\n\geq N+1}}
  \bigl( \ket{n+1}\bra{n}+\ket{n}\bra{n+1} \bigr)
  - \mu \sum_{\substack{n\leq 0,\\n\geq N+1}} \ket{n}\bra{n}.
\end{align*}
The left and right leads in $H_{QQ}$ do not couple directly but only
through the sample.  In order to compute $\Sigma_{PP}$ in equation
\eqref{eq:GPP} we need to know the Green function of the (left) lead
at the interface site $0$, $\bra{0} (zQ-H_{QQ})^{-1} \ket{0}$.  To
this end, consider a semi-infinite lead ranging from $-\infty$ up to
some site $j$, and denote the Green function at the rightmost site $j$
as $g_L(z) := (z-H_0)_{j,j}^{-1}$.  Consider adding one more site
$j+1$ to the right, with the same hopping amplitude $-t$.  Because the
lead is semi-infinite and homogeneous, the Green function $g_L'(z)$
at the new site $j+1$ should be the same as $g_L(z)$.  Again using
the same formula \eqref{eq:GPP} with $Q$ now denoting the states on
sites $-\infty,\dotsc,j$ and $P$ for $j+1$,
\begin{align*}
  [g_L'(z)]^{-1} & =
  (z-H_0)_{j+1,j+1} - (H_0)_{j+1,j} \,
  (z-H_0)_{j,j}^{-1} \, (H_0)_{j,j+1} \\
  & = (z+\mu) - t \, g_L(z) \, t \overset != [g_L(z)]^{-1}.
\end{align*}
This leads to a quadratic equation for $g_L(z)$,
\begin{align*}
  t^2 g_L^2(z) - (z+\mu) \, g_L(z) + 1 = 0
\end{align*}
with solution ($t=1$)
\begin{align}
  \label{eq:condGL}
  g_L(z) & = \frac 12 \left( z+\mu \mp i\sqrt{4-(z+\mu)^2} \right)
\end{align}
and $g_R(z)=g_L(z)$ because the right lead has the same structure as
the left lead.  $g_L(z)$ has a branch cut at the real axis $\Im\,z =
0$.  The sign $\pm$ is chosen such that the imaginary part of
$g_L(z)$ changes sign across the branch cut.  For the local density
of states of the leads we obtain
\begin{align*}
  \rho_L(\omega)
  & := -\frac{1}{\pi} \Im \, g_L(\omega+i0)
  = \frac{1}{2\pi} \sqrt{4-(\omega+\mu)^2},
\end{align*}
for $\abs{\omega+\mu} \leq 2$.  Now we have all ingredients for the
lead contributions to the propagator of the sample, equation
\eqref{eq:GPP}:
\begin{align}
  \Sigma_{PP}(z) & := H_{PQ} \frac{1}{zQ-H_{QQ}} H_{QP} \notag \\
  & = t_L \ket{1}\bra{0} (zQ-H_{QQ})^{-1} \ket{0}\bra{1} t_L \notag \\
  & \quad + t_R \ket{N}\bra{N+1} (zQ-H_{QQ})^{-1} \ket{N+1}\bra{N} t_R
  \notag \\
  \label{eq:condGrel0}
  & = t_L^2\, g_L(z)\, \ket{1}\bra{1} + 
  t_R^2\, g_R(z)\, \ket{N}\bra{N} \\[1ex]
  G_{j0}(z) & = \bra{j}G_{PQ}(z)\ket{0} \notag \\
  & = -\bra{j}G_{PP}(z)\ket{1}\, t_L\, \bra{0}(zQ-H_{QQ})^{-1}\ket{0}
  \notag \\
  \label{eq:condGrel1}
  & = -G_{j1}(z)\, t_L\, g_L(z) \qquad (j=1,\dots,N) \\[1ex]
  G_{N+1,j}(z) & = \bra{N+1}G_{QP}(z)\ket{j} \notag \\
  & = -\bra{N+1}(zQ-H_{QQ})^{-1}\ket{N+1}\, t_R\,
  \bra{N}G_{PP}(z)\ket{j}
  \notag \\
  \label{eq:condGrel2}
  & = -g_R(z)\, t_R\, G_{Nj}(z) \qquad (j=1,\dots,N) \\[1ex]
  G_{N+1,0}(z) & = \bra{N+1}G_{QQ}(z)\ket{0} \notag \\
  & = \bra{N+1}(zQ-H_{QQ})^{-1}\ket{0} \;\;
  \text{\small ($=0$ as left and right leads do not couple directly)}
  \notag \\
  & + \bra{N+1}(zQ-H_{QQ})^{-1}\ket{N+1}\, t_R\,
  \bra{N}G_{PP}(z)\ket{1}\, t_L\, \bra{0}(zQ-H_{QQ})^{-1}\ket{0}
  \notag \\
  \label{eq:condGrel3}
  & = g_R(z)\, t_R\, G_{N1}(z)\, t_L\, g_L(z).
\end{align}

In the non-interacting leads, $H_{QQ} = H_0$.  In the sample, the
quadratic (non-interacting) part of $H_{PP}$ shall include the
impurity:
\begin{align*}
  \xi_{jj'} := (H_0 + H_\imp)_{jj'}
  = -t (\delta_{j,j'+1} + \delta_{j,j'-1}) - \mu \delta_{jj'} + V_{jj'}.
\end{align*}
The lead contribution is also independent of the interaction,
\begin{align}
  \label{eq:condleads}
  \Sigma^\leads(z) & := \Sigma_{PP}(z) =
  t_L^2 \, g_L(z) \, \ket 1 \bra 1 +
  t_R^2 \, g_R(z) \, \ket N \bra N,
\end{align}
and all non-interacting contributions are combined into the inverse
\emph{bare} propagator in the sample,
\begin{align}
  \label{eq:condinvbareprop}
  Q_{jj'}(z) \equiv [G_0^{-1}(z)]_{jj'}
  := z \delta_{jj'} - \xi_{jj'} - \Sigma^\leads_{jj'}(z).
\end{align}
The interaction $H_\tint$ creates an effective one-particle potential
$\Sigma(z,T)$ at temperature $T$.  The \emph{full} sample propagator
is determined by the Dyson equation,
\begin{align}
  \label{eq:cond-dyson}
  [G^{-1}(z,T)]_{jj'} & = Q_{jj'}(z) - \Sigma_{jj'}(z,T).
\end{align}


\subsection{Bare vertices}
\label{sec:lutt:model:pathint}

The \fRG\ is formulated in terms of a path integral weighted by the
action, so the bare and interaction parts of the Hamiltonian have to
be translated into the Lagrangian language: the inverse bare
propagator $Q(z)$ defined in equation \eqref{eq:condinvbareprop}
becomes the quadratic part of the Lagrangian, and the bare interaction
$H_\tint$ becomes the interaction part:
\begin{align*}
  S[\psi,\bar\psi] = T \sum_{n,j,j'} Q_{jj'}(i\omega_n) \,
  \bar\psi_j(i\omega_n) \psi_{j'}(i\omega_n) +
  \sum_{1,1',2,2'} I_{1',2';1,2} \,
  \bar\psi(1') \bar\psi(2') \psi(2) \psi(1)
\end{align*}
where $I_{1',2';1,2}$ is the bare antisymmetrized interaction vertex
equivalent to $H_\tint$, and the sum over $1,1',\dotsc$ includes
Matsubara frequencies and lattice indices.  The nearest-neighbor
interaction conserves frequency but has no further frequency
dependence.  The remaining spatial dependence has the real-space form
\begin{align}
  \label{eq:bareintn}
  I_{j_1',j_2';j_1,j_2} = U_{j_1,j_2}
  [\delta_{j_1,j_2-1} + \delta_{j_1,j_2+1}]
  (\delta_{j_1,j_1'} \delta_{j_2,j_2'}
   - \delta_{j_1,j_2'} \delta_{j_2,j_1'}).
\end{align}
In the bulk $U_{j,j+1} \equiv U$ is homogeneous, hence the bare
interaction is translationally invariant and can be expressed in a
momentum basis:
\begin{align}
  \label{eq:bareintnmom}
  I_{k_1',k_2';k_1,k_2}
  = 2U[\cos(k_1'-k_1) - \cos(k_2'-k_1)]
  \delta_{k_1+k_2,k_1'+k_2'}^{(2\pi)}
\end{align}
where the Kronecker $\delta$ implements momentum conservation (modulo
$2\pi$).


\section{Functional RG flow equations}
\label{sec:lutt:flow}

For computing the properties of the \oneD\ fermion system, we use the
\onePI\ version of the \fRG, cf.\ section \ref{sec:rg:flow:gamma}.  We
cut off the infrared part of the free propagator on a scale $\Lambda$
and differentiate the generating functional for the vertex functions
with respect to this scale.  Thereby we obtain an exact hierarchy of
flow equations for the irreducible vertex functions.


\subsection[Truncated \onePI\ flow equations]{Truncated 1PI flow equations}
\label{sec:lutt:flow:gen}

As in \cite{AEMMSS04}, we choose a cutoff in Matsubara frequency.  The
cutoff bare propagator $G_0^\Lambda(i\omega)$ (with the notation $G_0
\equiv C$) is defined by
\begin{align}
  \label{eq:flowcutoff}
  G_0^\Lambda(i\omega)
  := \chi^\Lambda(\omega) \, G_0(i\omega),
\end{align}
where the characteristic function $\chi^\Lambda(\omega)$ is unity on
the high-energy modes and vanishes on the low-energy modes.  The exact
form of $\chi^\Lambda$ depends on whether $T=0$ or $T>0$, as explained
below.

Second, we truncate the infinite hierarchy of flow equations by
neglecting the flow of the three-particle and higher vertex functions.
This closes the hierarchy of flow equations for the one- and
two-particle vertex functions and is justified by a small renormalized
interaction.  The results agree remarkably well quantitatively with
known exact results (\DMRG, Bethe ansatz) not only for weak but also
for moderate interaction strength \cite{AEMMSS04,AndThesis}.

The truncated \onePI\ flow equations are written in terms of the
one-particle vertex $\Sigma^\Lambda$ at scale $\Lambda$ (self energy)
and the two-particle interaction vertex\footnote{Not to be confused
  with the full functional $\Gamma^\Lambda[\phi,\bar\phi]$!}
$\Gamma^\Lambda$ \cite{AEMMSS04},
\begin{align}
  \dl \Sigma^\Lambda(1',1)
  & = -T \sum_{2,2'} e^{i\omega_2 0^+} \, S^\Lambda(2,2') \,
  \Gamma^\Lambda(1',2';1,2) \\
  \dl \Gamma^\Lambda(1',2';1,2)
  & = T \sum_{3,3'} \sum_{4,4'} G^\Lambda(3',3) \, 
  S^\Lambda(4,4') \notag \\
  & \quad \times \Bigl[ \Gamma^\Lambda(1',2';3,4) \,
  \Gamma^\Lambda(3',4';1,2) \notag \\
  & \qquad - \Gamma^\Lambda(1',4';1,3) \, \Gamma^\Lambda(3',2';4,2) 
  - (3 \leftrightarrow 4, 3' \leftrightarrow 4') \nonumber \\
  \label{eq:genuflow}
  & \qquad + \Gamma^\Lambda(2',4';1,3) \, \Gamma^\Lambda(3',1';4,2)
  + (3 \leftrightarrow 4, 3' \leftrightarrow 4') \Bigr].
\end{align}
The indices $1,2,\dotsc$ label both frequency and spatial indices.
The full propagator is determined from the self energy by the Dyson
equation
\begin{align}
  \label{eq:cond-dyson2}
  G^\Lambda = [Q^\Lambda - \Sigma^\Lambda]^{-1},
\end{align}
where $Q^\Lambda \equiv [G_0^\Lambda]^{-1} = Q/\chi^\Lambda$ is the
inverse bare cutoff propagator.  $S^\Lambda$ denotes the
\emph{single-scale propagator} which will select only modes with
frequency near $\Lambda$ for the frequency cutoff,
\begin{align}
  \label{eq:singlescale}
  S^\Lambda & := G^\Lambda \, \dot Q^\Lambda \, G^\Lambda
  = -\dot \chi^\Lambda \, \frac{1}{Q-\chi^\Lambda \Sigma^\Lambda} Q 
  \frac{1}{Q-\chi^\Lambda \Sigma^\Lambda} \,,
\end{align}
where the dot denotes $\dl$.  The convergence factor $e^{i\omega_2
  0^+}$ in the $\Sigma^\Lambda$ flow equation is only necessary to
define the initial condition of the flow at $\Lambda =
\Lambda_0\to\infty$ (see below).


\subsubsection*{Parametrization of the interaction}

The two-particle interaction vertex $\Gamma^\Lambda(1',2';1,2)$ is in
general a very complicated function.  However, in the low-energy limit
the flow of $\Sigma^\Lambda$ and $\Gamma^\Lambda$ is dominated by very
few channels of the interaction \cite{AndThesis,AEMMSS04}.  At any
$\Lambda$, perturbation theory in the renormalized interaction
strength provides a guide for a simple parametrization of
$\Gamma^\Lambda$.  For a model of spinless fermions with
nearest-neighbor interaction and only few impurities (\ie, not
disordered), the following parametrization captures the qualitative as
well as the quantitative features of Luttinger liquids very well.  The
renormalized interaction vertex $\Gamma^\Lambda$ is assumed to be
frequency independent beyond the conservation of frequency, and the
\RHS\ of the flow equation is evaluated with all external frequencies
set to zero.  Likewise, its spatial dependence is approximated by a
renormalized nearest-neighbor interaction $U^\Lambda$, and in the flow
all external momenta are projected to $\pm\kf$.  The internal lines
carry propagators without the self energy or the impurity potential,
which would lead to corrections only at higher order in the
interaction.

In the bulk $U_{j,j+1} \equiv U$ is homogeneous, and $\Gamma^\Lambda$
has the momentum-space form (independent of frequency)
\begin{align*}
  \Gamma_{k_1',k_2';k_1,k_2}^\Lambda & =
  2U^\Lambda [\cos(k_1'-k_1) - \cos(k_2'-k_1)]
  \delta_{k_1+k_2,k_1'+k_2'}^{(2\pi)}
\end{align*}
which is just the bulk bare interaction \eqref{eq:bareintnmom},
\begin{align*}
  I_{k_1',k_2';k_1,k_2} & =
  2U [\cos(k_1'-k_1) - \cos(k_2'-k_1)]
  \delta_{k_1+k_2,k_1'+k_2'}^{(2\pi)}
\end{align*}
rescaled by $U^\Lambda/U$.  The particular flow equation for
$U^\Lambda$ depends on the cutoff chosen and will be given below for
several types of cutoff.  However, the general form is a consequence
of the above parametrization,
\begin{align}
  \label{eq:uflow}
  \dl U^\Lambda & = -(U^\Lambda)^2 \, T\sum_\omega
  \dot\chi^\Lambda(\omega) \oint \frac{dp}{2\pi} f(p,\omega)
\end{align}
where $f(p,\omega)$ is the sum of the three different channels (PP,
PH, PH') in equation \eqref{eq:genuflow} \cite{AEMMSS04}.  For
instance at half filling where $\mu=0$ and $\xi_p = -2\cos(p) =
\xi_{-p}$,
\begin{align}
  \label{eq:bubblecontrib}
  f(p,\omega) & =
  \frac{2\sin^2(p)}{(i\omega-\xi_p)(-i\omega-\xi_{-p})} -
  \frac{\cos^2(p)}{(i\omega-\xi_p)^2} -
  \frac{[1+\sin(p)]^2}{(i\omega-\xi_p)(-i\omega-\xi_p)} \\
  & = -\cos^2(p) \left[ \frac{1}{(i\omega-\xi_p)(-i\omega-\xi_p)} +
    \frac{1}{(i\omega-\xi_p)^2} \right].
\end{align}
Going from the bulk $U_{j,j+1} \equiv U$ back to the lattice, we apply
the $U^\Lambda$ flow equation for each $U_{j,j+1}^\Lambda$
locally.

Parametrizing $\Gamma^\Lambda$ by a renormalized nearest-neighbor
interaction $U_{j,j+1}^\Lambda$ has the great advantage that the self
energy is a tridiagonal matrix in real space: only the matrix elements
$\Sigma_{j,j(\pm 1)}^\Lambda$ are non-zero.  The tridiagonal flow
equations for a general cutoff are
\begin{align}
  \label{eq:sigmaflowlattice}
  \begin{split}
    \dl \Sigma_{j,j}^\Lambda & =
    T \sum_{\omega} \dot\chi^\Lambda \sum_{r=\pm 1}
    U_{j,j+r}^\Lambda \, S_{j+r,j+r}^\Lambda(i\omega) \\
    \dl \Sigma_{j,j\pm 1}^\Lambda & =
    -T \sum_{\omega} \dot\chi^\Lambda
    U_{j,j\pm 1}^\Lambda \, S_{j,j\pm 1}^\Lambda(i\omega).
  \end{split}
\end{align}
Note that the self energy remains independent of frequency and real
because the interaction does not depend on frequency in our
parametrization.


\subsection{Frequency cutoff at zero temperature}
\label{sec:lutt:flow:freqT0}

At zero temperature we choose the sharp cutoff
\begin{align}
  \label{eq:zerotempsharpcutoff}
  \chi^\Lambda(\omega) & := \Theta(\abs\omega-\Lambda), &
  \dot\chi^\Lambda(\omega) & = -\delta(\abs\omega - \Lambda)
\end{align}
which cuts off all modes with frequency smaller than $\Lambda$.  As
explained in section \ref{sec:rg:flow:gamma}, a sharp cutoff at $T=0$
allows to integrate over the $\Theta$ step functions analytically
which leaves only \emph{smooth} propagators $\tilde G$ in each loop
diagram \eqref{eq:Gshcutoff},
\begin{align}
  \label{eq:tildeG}
  \tilde G^\Lambda(i\omega) := [Q(i\omega) - \Sigma^\Lambda]^{-1},
\end{align}
as opposed to $G^\Lambda(i\omega)$ from equation
\eqref{eq:cond-dyson2} which has a step at $\abs\omega = \Lambda$.
The flow equations \eqref{eq:sigmaflowlattice} at $T=0$ for a sharp
frequency cutoff then read (with $T\sum_\omega \mapsto \frac{1}{2\pi}
\int d\omega$)
\begin{align}
  \label{eq:sigmaflowt0}
  \dl \Sigma_{j,j}^\Lambda & =
  -\frac{1}{2\pi} \sum_{\omega=\pm\Lambda} \sum_{r=\pm 1}
  U_{j,j+r}^\Lambda \tilde G_{j+r,j+r}^\Lambda(i\omega) \\
  \dl \Sigma_{j,j\pm 1}^\Lambda & =
  \frac{1}{2\pi} \sum_{\omega=\pm\Lambda}
  U_{j,j\pm 1}^\Lambda \tilde G_{j,j\pm 1}^\Lambda(i\omega)\,.
\end{align}

\subsubsection*{Flow of the interaction}

Inserting equation \eqref{eq:zerotempsharpcutoff} into
\eqref{eq:uflow}, the flow equation for $U^\Lambda$ is
\begin{align}
  \label{eq:ufloweqt0}
  \dl U^\Lambda & = (U^\Lambda)^2 \, 
  \frac{1}{2\pi} \sum_{\omega=\pm\Lambda}
  \oint \frac{dp}{2\pi} f(p,\omega).
\end{align}
For instance at half filling, the momentum integral over the bubble
$f(p,\omega)$ is
\begin{align*}
  \oint \frac{dp}{2\pi} \sum_{\omega=\pm\Lambda} f(p,\omega)
  & = -\oint \frac{dp}{2\pi} \cos^2(p) \, 2 \Re
  \Bigl[ \frac{1}{(i\Lambda-\xi_p)(-i\Lambda-\xi_p)} +
    \frac{1}{(i\Lambda-\xi_p)^2} \Bigr] \\
  & = -\oint \frac{dp}{2\pi}
  \frac{[2\cos(p)]^4} {(\Lambda^2+[2\cos(p)]^2)^2} \\
  & = -\left( 1 - \Lambda \frac{\Lambda^2+6}{(\Lambda^2+4)^{3/2}} \right).
\end{align*}
The differential equation \eqref{eq:ufloweqt0} separates,
\begin{align*}
  \dl \left( \frac{1}{U^\Lambda} \right)
  & = \frac{1}{2\pi} \left( 1 - \Lambda
    \frac{\Lambda^2+6}{(\Lambda^2+4)^{3/2}} \right).
\end{align*}
Integrating each side separately from $\Lambda=\infty$ down to
$\Lambda$ with initial condition $U^{\Lambda=\infty} = U$,
\begin{align*}
  \frac{1}{U^\Lambda} - \frac{1}{U}
  & = \frac{1}{2\pi} \left( \Lambda -
    \frac{2+\Lambda^2}{\sqrt{4+\Lambda^2}} \right)
\end{align*}
and finally
\begin{align}
  \label{eq:uflowt0}
  U^\Lambda & = \frac{U}
  {1+\left(\Lambda-\frac{2+\Lambda^2}{\sqrt{4+\Lambda^2}}\right)
    U/(2\pi)}
  \xrightarrow{\Lambda\to 0}
  \frac{U}{1-U/(2\pi)} \, .
\end{align}
Even away from half filling, the flow equation can be integrated
analytically using contour integration \cite{AndThesis}.

At $T=0$, the simple expression \eqref{eq:uflowt0} yields the correct
low-energy asymptotics to second order in the renormalized vertex and
moreover contains the second-order corrections from the lattice
dispersion to the vertex (with all external lines at the Fermi
surface) at any scale $\Lambda$.

\subsubsection*{Initial conditions}

The flow, being given by the solution of an \abb{ODE} in $\Lambda$, is
determined uniquely by the flow equation and the initial condition.
At the initial upper cutoff scale $\Lambda=\Lambda_0$, the initial
condition has contributions from the bare interaction and from the
bare impurity potential \cite{AEMMSS04}.  At $T=0$, only the
combination $\tilde G^{-1} = Q-\Sigma^\Lambda$ appears in the flow,
such that there is no difference whether one treats the impurity
potential as the initial condition for $\Sigma^\Lambda$ at the
beginning of the flow or, alternatively, as part of the bare
propagator $Q$ as in this work.  The contribution of the interaction
as $\Lambda=\Lambda_0\to\infty$ is
\begin{align}
  \label{eq:initcondt0s}
  \Sigma_{1,1'}^{\Lambda_0} & := \frac 12 \sum_2 I_{1',2;1,2} \\
  \label{eq:initcondt0g}
  \Gamma_{1',2';1,2}^{\Lambda_0} & := I_{1',2';1,2} \,,
\end{align}
where $I_{1',2';1,2}$ is the bare antisymmetrized interaction
\eqref{eq:bareintn}.  The initial condition for the self energy is
usually compensated by a local potential to avoid that the filling
changes (see section \ref{sec:lutt:flow:scmu} below).  Only if the
interaction extends to an open boundary without lead, a boundary term
$\Sigma_{\text{boundary}}^{\Lambda_0} := -U/2$ remains.


\subsection{Frequency cutoff at finite temperature}
\label{sec:lutt:flow:freqTg0}

At $T>0$, a sharp cutoff $\chi^\Lambda(\omega_n) =
\Theta(\abs{\omega_n}-\Lambda)$ analogous to the $T=0$ case would lead
to a $\delta$ peak on the \RHS\ of the flow equation of
$\Sigma^\Lambda$ at every fermionic Matsubara frequency $\Lambda =
\abs{\omega_n}$.  This creates finite jumps in the integrated
$\Sigma^\Lambda$, therefore, the $\Theta$ integration formula
\eqref{eq:lemmamorris} cannot be applied as it is only valid for
continuous functions.  For the \abb{ODE} integration, a smooth right-hand
side is best, but this has to be balanced against the number of
Matsubara frequencies $\omega_n$ on which
$\dot\chi^\Lambda(\omega_n)>0$ for every particular value of
$\Lambda$.  

\begin{figure}[ht!]%
    \centerline{\input{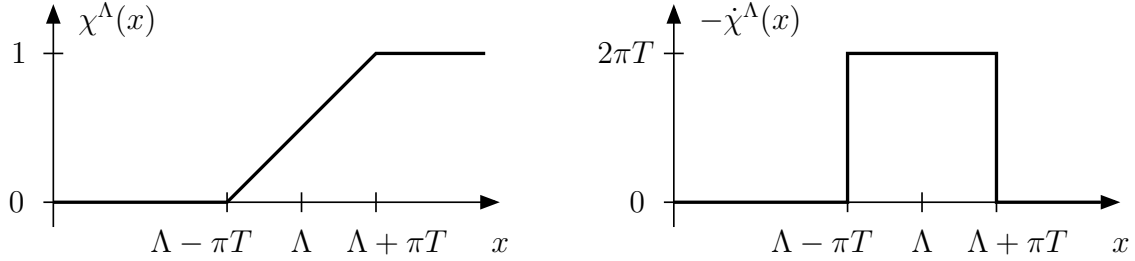}}
    \caption{\small\label{fig:tcutoff}The cutoff function used at finite
  temperature.}\end{figure}
As a compromise, we use a cutoff function (cf.\ figure
\ref{fig:tcutoff})
\begin{align*}
  \chi^\Lambda(\omega_n) & :=
  \begin{cases}
    0 & \abs{\omega_n} \leq \Lambda-\pi T \\
    \frac{1}{2} + \frac{\abs{\omega_n}-\Lambda}{2\pi T}
    & \Lambda-\pi T \leq \abs{\omega_n} \leq \Lambda+\pi T \\
    1 & \Lambda+\pi T \leq \abs{\omega_n}
  \end{cases}
\end{align*}
with the accompanying $\Lambda$ derivative
\begin{align*}
  -\dot\chi^\Lambda(\omega_n) & =
  \begin{cases}
    \frac{1}{2\pi T}
    & \Lambda-\pi T < \abs{\omega_n} < \Lambda+\pi T \\
    0 & \text{otherwise.}
  \end{cases}
\end{align*}
A Matsubara sum over the single-scale propagator
\eqref{eq:singlescale} with this cutoff contains exactly one term
$\pm\omega_n$ in the Matsubara sum,
\begin{align*}
  T \sum_n S^\Lambda(i\omega_n)
  & = T \sum_n -\dot\chi^\Lambda(\omega_n) \dotsm 
  = \frac{1}{2\pi} \sum_{\omega_n\approx\pm\Lambda} \dotsm
\end{align*}
The flow equation for the self energy $\Sigma^\Lambda$ has the same
general form as for $T=0$, but for $T>0$, the single-scale propagator
$S^\Lambda$ cannot be further simplified to $\tilde G$:
\begin{align}
  \dl \Sigma_{j,j}^\Lambda & =
  -\frac{1}{2\pi}
  \sum_{\omega_n\approx\pm\Lambda} \sum_{r=\pm 1} U_{j,j+r}^\Lambda
  \notag \\
  & \qquad \times \left[ \frac{1}{Q(i\omega_n)-\chi^\Lambda(\omega_n)
      \Sigma^\Lambda} \,
    Q(i\omega_n) \, \frac{1}{Q(i\omega_n)-\chi^\Lambda(\omega_n)
      \Sigma^\Lambda} \right]_{j+r,j+r} \notag \\
  \label{eq:sigmaflowtg0}
  \dl \Sigma_{j,j\pm 1}^\Lambda & =
  \frac{1}{2\pi}
  \sum_{\omega_n\approx\pm\Lambda} U_{j,j\pm 1}^\Lambda \\
  & \qquad \times 
  \left[ \frac{1}{Q(i\omega_n)-\chi^\Lambda(\omega_n) \Sigma^\Lambda} \,
    Q(i\omega) \, \frac{1}{Q(i\omega_n)-\chi^\Lambda(\omega_n)
      \Sigma^\Lambda} \right]_{j,j\pm 1} \,. \notag
\end{align}
At every $\Lambda=\omega_n-\pi T$, a switch from $\omega_n$ to
$\omega_{n-1}$ occurs and the integrated flow has a kink, instead of a
jump for a sharp cutoff.  At large frequencies, the self energy is
still so small that an adaptive step-size integration algorithm
efficiently takes steps of many Matsubara frequencies at once, while
at small frequencies it inefficiently takes many small steps to
resolve the kink accurately.  But since the positions of the kinks are
known, we can instead integrate only in the interval between each pair
of kinks.  Empirically, the switch between continuous integration and
integration only between kinks is best done around $\omega_n\approx
t$.  As there are $\mathcal O(1/T)$ such intervals, the runtime for
the complete flow scales as $\mathcal O(N/T)$.

\subsubsection*{Flow of the interaction}

The flow of the interaction is simpler: as self-energy corrections on
the internal lines of the interaction flow equation would be of
$\Ord((\Gamma^\Lambda)^3)$, we take only bare propagators on the
internal lines, such that the bare single-scale propagator is (setting
$\Sigma^\Lambda=0$)
\begin{align*}
  S_0^\Lambda(i\omega_n) =
  \frac{-\dot\chi^\Lambda(\omega_n)}{Q(i\omega_n)}
\end{align*}
which is the same as for $T=0$, except that for each $\Lambda$ one has
to insert instead the nearest discrete $\omega_n$ into the propagator.
At half filling,
\begin{align*}
  \dl \left( \frac{1}{U^\Lambda} \right)
  & = \frac{1}{2\pi} \left( 
    1 - \omega_n \frac{\omega_n^2+6}{(\omega_n^2+4)^{3/2}} \right)
  \at_{\omega_n\approx\Lambda} \,.
\end{align*}
At $T=0$, $\Sigma^\Lambda$ and $\Gamma^\Lambda$ flow substantially
only in the range $0.1t \lesssim \omega \lesssim t$.  At $T>0$, since
the lowest fermionic Matsubara frequency is $\omega_0=\pi T$,
$\Sigma^\Lambda$ and $\Gamma^\Lambda$ are renormalized substantially
only for $T<1$.  The interaction is still correct to second order in
$U$ if all external legs lie on the Fermi surface, but the frequency
dependence at second order becomes important with inelastic
scattering.

\subsubsection*{Initial conditions}

At $T>0$, the combination $Q-\chi^\Lambda\Sigma^\Lambda$ appears in
the single-scale propagator in equation \eqref{eq:sigmaflowtg0}, \ie,
$Q$ and $\Sigma^\Lambda$ do not enter on an equal footing.  For the
exact theory and full hierarchy of flow equations, the final result
should be independent of the specific form of the cutoff, but in the
truncated flow it does make a difference.  Both the bare impurity
potential and the lead contribution are contained in the bare
propagator \eqref{eq:condinvbareprop}, such that the self energy
receives only contributions from the interaction, and the same initial
conditions \eqref{eq:initcondt0s} and \eqref{eq:initcondt0g} as for
$T=0$ apply.


\subsection{Temperature flow}
\label{sec:lutt:flow:temp}

In the temperature flow (see section \ref{sec:rg:flow:regu} and
\cite{HS01}), the kinetic term in the action,
\begin{align*}
  \inner{\bar\psi}{Q\psi} =
  T \sum_n \sum_{\vec{k}} \bar\psi_{\omega_n,\vec{k}}
  (i\omega_n-\xi_{\vec{k}}) \psi_{\omega_n,\vec{k}}\,,
\end{align*}
is replaced after a rescaling of the fields, $\psi \mapsto T^{-3/4}
\psi$, by
\begin{align*}
  \inner{\bar\psi}{Q\psi} =
  \sum_n \sum_{\vec{k}} \bar\psi_{\omega_n,\vec{k}}
  \frac{i\omega_n-\xi_{\vec{k}}}{\sqrt T} \psi_{\omega_n,\vec{k}} \,.
\end{align*}
We shall postpone the leads for a moment but include the impurity
potential in $\xi$ as usual.  With the new convention that the scalar
product shall contain no further factor of $T$, we define the
\emph{rescaled} inverse bare propagator (indicated by the superscript
$T$) as
\begin{align*}
  Q^T = \frac{i\omega_n-\xi}{\sqrt T}
\end{align*}
which yields the rescaled full propagator (with rescaled self energy
$\Sigma^T$)
\begin{align*}
  G^T = \frac{1}{Q^T - \Sigma^T}
  = \frac{\sqrt T}{i\omega_n - \xi - \sqrt T \Sigma^T} \,.
\end{align*}
The $T$ derivatives of these quantities are
\begin{align*}
  & \dT (i\omega_n) = \frac{i\omega_n}{T} \\
  & \dT Q^T
  = \frac{i\omega_n}{T^{3/2}} - \frac{i\omega_n-\xi}{2T^{3/2}}
  = \frac{i\omega_n \boldsymbol{+} \xi}{2T^{3/2}} 
  \qquad \text{(note the $+$ sign!)} \\
  \Rightarrow \quad & S^T = \frac{1}{2\sqrt T}
  \frac{1}{i\omega_n-\xi-\sqrt T \Sigma^T}
  \left(i\omega_n \boldsymbol{+} \xi\right)
  \frac{1}{i\omega_n-\xi-\sqrt T \Sigma^T} \,.
\end{align*}
Note the unusual $+$ sign in $dQ^T/dT$ which is due to the fact
that one does not just have a multiplicative cutoff function
$\chi^T=\sqrt T$ but the temperature also enters in the Matsubara
frequency $\omega_n$.

To understand the structure of the flow equation, consider first the
\RG\ flow of the self energy without leads, feedback of the self energy
or vertex renormalization, \ie, first-order perturbation theory:
\begin{align*}
  -S^T & = \dT C^T = \dT \left( \frac{\sqrt T}{i\omega_n-\xi} \right) \\
  -\sum_n S^T & = \sum_n \dT C^T = \dT \left( \frac{1}{\sqrt T} \;
    T\sum_n \frac{e^{i\omega_n0^+}}{i\omega_n-\xi} \right)
  = \dT \left( \frac{f(\xi)}{\sqrt T} \right).
\end{align*}
The flow equation of the rescaled self energy $\Sigma^T$ is thus
\begin{align*}
  \dT \Sigma^T & = -\sum_n \tr (U S^T)
  = \dT \left[ \frac{1}{\sqrt T} \tr \bigl(U f(\xi) \bigr) \right].
\end{align*}
With the initial condition $\sqrt{T_0} \,
\Sigma^{T_0}\at_{T_0\to\infty} = \tr\bigl(U
f(\xi)\bigr)\at_{T_0\to\infty} \to \frac 12 \tr(U)$ as in equation
\eqref{eq:initcondt0s} we can integrate from $T=T_0$ down to $T$ and
obtain
\begin{align*}
  \sqrt T \, \Sigma^T & = \tr (U f(\xi))
\end{align*}
as expected from the relation between rescaled and original Green
functions, equation \eqref{eq:Tscaling}.

Notice one peculiarity of the temperature flow: since the Green
functions of the leads depend on the frequency $i\omega_n$, they also
depend on temperature and have to be differentiated appropriately in
$dQ^T/dT$.  At half filling the expressions are particularly simple,
\begin{align*}
  Q^T(i\omega_n)
  & = \frac{i\omega_n - \Sigma^\leads(i\omega_n) - \xi} {\sqrt T} \\
  g_L(i\omega_n)
  & = \frac 12 \left( i\omega_n - i\sgn(\omega_n) \sqrt{4+\omega_n^2}
  \right) \\
  \dT g_L(i\omega_n) & = \frac{1}{T} \left( g_L(i\omega_n)
    +2i\frac{\sgn(\omega_n)}
    {\sqrt{4+\omega_n^2}} \right) \\
  \dT Q^T(i\omega_n) & = \frac{1}{2T^{3/2}} \Bigl[ i\omega_n -
  \Sigma^\leads(i\omega_n) \boldsymbol{+} \xi
  -4i\frac{\sgn(\omega_n)}
  {\sqrt{4+\omega_n^2}}
  \bigl( t_L^2 \ket 1 \bra 1 + t_R^2 \ket N \bra N \bigr) \Bigr].
\end{align*}

\subsubsection*{Temperature flow of the interaction}

The temperature flow of the interaction $U^T$ has the structure of the
usual one-loop diagrams of the perturbation expansion in the original
$\psi$ fields, but with a temperature derivative of the particle-hole
and particle-particle bubbles \cite{HS01}.  In our \oneD\ case,
\begin{align}
  \label{eq:tempflowuflow}
  \frac{dU^T}{dT} & = (U^T)^2 \, \dT B^T
\end{align}
where
\begin{align}
  \label{eq:bubbleint}
  B^T & := \oint \frac{dp}{2\pi} \, T\sum_n f(p,\omega)
\end{align}
is the sum of the three bubble contributions.  For instance at half
filling,
\begin{align*}
  B^T & = -\oint \frac{dp}{2\pi} \, \cos^2(p) \Bigl[ 
    \frac{f(\xi_p)-\frac 12}{\xi_p} - \frac{f'(\xi_p)}{T} \Bigr] \\
  & = \oint \frac{dp}{2\pi} \, \Bigl[ \frac{\cos(p)}{4} 
    \tanh\left(\frac{\cos(p)}{T}\right) + \frac{\cos^2(p)}{4T} 
    \left(\tanh^2\left(\frac{\cos(p)}{T}\right) - 1\right) \Bigr].
\end{align*}
In the limit $T\to 0$, $B^T \to \frac{1}{2\pi}$, while for
$T\to\infty$, $B^T\to 0$ vanishes.  Equation \eqref{eq:tempflowuflow}
can be written as
\begin{align*}
  d\left(\frac{1}{U^T}\right) = -d B^T.
\end{align*}
Integrating from $T=\infty$ (with $U^{T=\infty}=U$) down to $T$, we
obtain
\begin{align*}
  U^T = \frac{U}{1-U B^T}.
\end{align*}
For $T=0$, the result $U^{T=0} = U/\bigl(1-U/(2\pi)\bigr)$ agrees with
the frequency-cutoff result \eqref{eq:uflowt0} at $\Lambda=0$.

\subsubsection*{Initial conditions}

The initial conditions for $\Sigma = \sqrt T \, \Sigma^T$ and $U =
U^T$ in the limit $T=T_0\to=\infty$ are the same as for the frequency
cutoff, equations \eqref{eq:initcondt0s} and \eqref{eq:initcondt0g}.


\subsection{Interaction flow}
\label{sec:lutt:flow:intn}

In the interaction flow scheme (cf.\ section \ref{sec:rg:flow:regu} or
\cite{HRAE04}) the propagators are slowly switched on by a global
scale factor $\chi^\Lambda = g$, $g=0\dotsc 1$, irrespective of
frequency or momentum.  We define the inverse bare propagator
\begin{align*}
  Q^g(i\omega) := \frac{Q(i\omega)}{g}
  = \frac{i\omega - \Sigma^\leads(i\omega) - \xi}{g}
\end{align*}
and the full propagator
\begin{align*}
  G^g(i\omega) & := \frac{1}{Q^g - \Sigma^g}
  = \frac{g}{i\omega - \Sigma^\leads(i\omega) - \xi - g\Sigma^g}
\end{align*}
where $\xi$ shall include the impurity potential $V$.  Then the
``single-scale'' propagator (which is not at all single-scale) is
\begin{align*}
  S^g & := G^g \dot Q^g G^g
  = - \frac{1}{Q(i\omega)-g\Sigma^g} Q(i\omega)
  \frac{1}{Q(i\omega)-g\Sigma^g}.
\end{align*}
Furthermore, because the bare $G_0^g = g G_0$ contains one $g$ factor
and $S_0^g=G_0$ contains none, the flow equation for the interaction
vertex contains the combination $G_0^g S_0^g + S_0^g G_0^g = 2g G_0
G_0$ in the bubble,
\begin{align*}
  \frac{d}{dg} U^g & = (U^g)^2 \; 2g B^T
\end{align*}
with the bubble integral $B^T$ at temperature $T$, equation
\eqref{eq:bubbleint}.  Integrating from $g=0$ (with $U^{g=0}=U$) up to
$g$,
\begin{align*}
  U^g & = \frac{U}{1-g^2 U B^T} \\
  & = \frac{U}{1-g^2 U/(2\pi)} \quad (T=0).
\end{align*}
At the end of the flow,
\begin{align*}
  U^{g=1} = \frac{U}{1-U B^T} = U^T.
\end{align*}
The initial condition for the self energy is $\Sigma^{g=0} = 0$
because all propagators vanish, but $\Gamma^{g=0}$ is the same as for
the frequency cutoff, equation \eqref{eq:initcondt0g}.


\subsection{Initial conditions for general filling}
\label{sec:lutt:flow:scmu}

At any filling $n$, we wish to fulfill two conditions for the model
without impurities: $(i)$ at the end of the flow, the density profile
should be uniformly $n_j \equiv n$, both in the interacting region and
in the leads, such that power-law exponents depending on the density
can be read off reliably; $(ii)$ no backscattering should occur at the
ends of the wire where the interaction is switched on, \ie, the
transmission without impurity should be perfect.  In order to achieve
this, we allow the freedom to add to our microscopic model a local
potential in the wire that depends only on the interaction strength
but not on the temperature, while with changing temperature only the
global chemical potential $\mu(T)$ may be adjusted.

At half filling, these conditions are met without a local potential
and by setting $\mu=0$.  Away from half filling, we use the following
procedure:
\begin{enumerate}
\item At zero temperature, the global chemical potential is defined as
  $\mu := -2t\cos(\kf) \neq 0$ such that $\xi_{\kf} = 0$, where the
  Fermi wave vector $\kf \equiv n \pi$ at filling $n$.  Second,
  finding the local potential at all $N$ lattice sites such that at
  the end of the flow $n_j \equiv n$ (without impurities) is a
  complicated $N$-parameter optimization problem, but we obtain a very
  good guess in the following way: the problem is first solved
  in the bulk where the global potential, implemented as the initial
  condition of the self energy $\Sigma_{\text{bulk}}^{\Lambda_0}$, is
  tuned self-consistently until $\Sigma_{\kf}^{\Lambda=0} = 0$.  Then
  the initial condition of the self energy on the lattice (the local
  potential) is defined as
  \begin{align*}
    \Sigma_{jj}^{\Lambda_0} := \frac{U_{j-1,j} + U_{j,j+1}}{2U}
    \Sigma_{\text{bulk}}^{\Lambda_0}
  \end{align*}
  such that in the middle of the system where $U_{j,j+1} \equiv U$ is
  homogeneous, the bulk initial condition holds, while towards the
  ends when the system becomes non-inter\-act\-ing $U_{j,j+1} \to 0$,
  no local potential is necessary.  Equivalently, the local potential
  may also be written as part of the interaction term \eqref{eq:Hint},
  \begin{align*}
    H_\tint = \sum_j U_{j,j+1} (n_j-\nu) (n_{j+1}-\nu)
  \end{align*}
  where $\nu = (1-\Sigma_{\text{bulk}}^{\Lambda_0}/U)/2$.
\item Now the temperature $T>0$ is switched on and $\mu$ is tuned
  self-consistently until the bulk density $n(\mu) \equiv
  \frac{1}{2\pi} \int dk \, f(\epsilon_k - \mu) = n$ at the end of the
  flow, with the non-interacting bulk dispersion $\epsilon_k$ of the
  leads.  By the above construction also the density in the
  interacting region, obtained using the dispersion $\epsilon_k +
  \Sigma_k$ with renormalized hopping and potential, has the same
  value to a high accuracy, and the density computed using a flowing
  density vertex (see below) deviates by less than 1\% for $\abs U\leq
  1$.
\end{enumerate}
Only at this stage, impurities are inserted into the lattice and
change the homogeneous density profile.


\subsection{Algorithm for tridiagonal matrices}
\label{sec:lutt:flow:tridiag}

With the parametrization of the interaction by one single bulk
parameter $U^\Lambda$, the most time-consuming part of the \abb{ODE} flow
is the flow of the self energy.  On the \RHS\ of the $\Sigma^\Lambda$
flow equation, both $Q$ and $\Sigma^\Lambda$ are tridiagonal matrices,
hence it involves the inversion of tridiagonal matrices.  I have
developed an efficient inversion algorithm to compute the \RHS\ in
$\Ord(N)$ time at any temperature.  Therefore, the runtime scales only
linearly with the system size $N$, and the self energy of systems as
large as $N=10^7$ sites (at $T=0$) can be computed.  At $T=0$, only
the tridiagonal part of the inverse tridiagonal matrix $\tilde
G^\Lambda$ is needed in \eqref{eq:sigmaflowt0}, and an algorithm by
the author was presented already in \cite{AEMMSS04} and is reprinted
in appendix \ref{sec:app:loop:prop}.  For $T>0$, the \RHS\ of the flow
equation \eqref{eq:sigmaflowtg0} is more complicated: a matrix product
of an inverse tridiagonal matrix (which is a full matrix), a
tridiagonal matrix $Q$, and another inverse tridiagonal matrix.  In a
forthcoming article and in appendix \ref{sec:app:loop:bubble}, I
present an algorithm to compute even this product of full $N\times N$
matrices in $\Ord(N)$ time \cite{Enss04a}.  However, the \RHS, which has
jumps in $\Lambda$ at every Matsubara frequency (see section
\ref{sec:lutt:flow:freqTg0}), makes the integration at finite
temperature scale as $\Ord(N/T)$, such that only systems up to $N=10^4$
have been computed so far for low temperatures $T=10^{-4}$.

The flow equation \eqref{eq:Rflow} of the density-response vertex used
below for the Friedel oscillations at $T=0$ is of the same
computational complexity as the self-energy flow at $T>0$: the \RHS\ 
involves the matrix product of an inverse tridiagonal, a tridiagonal,
and another inverse tridiagonal matrix.  This can also be done in
$\Ord(N)$ time by the same algorithm in appendix
\ref{sec:app:loop:bubble}, and at $T=0$ the density profile $n_j$ of a
system of size $N=10^7$ can be computed in a few hours for each $j$.


\section{Flow of the density-response vertex}
\label{sec:lutt:density}

The expectation value of the local density $n_j$ could be computed
from the local one-particle propagator $G_{jj}$ if $G$ was known
exactly. However, the approximate flow equations for $\Sigma$ can be
expected to describe the asymptotic behavior of $G$ correctly only at
long distances between creation and annihilation operators in time
and/or space, while in the local density operator time and space
variables coincide. In the standard \RG\ terminology $n_j$ is a
\emph{composite} operator, which has to be renormalized separately.

To derive a flow equation for $n_j$, we follow the usual procedure for
the renormalization of correlation functions involving composite
operators: one adds a term $\phi_j n_j$ with a small field $\phi_j$ to
the Hamiltonian and takes derivatives with respect to $\phi_j$ in the
flow equations.  The local density is given by
\begin{align*}
  n_j = \left. \frac{\partial\Omega(\phi_j)}{\partial\phi_j}  
  \right|_{\phi_j = 0} \, ,
\end{align*}
where $\Omega(\phi_j)$ is the grand canonical potential of the system
in the presence of the field $\phi_j$.  Note that I use the same
symbol $n_j$ for the density operator and its expectation value.  In
the presence of a cutoff $\Lambda$ the grand canonical potential obeys
the exact flow equation
\begin{align*}
  \dl \Omega^\Lambda = T \, \sum_{\omega} \, \tr \left\{ e^{i\omega
      0^+} \, [\partial_\Lambda Q^\Lambda(i\omega)] \,
    [G^\Lambda(i\omega) - G_0^\Lambda(i\omega)] \right\} \,,
\end{align*}
which follows from the flow equations for the vertex functions,
equation \eqref{eq:gammaflow3}, and the relation between the grand
canonical potential and the zero-particle vertex, $\Omega^\Lambda =
-\gamma_0^\Lambda$.  At zero temperature (which is the only case I
consider here) the Matsubara frequency sum becomes an integral which,
for the sharp frequency cutoff \eqref{eq:zerotempsharpcutoff}, can be
carried out analytically. This yields
\begin{align}
  \label{eq:Omegaflow}
  \dl \Omega^\Lambda = 
  \frac{1}{2\pi}
  \sum_{\omega=\pm\Lambda} \tr \left\{ e^{i\omega 0^+}
    \ln [ 1 - G_0(i\omega) \, \Sigma^\Lambda(i\omega) ] \right\} \;,
\end{align}
which is the first term of equation \eqref{eq:Gshcutoff}.  Because at
$T=0$ any perturbation of the Hamiltonian can be shifted between the
bare propagator and the self energy, in this section I choose to
attribute $\phi_j$ to the interaction part of the Hamiltonian, not to
$H_0$, such that $G_0$ remains independent of $\phi_j$.  The self
energy is modified via the additional local and frequency-independent
contribution $\phi_j \, \delta_{jj'}$ to its initial value
$\Sigma^{\Lambda_0}_{jj'}$ at scale $\Lambda_0$.

The density profile can be obtained from the above equations and the
flow equation for $\Sigma^\Lambda$ by computing the shift of
$\Omega^\Lambda$ generated by a small finite perturbation $\phi_j$,
\ie, by numerical differentiation (cf.\ section
\ref{sec:ward:sc:import}).  Alternatively, one may carry out the
$\phi_j$ derivative analytically in the flow equations, which yields a
flow equation for the density in terms of the density response vertex.
Taking the $\phi_j$ derivative in equation \eqref{eq:Omegaflow} yields
\begin{align}
  \label{eq:njflow}
  \dl n_j^\Lambda = 
  - \frac{1}{2\pi} \sum_{\omega = \pm\Lambda} 
  \tr \left[ e^{i\omega 0^+}
    \tilde G^\Lambda(i\omega) \, R_j^\Lambda(i\omega) \right]
\end{align}
with the density-response vertex
\begin{align*}
  R_j^\Lambda(i\omega) = 
  \left. \frac{\partial\Sigma^\Lambda(i\omega)}{\partial\phi_j}
  \right|_{\phi_j = 0}
\end{align*}
and the propagator $\tilde G^\Lambda$ as defined in equation
\eqref{eq:tildeG}, \ie, in the absence of $\phi_j$.  We compute the
self energy $\Sigma^\Lambda$ in the presence of $\phi_j$ within the
same approximation as previously.  It is thus determined from the flow
equation \eqref{eq:sigmaflowt0} with a frequency-independent
two-particle vertex $\Gamma^\Lambda$. Taking a derivative of that
equation with respect to $\phi_j$ at $\phi_j = 0$ yields the flow
equation for the response vertex
\begin{align*}
  \dl R_{j;1',1}^\Lambda = 
  - \frac{1}{2\pi} \sum_{\omega = \pm\Lambda} 
  \sum_{2,2'} \sum_{3,3'} 
  \tilde G_{2,3'}^\Lambda(i\omega) \, R_{j;3',3}^\Lambda \, 
  \tilde G_{3,2'}^\Lambda(i\omega) \,
  \Gamma_{1',2';1,2}^\Lambda \; .
\end{align*}
Note that $R_j^\Lambda$ is frequency independent in our approximation
and that there is no contribution from the $\phi_j$ derivative of
$\Gamma^\Lambda$ since we neglect self-energy contributions in the
flow of the two-particle vertex.

For spinless fermions with a (renormalized) nearest-neighbor
interaction, the matrix $R_j^\Lambda$ is tridiagonal, \ie, only the
components $R_{j;l,l}^\Lambda$ and $R_{j;l,l\pm 1}^\Lambda$ are
non-zero, and their flow is given by
\begin{align}
  \dl R_{j;l,l}^\Lambda &= -
  \frac{1}{2\pi} \sum_{\omega=\pm\Lambda} \sum_{l'}
  \sum_{r=\pm 1} \sum_{r' = 0,\pm 1} U_{l,l+r}^\Lambda
  \tilde G_{l+r,l'}^\Lambda(i\omega) \, R_{j;l',l'+r'}^\Lambda \,
  \tilde G_{l'+r',l+r}^\Lambda(i\omega) \notag \\
  \label{eq:Rflow}
  \dl R_{j;l,l\pm 1}^\Lambda &=
  \frac{1}{2\pi} \sum_{\omega=\pm\Lambda} \sum_{l'} \sum_{r' =
    0,\pm 1} U_{l,l\pm 1}^\Lambda \tilde G_{l,l'}^\Lambda(i\omega) \,
  R_{j;l',l'+r'}^\Lambda \, \tilde G_{l'+r',l\pm 1}^\Lambda(i\omega) \,.
\end{align} 
Then, the flow of $n_j^\Lambda$ is
\begin{align*}
  \dl n_j^\Lambda &=
  - \frac{1}{2\pi} \sum_{\omega=\pm\Lambda}
  \sum_{l'} \sum_{r' = 0,\pm 1}
  \tilde G_{l'+r',l'}^\Lambda(i\omega) \, R_{j;l',l'+r'}^\Lambda \,.
\end{align*}
Although naively the flow of $R^\Lambda$ scales as $\Ord(N^2)$ in time
because of the unrestricted loops over $l,l'$, it can be performed in
$\Ord(N)$ by the method described in the appendix
\ref{sec:app:loop:bubble}.  The initial condition for the response
vertex is $R_{j;l,l'}^{\Lambda_0} = \delta_{jl} \delta_{ll'}$.  The
initial condition for the density is $n_j^{\Lambda_0} = \frac 12$, for
any filling, due to the slow convergence of the flow equation
\eqref{eq:njflow} at large frequencies, which yields a finite
contribution to the integrated flow from $\Lambda = \infty$ to
$\Lambda_0$ for arbitrarily large finite $\Lambda_0$, as in the case
of the self energy $\Sigma^{\Lambda_0}$.

To avoid the interference of Friedel oscillations emerging from the
impurity or one boundary with those coming from the (other) boundaries
of the system one suppresses the influence of the latter by coupling
the finite chain to semi-infinite non-interacting leads, with a smooth
decay of the interaction at the contacts (cf.\ section
\ref{sec:lutt:model:wireproj}).


\section{Computation of the conductance}
\label{sec:lutt:condtech}

The linear-response conductance is defined via the infinitesimal
current induced by an infinitesimal voltage drop at zero bias voltage,
$G=dI/dV$.  It is a global quantity defined over the whole wire, from
one lead through the interacting (scattering) region to the other
lead.  Even for a perfectly clean wire (interacting or not), the
conductance is limited to $e^2/h$ per channel.  For a \oneD\ system of
spinless fermions there is exactly one channel, while for spin-$\frac
12$ fermions there are two channels.

I choose the conductance as the appropriate observable for transport,
as opposed to the local quantity conductivity, because I am interested
not in bulk properties but the effect of a specific spatial setup of
impurities at defined positions, the double barrier, for which there
are experimental data available \cite{Pos01}.


\subsection{Kubo formula}

The conductance is computed, just as the linear-response conductivity,
via the Kubo formula from the current-current correlation function,
see for instance \cite[chapters 3 and 7]{Mah00}.  This correlation
function at finite frequency and temperature is defined as
\begin{align*}
  \pi(i\omega) := \int_0^\beta d\tau\, e^{i\omega\tau}\,
  \vev{T_\tau J_R(\tau) J_L(0)}
\end{align*}
with $J_{L,R}$ the current operators at the left and right ends of
the system.  The retarded correlation function $\pi_\ret(\omega)$ is
obtained by analytical continuation $i\omega \mapsto \omega+i0$, and
the dc conductance is given by finally taking the limit $\omega \to
0$,
\begin{align}
  \label{eq:Gdef}
  G & := \frac{e^2}{\hslash} \lim_{\omega \to 0}
  \frac{\pi_\ret(\omega) - \pi_\ret(0)}{i\omega}\,.
\end{align}
It is important that one does not set $\omega=0$ from the beginning
because the limits $\omega\to 0$ and $q\to 0$ (macroscopic transport
from one end of the system to the other) do not commute \cite{Lut64}.
Another way to compute the conductance at zero temperature via
persistent currents is explained in \cite{MS03a,MS03b}.

Consider the following microscopic setup (cf.\ figure~\ref{fig:leads}
on page \pageref{fig:leads}).  The scattering region on the lattice
sites $1,\dotsc,N$ (``sample'') is interacting; it is connected to
semi-infinite, non-interacting leads at the interfaces at sites $1$
and $N$.  The current and total number operators are
\begin{align}
  \label{eq:Jdef}
  \begin{split}
    J_L & := i t_L (c_1^\dagger c_0^{\phantom\dagger} - c_0^\dagger
    c_1^{\phantom\dagger}) \\
    J_R & := i t_R (c_{N+1}^\dagger c_N^{\phantom\dagger} -
    c_N^\dagger c_{N+1}^{\phantom\dagger}) \\
    n_C & := \sum_{j=1}^N n_j
  \end{split}
\end{align}
with $J_L$ the current flowing from the left lead into the sample,
$J_R$ the current flowing out from the sample into the right lead,
and $n_C$ the total particle number in the interacting
region.\footnote{The components of the gauge potential $A_\mu =
  \big(\varphi(t,j), A(t,[j,j+1])\big)$ are the scalar potential
  $\varphi$ at time $t$ and on site $j$, and the vector potential $A$
  on the (directed) bond from site $j$ to site $j+1$, respectively,
  where we have introduced the notation $[j,j+1]$ for the bond.
  Coupling to the potential $A_\mu$ changes the bare dispersion $\xi$
  to \cite[chapter 34]{Zin02}
  \begin{align*}
    \xi_{j,j} & = -\mu - e\varphi(t,j) \\
    \xi_{j,j+1} & = -t e^{ieA(t,[j,j+1])} = \xi_{j+1,j}^*.
  \end{align*}
  Derivatives with respect to a $A([0,1])$, $A([N,N+1])$ and a global
  $\varphi$ yield the expressions \eqref{eq:Jdef}.}

\fig[width=0.9\linewidth]{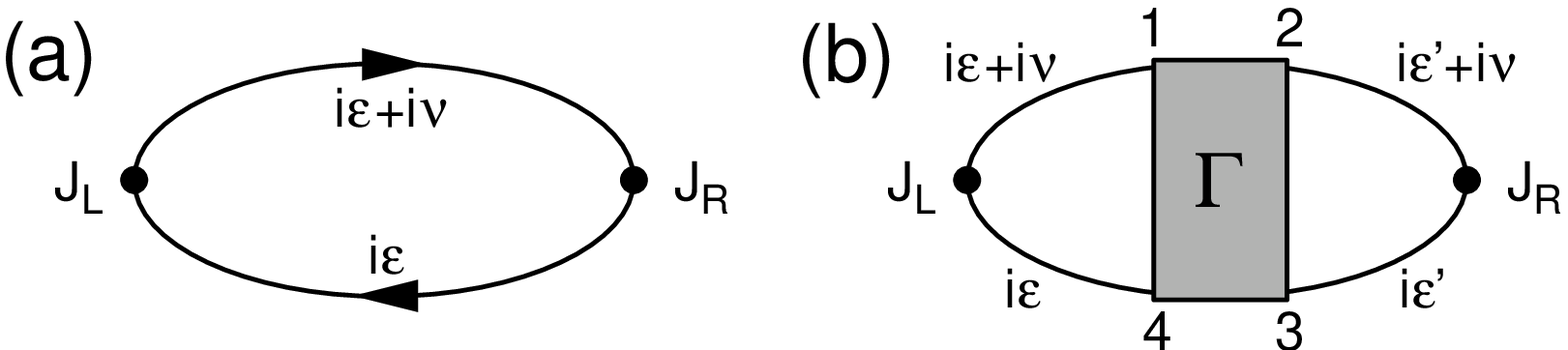}{fig:kubo}{Two contributions to the
  current-current correlation function $\pi(i\omega)$.  The shaded
  region represents the vertex part $\Gamma_{1,2;3,4}
  (i\epsilon,i\epsilon';i\omega)$.}
In the exact theory, the current-current correlation function can be
expressed in terms of the bare current operators $J_{L,R}$ and the
\onePI\ two-particle interaction vertex $\Gamma$, see
figure~\ref{fig:kubo}.  The expressions for these two contributions to
$\pi(i\omega)$ are
\begin{align*}
  \pi^{(a)}(i\omega) & = 
  -(it_L)(it_R) T\sum_{i\epsilon}
  \Bigl[ G_{N,1}(i\epsilon+i\omega) \, G_{0,N+1}(i\epsilon)
  + G_{N+1,0}(i\epsilon+i\omega) \, G_{1,N}(i\epsilon) \\
  & \qquad \qquad \qquad \qquad
  - G_{N+1,1}(i\epsilon+i\omega) \, G_{0,N}(i\epsilon)
  - G_{N,0}(i\epsilon+i\omega) \, G_{1,N+1}(i\epsilon) \Bigr], \\
  \pi^{(b)}(i\omega) & = 
  -(it_L)(it_R) T^2 \sum_{i\epsilon,i\epsilon'}
  \sum_{j_1,\dotsc,j_4=1}^N \\
  & \qquad \Bigl[ G_{j_1,1}(i\epsilon+i\omega) \, G_{0,j_4}(i\epsilon)
  - G_{j_1,0}(i\epsilon+i\omega) \, G_{1,j_4}(i\epsilon) \Bigr] \\
  & \quad \times
  \Gamma_{j_1,j_2;j_3,j_4}(i\epsilon,i\epsilon';i\omega) \\
  & \quad \times
  \Bigl[ G_{N,j_2}(i\epsilon'+i\omega) \, G_{j_3,N+1}(i\epsilon')
  - G_{N+1,j_2}(i\epsilon'+i\omega) \, G_{j_3,N}(i\epsilon') \Bigr].
\end{align*}


\subsubsection*{Kubo formula projected into the interacting region}

Using equations \eqref{eq:condGrel1}, \eqref{eq:condGrel2}, and
\eqref{eq:condGrel3}, we can express $\pi^{(a,b)}(i\omega)$ in terms
of Green functions with indices in the interacting region
$1,\dotsc,N$,
\begin{align}
  \label{eq:condpia}
  \pi^{(a)}(i\omega) & = T \sum_{i\epsilon}
  \lambda_L(i\epsilon,i\epsilon+i\omega) \, G_{1,N}(i\epsilon) \,
  \lambda_R(i\epsilon,i\epsilon+i\omega) \,
  G_{N,1}(i\epsilon+i\omega) \\
  \pi^{(b)}(i\omega) & = T \sum_{i\epsilon} \sum_{j_1,j_4=1}^N
  \lambda_L(i\epsilon,i\epsilon+i\omega) \, G_{1,j_4}(i\epsilon) \,
  P_{R;j_4,j_1}(i\epsilon,i\epsilon+i\omega) \,
  G_{j_1,1}(i\epsilon+i\omega) \notag
\end{align}
with bare current vertices on sites $1$ and $N$, respectively,
\begin{align*}
  \lambda_L(i\epsilon,i\epsilon+i\omega) & =
  - i t_L^2 [g_L(i\epsilon+i\omega) - g_L(i\epsilon)] \\
  \lambda_R(i\epsilon,i\epsilon+i\omega) & =
  + i t_R^2 [g_R(i\epsilon+i\omega) - g_R(i\epsilon)]
\end{align*}
and the current-vertex correction
\begin{multline*}
  P_{R;j_4,j_1}(i\epsilon,i\epsilon+i\omega) :=
  T \sum_{i\epsilon'} \sum_{j_2,j_3=1}^N
  \Gamma_{j_1,j_2;j_3,j_4} (i\epsilon,i\epsilon';i\omega) \\
  \times G_{j_3,N}(i\epsilon') \, 
  \lambda_R(i\epsilon'; i\epsilon'+i\omega) \,
  G_{N,j_2}(i\epsilon'+i\omega)
\end{multline*}
and likewise for $P_L$.


\subsubsection*{Analytical continuation of the Kubo formula}

The Green function \eqref{eq:condGL} of the lead has a branch cut at
the real axis,
\begin{align*}
  g_{L,R}(\epsilon\pm i\delta) & = \frac 12 \left( 
    \epsilon+\mu \pm i\delta
    \mp i\sqrt{4 - (\epsilon+\mu\pm i\delta)^2} \right) \\
  & = \frac 12 \left( \epsilon+\mu
    \mp i\sqrt{4 - (\epsilon+\mu)^2} \right) + \Ord(\delta).
\end{align*}
Thus, for small $\omega$,
\begin{align*}
  \lambda_{L,R}(\epsilon\pm i\omega,\epsilon\pm i0) & = \Ord(\omega) \\
  \lambda_L(\epsilon+i\omega,\epsilon-i0)
  & = + t_L^2 \, \sqrt{4 - (\epsilon+\mu)^2} + \Ord(\omega) \\
  \lambda_R(\epsilon+i\omega,\epsilon-i0)
  & = - t_R^2 \, \sqrt{4 - (\epsilon+\mu)^2} + \Ord(\omega),
\end{align*}
and only the current vertices with incoming and outgoing legs on
opposite sides of the branch cut contribute in the limit $\omega\to
0$.  The Matsubara sum in the bubble term \eqref{eq:condpia} is
evaluated by a contour integral as explained in \cite[chapter
7]{Mah00},
\begin{align*}
  \pi^{(a)}(i\omega) & = i \int \frac{d\epsilon}{2\pi} f(\epsilon) \\
  \times \bigl(
  & +\lambda_L(\epsilon+i0,\epsilon+i\omega) \, G_{1,N}(\epsilon+i0) \,
  \lambda_R(\epsilon+i0,\epsilon+i\omega) \, G_{N,1}(\epsilon+i\omega) \\
  & -\lambda_L(\epsilon-i0,\epsilon+i\omega) \, G_{1,N}(\epsilon-i0) \,
 \lambda_R(\epsilon-i0,\epsilon+i\omega) \, G_{N,1}(\epsilon+i\omega) \\
  & +G_{1,N}(\epsilon-i\omega) \, \lambda_R(\epsilon-i\omega,\epsilon+i0) \,
  G_{1,N}(\epsilon+i0) \, \lambda_L(\epsilon-i\omega,\epsilon+i0) \\
  & -G_{1,N}(\epsilon-i\omega) \, \lambda_R(\epsilon-i\omega,\epsilon-i0) \,
  G_{1,N}(\epsilon-i0) \, \lambda_L(\epsilon-i\omega,\epsilon-i0) \bigr)\,.
\end{align*}
The first and fourth term on the \RHS\ have frequency arguments on the
same side of the branch cut and are, therefore, of $\Ord(\omega^2)$, hence
they will vanish in the limit $\omega\to 0$ in equation
\eqref{eq:Gdef}.  We retain the other two terms and perform the
analytical continuation of the external frequency, $i\omega \mapsto
\omega+i0$,
\begin{align*}
  \pi_\ret^{(a)} & (\omega) = i \int \frac{d\epsilon}{2\pi}
  f(\epsilon) \\
  \times \bigl(
  & -\lambda_L(\epsilon-i0,\epsilon+\omega+i0) \,
  G_{1,N}(\epsilon-i0)
  \lambda_R(\epsilon-i0,\epsilon+\omega+i0) \,
  G_{N,1}(\epsilon+\omega+i0) \\
  & +\lambda_L(\epsilon-\omega-i0,\epsilon+i0) \,
  G_{1,N}(\epsilon-\omega-i0)
  \lambda_R(\epsilon-\omega-i0,\epsilon+i0) \,
  G_{N,1}(\epsilon+i0) \bigr).
\end{align*}
We then substitute $\epsilon \mapsto \epsilon+\omega$ in the second
term,
\begin{align*}
  \pi_\ret^{(a)} & (\omega) = i \int \frac{d\epsilon}{2\pi}
  [f(\epsilon+\omega) - f(\epsilon)] \\
  \times & \lambda_L(\epsilon-i0,\epsilon+\omega+i0) \,
    G_{1,N}(\epsilon-i0) \,
    \lambda_R(\epsilon-i0,\epsilon+\omega+i0) \,
    G_{N,1}(\epsilon+\omega+i0) \, .
\end{align*}
As the final step we take the limit $\omega\to 0$,
\begin{align}
  G^{(a)} & = \frac{e^2}{\hslash} \lim_{\omega\to 0}
  \frac{\pi_\ret^{(a)}(\omega)}{i\omega} \notag \\
  & = \frac{e^2}{\hslash} \int \frac{d\epsilon}{2\pi}
  f'(\epsilon) \lambda_L(\epsilon-i0,\epsilon+i0) \,
  G_{1,N}(\epsilon-i0) \,
  \lambda_R(\epsilon-i0,\epsilon+i0) \,
  G_{N,1}(\epsilon+i0) \notag \\
  \label{eq:cond}
  & = \frac{e^2}{h} \int d\epsilon
  \bigl( -f'(\epsilon) \bigr) \, \T^{(a)}(\epsilon,T).
\end{align}
$\T^{(a)}(\epsilon,T)$ is the transmission probability at temperature
$T$ without vertex corrections,
\begin{align}
  \label{eq:trans}
  \T^{(a)}(\epsilon,T)
  & := t_L^2 \, t_R^2 \, [4 - (\epsilon+\mu)^2] \,
  \abs{G_{N,1}(\epsilon+i0)}^2
\end{align}
where we have used $G_{1,N}(\epsilon-i0) \, G_{N,1}(\epsilon+i0) =
\abs{G_{N,1}(\epsilon+i0)}^2$: because the Hamiltonian is
time-reversal invariant, the amplitude from site $1$ to $N$ is the
same as from $N$ to $1$, and $G_{jj'}$ is symmetric (not hermitean).
The energy integration extends over the band of the non-interacting
leads, while the factor $[4-(\epsilon+\mu)^2]$ from the density of
states of the leads suppresses the transmission towards the edge of
the band.

The \fRG\ provides an approximation of the frequency-independent self
energy $\Sigma_{jj'}(T)$ at zero or finite temperature.  $\Sigma$ acts
as an effective static potential by which non-interacting electrons
are scattered.  The full propagator is determined via the Dyson
equation \eqref{eq:cond-dyson}.


\subsection{Vertex corrections}

The second contribution $\T^{(b)}(\epsilon,T)$ to the transmission is
due to current-vertex corrections.  It is obtained from $\pi^{(b)}$
following the same steps as for $\T^{(a)}$.  We have approximated the
full effective two-particle interaction vertex $\Gamma$ by a
renormalized nearest-neighbor interaction with all external
frequencies set to zero, \ie, without branch cuts.  Then the vertex
corrections are, omitting the lattice indices and denoting the loop
summation by the trace,
\begin{align*}
  P_R(i\omega) = T \sum_{i\epsilon'} \tr \bigl(
  G(i\epsilon') \lambda_R & (i\epsilon',i\epsilon'+i\omega) \,
  G(i\epsilon'+i\omega) \, \Gamma \bigr) \\
  = i \int \frac{d\epsilon'}{2\pi} f(\epsilon') \tr \bigl(
  & + G(\epsilon'+i0) \, \lambda_R(\epsilon'+i0,\epsilon'+i\omega) \,
  G(\epsilon'+i\omega) \, \Gamma \\
  & - G(\epsilon'-i0) \, 
  \lambda_R(\epsilon'-i0,\epsilon'+i\omega) \,
  G(\epsilon'+i\omega) \, \Gamma \\
  & + G(\epsilon'-i\omega) \, 
  \lambda_R(\epsilon'-i\omega,\epsilon'+i0) \, G(\epsilon'+i0) \,
  \Gamma \\
  & - G(\epsilon'-i\omega) \, 
  \lambda_R(\epsilon'-i\omega,\epsilon'-i0) \,
  G(\epsilon'-i0) \, \Gamma \bigr).
\end{align*}
Again, the first and fourth term on the \RHS\ are by an $\Ord(\omega)$
smaller than the other two, so we retain only the second and third
term and perform the analytical continuation $i\omega \mapsto
\omega+i0$.  Substituting $\epsilon' \mapsto \epsilon'+\omega$ in the
second term,
\begin{multline*}
  P_R(\omega+i0) = i \int \frac{d\epsilon'}{2\pi}
  \underbrace{[f(\epsilon'+\omega) - f(\epsilon')]}_{\Ord(\omega)} \\
  \times \tr \Bigl( G(\epsilon'-i0) \,
  \lambda_R(\epsilon'-i0,\epsilon'+\omega+i0) \,
  G(\epsilon'+\omega+i0) \, \Gamma \Bigr) \xrightarrow{\omega\to 0} 0
\end{multline*}
vanishes since there is no division by $\omega$ as in equation
\eqref{eq:Gdef}.  Because $\Gamma$ is frequency independent, there are
no vertex corrections, hence equation \eqref{eq:trans} is the complete
transmission probability in our approximation.


\subsubsection*{Conformance with Ward identities}

The approximation that $\Gamma$ is frequency independent has another
consequence: by the flow equation, it follows that $\Im \, \Sigma =
0$, \ie, we do not capture inelastic processes at second order in the
interaction.  They could be included in the flow equation by retaining
the frequency dependence and imaginary part of $\Gamma$.

However, the fact that the vertex corrections and $\Im \, \Sigma$
vanish simultaneously shows that our approximation is at least
consistent with the (non-perturbative) Ward identity associated with
global particle number (charge) conservation.  The global continuity
equation for the interacting region is $\partial n_C / \partial t +
J_R - J_L = 0$.  Following \cite{Ogu01}, we define the number and
current response functions as the time-ordered expectation values of
the current and number operators \eqref{eq:Jdef} with two extra
electron legs ($1 \leq j,j' \leq N$),
\begin{align*}
  \Phi_{C;jj'}(\tau;\tau_1,\tau_2) & =
  \vev{T_\tau [n_C(\tau) - \vev{n_C}] \,
    c_j^{\phantom\dagger}(\tau_1) \, c_{j'}^\dagger(\tau_2)} \\
  \Phi_{L;jj'}(\tau;\tau_1,\tau_2) & =
  \vev{T_\tau J_L(\tau) \, c_j^{\phantom\dagger}(\tau_1) \,
    c_{j'}^\dagger(\tau_2)} \\
  \Phi_{R;jj'}(\tau;\tau_1,\tau_2) & =
  \vev{T_\tau J_R(\tau) \, c_j^{\phantom\dagger}(\tau_1) \,
    c_{j'}^\dagger(\tau_2)}.
\end{align*}
Performing a Fourier transform on each of these ($x=L,R,C$),
\begin{align*}
  \Phi_x(\tau;\tau_1,\tau_2) & =
  T^2 \sum_{i\epsilon,i\omega} \Phi_x(i\epsilon,i\epsilon+i\omega)
  \, e^{-i\epsilon(\tau_1-\tau)}
  \, e^{-i(\epsilon+\omega)(\tau-\tau_2)}.
\end{align*}
The \onePI\ response vertices are obtained by amputating full
propagators,
\begin{align*}
  {\Lambda}_x(i\epsilon,i\epsilon+i\omega) & :=
  [G(i\epsilon)]^{-1} \, \Phi_x(i\epsilon,i\epsilon+i\omega) \,
  [G(i\epsilon+i\omega)]^{-1}
\end{align*}
which can be written in terms of the bare current and the vertex
corrections as
\begin{align}
  \label{eq:condLambdaR}
  \begin{split}
    \Lambda_{L,jj'}(i\epsilon,i\epsilon+i\omega) & =
    \lambda_L(i\epsilon,i\epsilon+i\omega) \, \delta_{j',1} \delta_{j,1}
    + P_{L,jj'}(i\epsilon,i\epsilon+i\omega)    \\
    \Lambda_{R,jj'}(i\epsilon,i\epsilon+i\omega) & =
    \lambda_R(i\epsilon,i\epsilon+i\omega) \, \delta_{j',N} \delta_{j,N}
    + P_{R,jj'}(i\epsilon,i\epsilon+i\omega) \;.
  \end{split}
\end{align}
The \onePI\ Ward identity \eqref{eq:gammaward} in terms of these
response vertices then reads
\begin{align*}
  & i\omega \Lambda_C(i\epsilon,i\epsilon+i\omega)
  + i\Lambda_R(i\epsilon,i\epsilon+i\omega) 
  - i\Lambda_L(i\epsilon,i\epsilon+i\omega) \\
  & = G^{-1}(i\epsilon+i\omega) - G^{-1}(i\epsilon).
\end{align*}
Clearly, the knowledge of the \RHS\ is not sufficient to determine
$\Lambda_C$, $\Lambda_R$, and $\Lambda_L$ separately but fixes only
the difference.  We use the Dyson equation \eqref{eq:cond-dyson} on
the \RHS, continue analytically to $i\epsilon+i\omega \mapsto
\epsilon+\omega+i0$ and $i\epsilon \mapsto \epsilon-i0$, and take the
limit $\omega\to 0$.  Thereby, the density-response term is suppressed, and
we are left with a relation between current and self energy, using
equation \eqref{eq:condleads} for the lead contribution,
\begin{align*}
  \Lambda_R&(\epsilon-i0,\epsilon+i0)
  - \Lambda_L(\epsilon-i0,\epsilon+i0) \\
  & = - \Im \, \Sigma(\epsilon+i0) - \Im \,\Sigma^\leads(\epsilon+i0)
  + \Im \, \Sigma(\epsilon-i0) + \Im \,\Sigma^\leads(\epsilon-i0) \\
  & = \lambda_R\ket N \bra N - \lambda_L\ket 1 \bra 1
  -2\Im\,\Sigma(\epsilon+i0)\,.
\end{align*}
By equation \eqref{eq:condLambdaR},
\begin{align}
  \label{eq:condward}
   P_R(\epsilon-i0,\epsilon+i0) -
   P_L(\epsilon-i0,\epsilon+i0)
  & = -2\, \Im \, \Sigma(\epsilon+i0)\,.
\end{align}
Thus, the current-vertex corrections are related to the imaginary part
of the self energy.  At $T=0$, both sides vanish exactly because there
is no inelastic scattering, while at higher temperatures, neglecting
inelastic processes is a consistent approximation we made.  If one
tried to improve the approximation by computing vertex corrections but
keeping a real $\Sigma$, it might be no improvement at all because one
would violate number conservation, as explained in \cite{BK61}.


\subsection{Different flow schemes}

In order to check the robustness of our \fRG\ results, I have
considered also the temperatu\-re-flow scheme (cf.\ section
\ref{sec:lutt:flow:temp}) where the temperature is successively
lowered during the \fRG\ flow.  Another option is the recently
introduced interaction-flow scheme (cf.\ section
\ref{sec:lutt:flow:intn}) which slowly switches on the interaction
strength during the flow (at finite temperature, or even at zero
temperature because in the model at hand, the finite size already
provides a regularization).  Using these schemes, I have obtained the
same results as for the Matsubara-frequency cutoff presented in
section \ref{sec:lutt:flow:freqTg0} for a few test cases (cf.\ figure
\ref{fig:dbrgcomp}).  The frequency cutoff scheme at finite
temperature remains numerically the most efficient method.
\fig{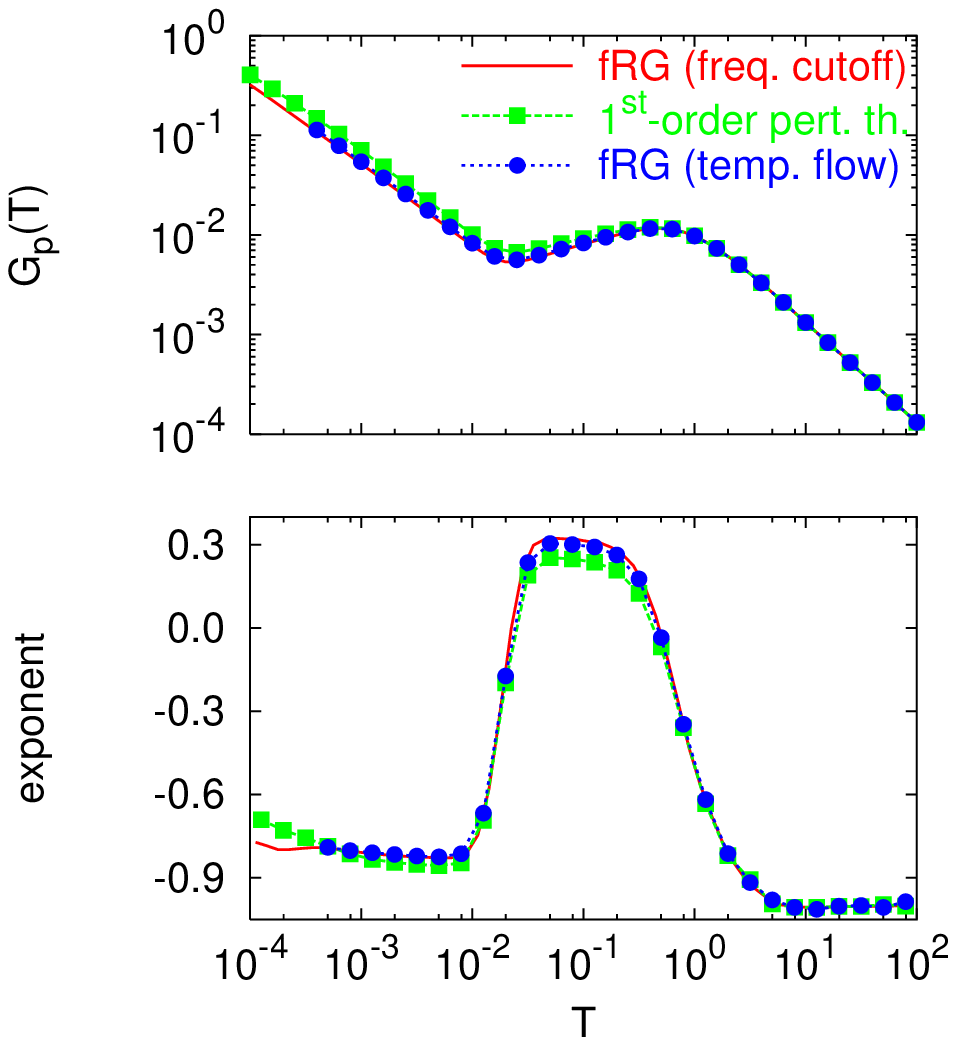}{fig:dbrgcomp}{Comparison of frequency cutoff and
  temperature flow \fRG\ schemes to first-order perturbation theory for
  the resonant conductance through a double barrier ($N=10^4$,
  $U=0.5$, $V=10$, $N_\tdot=100$).  This shows that different \fRG\
  schemes essentially give the same results, indicating the robustness
  of the method.}



\chapter{Luttinger liquids with impurities}
\label{sec:results}

The behavior of three-dimensional metals can be described by
Fermi-liquid theory, where the elementary excitations are fermionic
quasi-particles.  In one dimension, however, the situation is
completely different (for reviews see \cite{Voi95,Gia03}).

A model for interacting \oneD\ spinless fermions with linear
dispersion was introduced by \cite{Tom50} and \cite{Lut63} and has
been solved by \cite{ML65}.  It is characterized by a gapless spectrum
of collective bosonic, density-wave elementary excitations with a
linear dispersion, as well as charge and current excitations, and
correlation functions decaying at zero temperature algebraically in
space and time.  \cite{LP74a} and \cite{Mat74} introduced the
technique of bosonization as a particularly simple method to describe
the low-energy properties of this model.  Any non-linearity of the
fermion dispersion, for instance on the lattice, gives rise to
interactions between the collective modes.  For spinless fermions, a
perturbative expansion of these bosonic couplings is completely
regular in the infrared, and the low-energy excitation spectrum is
similar to the Luttinger-model spectrum \cite{Hal81a}.  Therefore, the
term \emph{Tomonaga-Luttinger liquid} (\LL) was coined for the generic
low-energy phase of interacting \oneD\ fermions.  The correlation
functions obey power laws with exponents that are functions of a
single interaction parameter $K$ with $K=1$ in the non-interacting
case, $0<K<1$ for repulsive and $K>1$ for attractive interaction.

In the presence of impurities or boundaries there are further
characteristic power laws.  For example, the local density of states
$\rho_j(\omega) = -\frac{1}{\pi} \Im \, G_{jj}(\omega+i0)$ near a
single impurity or boundary is suppressed for repulsive interaction as
\cite{KF92c}
\begin{align*}
  \rho_j(\omega) \sim \abs\omega^\ab
  \quad \text{ for } \omega \to 0
\end{align*}
where
\begin{align*}
  \ab = 1/K - 1
\end{align*}
is the \emph{boundary exponent}.  The linear conductance across an
impurity of arbitrary strength scales as \cite{KF92a}
\begin{align*}
  G(T) \sim T^{2\ab}
  \quad \text{ for } T \to 0.
\end{align*}
For repulsive interaction ($\ab>0$), a system containing even a very
weak impurity is insulating at $T=0$, effectively cutting the chain
into two disconnected parts.  Another \LL\ signature is the decay of
Friedel oscillations: an impurity or boundary induces oscillations in
the density profile $n_j$ whose amplitude scales with the distance $x$
as \cite{EG95}
\begin{align*}
  \Delta n_j \sim x^{-K}
  \quad \text{ for } x \to \infty,
\end{align*}
in contrast to $\Delta n_j \sim x^{-1}$ for a \oneD\ Fermi gas.  These
power laws are strictly valid only in the low-energy limit.  This
raises the question at which scales $N$, $T$ the asymptotic \LL\ 
behavior sets in for a specific model, and how the system behaves
before reaching the asymptotic limit.

One way to study \LL\ behavior is to consider the \oneD\ model of
spinless fermions on the lattice with nearest-neighbor interaction
$U$, which we have defined already in section \ref{sec:lutt:model}.
Without impurity, this model can be solved exactly by the Bethe ansatz
\cite{YY66}; it is a Luttinger liquid at any interaction strength $U$
and any filling, except for $\abs{U}>2$ at half filling.  The \LL\ 
interaction parameter $K$ is given for $\abs{U}\leq 2$ at half filling
by \cite{Hal80}
\begin{align*}
  K^{-1} = \frac{2}{\pi} \arccos\left( -\frac{U}{2} \right).
\end{align*}
For $U>2$ at half filling, a phase transition towards a charge-density
wave occurs, while for $U<-2$ the system undergoes phase separation.
In this work we shall concentrate on \LL\ physics, so parameters have
to be chosen to stay away from these phase transitions.

The local density of states has been studied in depth by
\cite{AndThesis}; I will present new results for Friedel oscillations
(section \ref{sec:results:friedel}) and transport through double
barriers (section \ref{sec:results:cond}).  The \fRG\ turns out to be
a versatile tool to study physical effects on energy scales ranging
over several orders of magnitude, at weak to intermediate interaction
strength.


\section{Friedel oscillations}
\label{sec:results:friedel}

In a normal Fermi-liquid metal, impurities induce Friedel oscillations
in the density profile \cite{Fri58,TZ85} which far away from the
impurity have the form
\begin{align*}
  \Delta n(x) \sim \cos(2\kf x+\delta) \, x^{-d}
\end{align*}
where $\Delta n(x) = \vev{\hat n(x) - n_0}/n_0$ is the normalized
density profile, $\kf$ is the Fermi wave vector, $x$ the distance from
the impurity, $d$ the dimension of space, and $\delta$ a phase shift.
This raises the question which density profile is generated in a
Luttinger liquid.

The continuum \oneD\ Luttinger model with a single impurity is
integrable, hence Friedel oscillations should be computable exactly
for any coupling strength.  However, this is technically difficult,
and only approximate results have been obtained except for an exact
solution at $K=\frac 12$ \cite{LLS96}.  The asymptotic behavior for
very \emph{weak} and very \emph{strong} impurities was studied by
\cite{KF92a,KF92b} using the \RG\ method described below.  On the
other hand, the crossover at \emph{intermediate} impurity strength is
particularly important for understanding transport in \oneD\ wires and
was analyzed for weak interaction by \cite{MYG93}.

\cite{EG95} first studied arbitrary impurities for repulsive
interaction using bosonization, which is valid at low temperatures.
For a strong scatterer, the amplitude of the Friedel oscillations
decays as $x^{-K}$.  For a weak scatterer, there is a crossover from
the asymptotic $x^{-K}$ decay for large distances ($x\gg x_0$) to
linear-response decay \cite{Voi95} as $x^{1-2K}$ for short distances
($x\ll x_0$).  The crossover scale diverges as $x_0 \sim V^{-1/(1-K)}$
for $V\to 0$.

Shortly after, \cite{LLS96} obtained exact results for the density
profile in the continuum for $K=\frac 12$ at arbitrary temperature.
For this particular interaction, the problem can be mapped to a
free-fermion model, which simplifies the calculation.  At $T=0$,
\begin{align*}
  \Delta n(x) = 4h\cos(2\kf x+\etaf) \, e^{8\pi h^2 x} K_0(8\pi h^2 x)
\end{align*}
with impurity strength $h=\lambda/\sqrt 2$ ($\lambda$ is the
coefficient of the $\cos[\phi(0)]$ term in the bosonized Hamiltonian),
$K_0(x)$ is a modified Bessel function and $\etaf = -K\pi \lambda/\kf$
a phase shift.  For large distances, the amplitude decays like
$x^{-\frac 12}$ as expected.  At $T>0$,
\begin{align*}
  \Delta n(x) = \cos(2\kf x+\etaf) \, \sqrt{\frac{4\pi T}{\sinh(2\pi Tx)}} \,
  F\left(\frac 12,\frac 12;1+2\frac{h^2}{T},\frac{1-\coth(2\pi Tx)}{2}\right)
\end{align*}
with the hypergeometric function $F(a,b;c,d)$.  For $x\gg 1/T$ the
amplitude of the Friedel oscillations decays exponentially.

Recently, \cite{GYL04} used the functional bosonization technique and
a self-consistent harmonic approximation at low temperature and weak
impurity strength to obtain the Green function at arbitrary
interaction $K$.  It shows the full crossover from the
impurity-dominated behavior at short distances to the pure \LL\ 
behavior at large distances.  From the Green function, one can extract
the density profile as
\begin{align*}
  \Delta n(x) = \cos(2\kf x+\etaf) \,
  \abs{\sinh(2\pi TxK/\vf)}^{-K}.
\end{align*}
For distances shorter than the thermal coherence length, $x\ll
\pi\vf/T$, the amplitude decreases as $x^{-K}$, while for larger
distances it is suppressed exponentially.

In section \ref{sec:lutt:density} above I have described how to
compute the density profile in the \fRG\ framework, illustrating the
renormalization of a composite operator and the flow of response
functions in a concrete example.  In the following section I will
present my results for the density profile for a wide range of
parameters and compare them with exact \DMRG\ results and the
asymptotic formulae from bosonization presented here.


\subsection{Results}
\label{sec:results:friedel:results}

Figure~\ref{fig:densdmrg} shows \fRG\ and \DMRG\ results for the
density profile $n_j$ for a spinless-fermion chain with 128 sites and
interaction strength $U=1$ at half filling.  The Friedel oscillations
emerge from both boundaries and interfere in the center of the chain.
The accuracy of the \fRG\ results is excellent for all $j$.

For incommensurate filling factors the density profile looks more
complicated. This can be seen in figure~\ref{fig:incomm}, where \fRG\ 
results are shown for the density modulation $\abs{n_j - n}$ near the
boundary of a system with an average density $n=0.393$ and $8192$
sites. For long distances from the boundary the oscillation amplitude
has a well-defined envelope which fits to a power law as a function of
$j$.  In the following I will examine the large-distance behavior of
the amplitudes more closely for the half-filled case.

Figure~\ref{fig:friedbnd} shows \fRG\ results for the amplitude of
density oscillations emerging from an \emph{open boundary}, for a very
long spinless fermion chain with $2^{19}+1$ sites and various
interaction strengths $U$ at half filling.  The other end of the
chain (opposite to the open boundary) is smoothly connected to a
non-interacting lead.  In a log-log plot (upper panel of
figure~\ref{fig:friedbnd}) the amplitude follows a straight line for
almost all $j$, corresponding to a power-law dependence.  Deviations
from a perfect power law can be seen more clearly by plotting the
effective exponent $\alpha_j$, defined as the negative logarithmic
derivative of the amplitude with respect to $j$ (see the lower panel
of figure~\ref{fig:friedbnd}). The effective exponent is almost
constant except at very short distances or when $j$ approaches the
opposite end of the interacting chain, which is not surprising.  From
a comparison with the exact exponent (horizontal lines in the figure)
one can assess the quantitative accuracy of the \fRG\ results.

Effective exponents describing the decay of Friedel oscillations
generated by \emph{site impurities} of various strengths are shown in
figure~\ref{fig:friedimp}, for a half-filled spinless fermion chain
with $2^{18}+1$ sites and interaction $U=1$.  Both ends of the
interacting chain are coupled to non-interacting leads to suppress
oscillations otherwise induced by the boundaries.  For strong
impurities the results are close to the boundary result (cf.\ 
figure~\ref{fig:friedbnd}), as expected.  For weaker impurities the
oscillations decay more slowly, \ie, with a smaller exponent, and
do not reach the boundary behavior within the range of our chain for
$V<1$.  For very weak impurities ($V=0.01$ in
figure~\ref{fig:friedimp}) the oscillation amplitude follows a power
law corresponding to the linear-response behavior with exponent $2K -
1$ at intermediate distances.

The same crossover between linear-response behavior for very weak
impurities and \LL\ behavior for strong renormalized impurities is
observed in the oscillations of the effective impurity potential
$\Sigma_{jj}$ (cf.\ figure~\ref{fig:sigmaimp}).  When the density
oscillations decay with exponent $K$, the $\Sigma_{jj}$ oscillations
decay with exponent $1$.  In the linear-response regime, however, the
respective exponents are $2K-1$ and $K$.  This was explained in
\cite{MMSS02a,MMSS02b}:

\fig{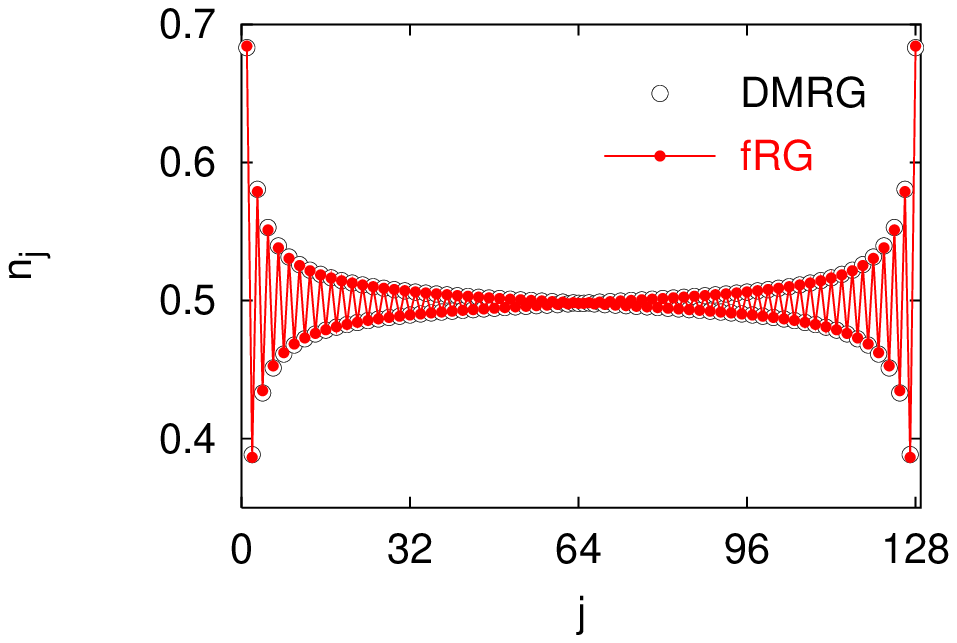}{fig:densdmrg}{Density profile $n_j$ for a spinless
  fermion chain with $128$ sites and interaction strength $U=1$ at
  half filling.  \fRG\ results show an excellent agreement with
  numerically exact \DMRG\ data.}
\fig{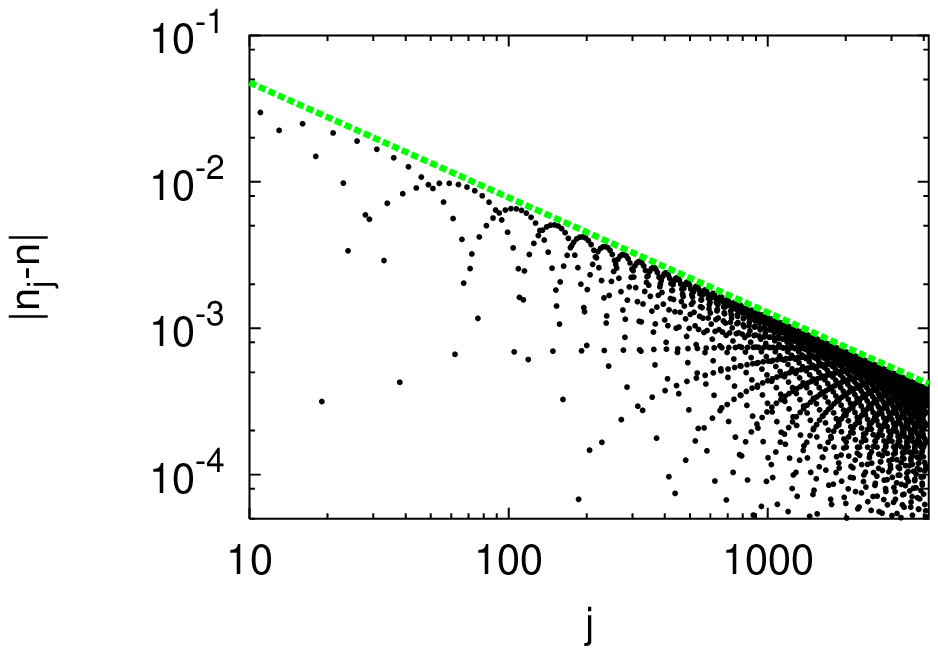}{fig:incomm}{Density modulation $\abs{n_j-n}$ as a
  function of the distance from a boundary, for spinless fermions with
  interaction strength $U=1$ and average density $n=0.393$ on a chain
  with $8192$ sites; the dashed line is a power-law fit to the
  envelope of the oscillation amplitudes with exponent $K=0.785$
  corresponding to $\ab \approx 0.274$.  It shows that Friedel
  oscillations display the expected behavior also away from half
  filling.}

\noindent since the linear-response backscattering
amplitude scales with the cutoff $\Lambda$ as $V_{\kf,-\kf} \sim
(1/\Lambda)^{1-K}$, this provides the generic scaling law in the
low-energy regime, and also the self energy scales away from $2\kf$
with the same exponent, $\Sigma_{k,k'} \sim (k-k'-2\kf)^{1-K}$.
Performing a Fourier transform, the real-space decay scales as
$\abs{j-j_0}^{-K}$.
\fig{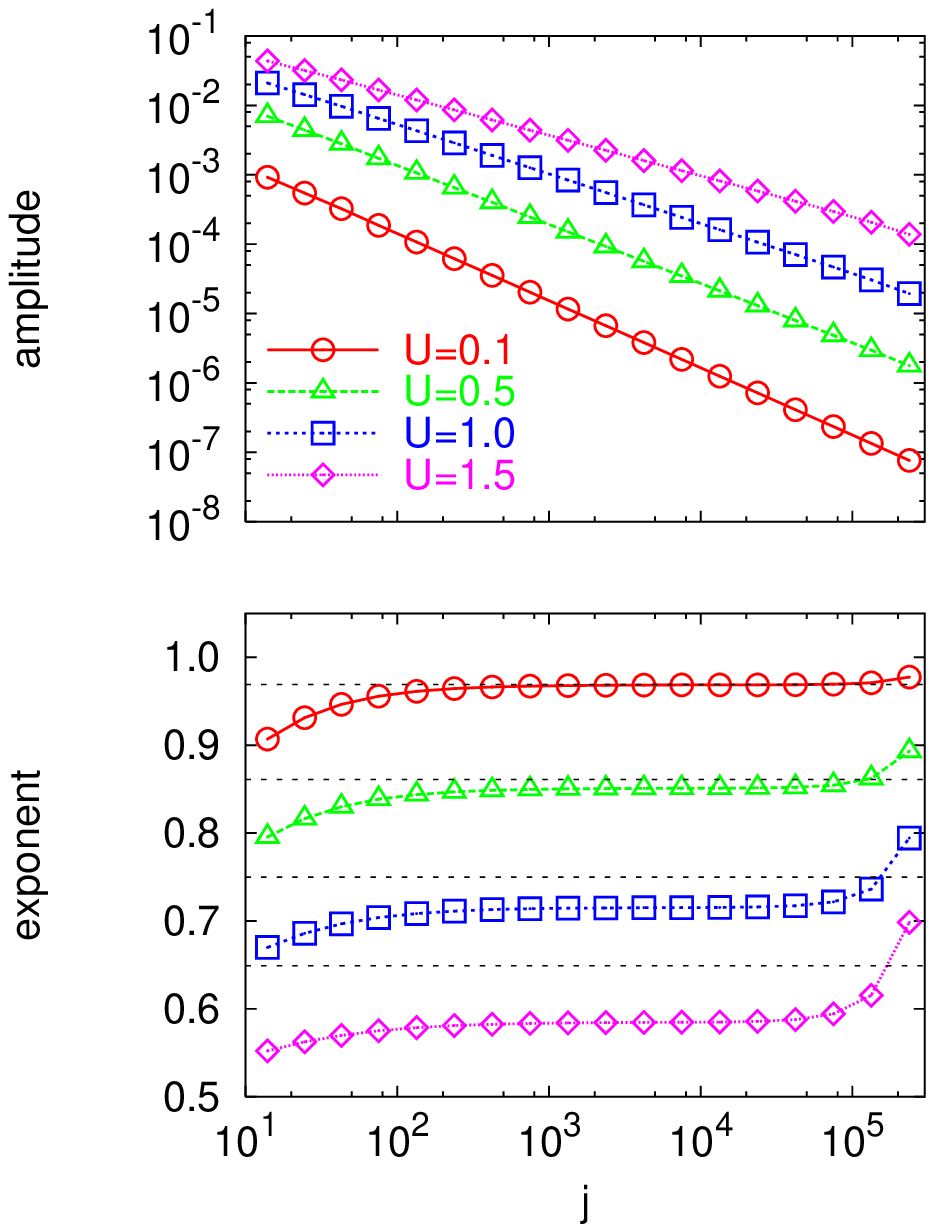}{fig:friedbnd}{Amplitude (envelope) of oscillations of
  the density profile $n_j$ induced by a \emph{boundary} as a function
  of the distance from the boundary, for spinless fermions with
  various interaction strengths $U$ at half filling; the interacting
  chain with $2^{19}+1$ sites is coupled to a semi-infinite
  non-interacting lead at the end opposite to the boundary.
  \emph{Upper panel:} log-log plot of the amplitude. \emph{Lower
    panel:} effective exponents for the decay, and the exact
  asymptotic exponents (Bethe ansatz) as horizontal lines.  The
  plateaux allow to read off the exponents very accurately.}

\fig{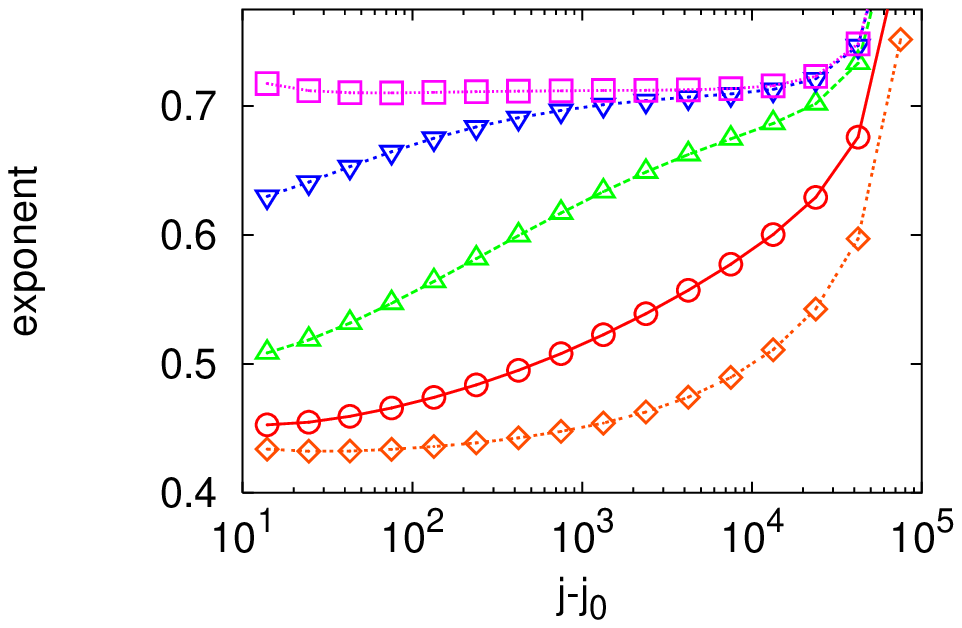}{fig:friedimp}{Effective exponent for the decay of
  density oscillations as a function of the distance from a \emph{site
    impurity} of strengths $V=0.01$, $0.1$, $0.3$, $1$, $10$ (from
  bottom to top); the impurity is situated at the center of a spinless
  fermion chain with $2^{18}+1$ sites and interaction strength $U=1$
  at half filling; the interacting chain is coupled to semi-infinite
  non-interacting leads at both ends.  This shows the crossover from
  the density response regime $(V\ll 1)$ to the boundary behavior
  $(V\gg 1)$.}
\fig{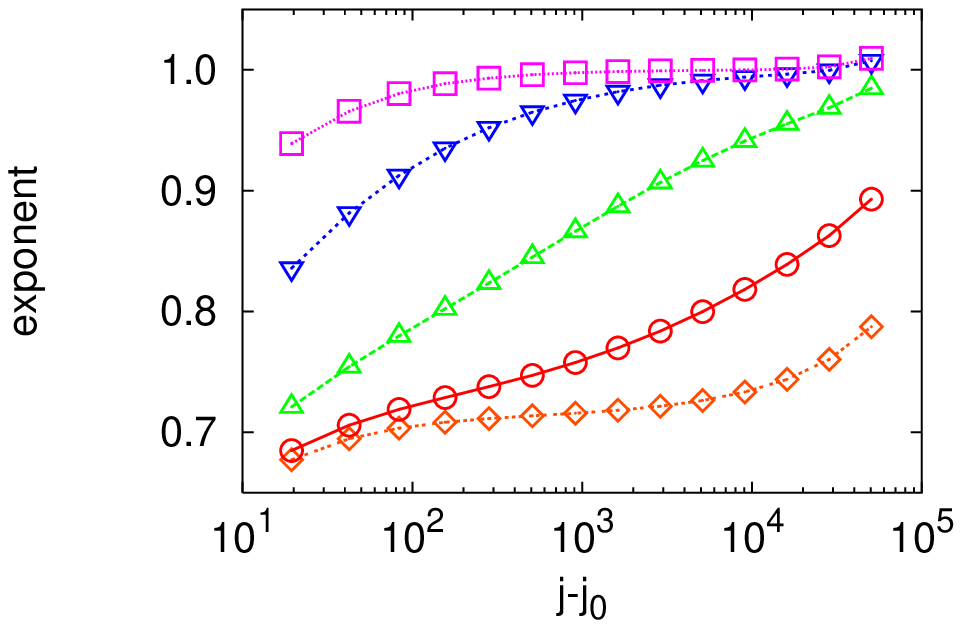}{fig:sigmaimp}{Effective exponent for the decay of
  oscillations of $\Sigma_{j,j}$ as a function of the distance from a
  site impurity of strengths $V=0.01$, $0.1$, $0.3$, $1$, $10$ (from
  bottom to top), for the spinless fermion model at half filling and
  interaction strength $U=1$; the impurity is situated at the center
  of a chain with $L=2^{18}+1$ sites.}

Finally, I present results for the effective exponent of the
density-oscillation decay in the case of an \emph{attractive}
interaction $U=-1$, see figure~\ref{fig:friedatt}. In that case the
effective impurity strength should scale to zero at low energies and
long distances \cite{KF92a}.  Indeed, for weak and moderate bare
impurity potentials the effective exponent in
figure~\ref{fig:friedatt} approaches the linear-response exponent $2K
- 1$.  Only for very strong impurities the density oscillations decay
with the smaller exponent $K$ over several orders of magnitude.
\fig{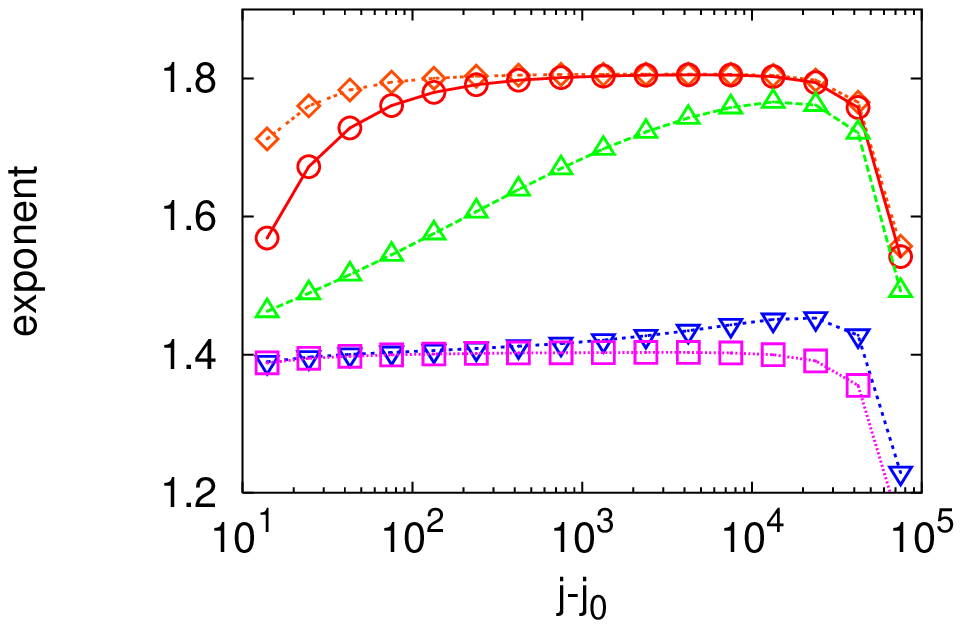}{fig:friedatt}{Effective exponent for the decay of
  density oscillations as a function of the distance from a site
  impurity of strengths $V=0.1$, $1$, $10$, $100$, $1000$ (from top to
  bottom) for the same chain as in figure~\ref{fig:friedimp} but now
  with an \emph{attractive} interaction, $U=-1$.  This demonstrates
  that our method works well also for negative $U$.}


\section{Transport through double barriers}
\label{sec:results:cond}

The motivation to study double barriers using the \fRG\ is twofold:
$(i)$ the double barrier is physically interesting and experimentally
accessible, for example in specifically fabricated quantum wires or
double kinks in carbon nanotubes, see for instance \cite{CW93,Pos01},
and $(ii)$ it is a multi-scale problem; in addition to the temperature
$T$, there are additional scales such as the size of the dot region
$N_\tdot$ between the barriers and the strength of the barriers
$V_{L,R}$, cf.\ figure \ref{fig:doublebar}.  The double barrier
exhibits universal scaling in certain limits---we observe several
different power laws---but also non-universal crossover behavior in
between, as different physical processes become relevant.  This is
particularly important because in experimental setups, the parameters
are usually in the intermediate range.  Therefore, the double barrier
provides a showcase for the power of the \fRG\ to treat all scales on
an equal footing.  Nevertheless, as a note of caution, several
important ingredients for a realistic description are still missing in
our model, although they can in principle be handled by our method,
such as the spin degree of freedom, higher-dimensional leads, and
realistic contacts.  We therefore refrain from a detailed comparison
of our findings to experiments.

For a weak \emph{single} impurity, the conductance in the limit $T\to
0$ is suppressed for repulsive interaction and enhanced for attractive
interaction as compared to the non-interacting case \cite{LP74b,AR82}.
\cite{KF92a} showed that this is a universal result independent of the
impurity strength: for repulsive interaction, the conductance across
the impurity scales as $G(T)\sim T^{2\ab}$, such that asymptotically
for $T\to 0$ the wire becomes insulating.  This was derived using an
\RG\ method: a weak barrier, or rather its backscattering component
$V(2\kf)$, is a relevant perturbation of the clean system and grows
stronger in the \RG\ flow.  On the other hand, a weak link between two
otherwise separate semi-infinite chains is an irrelevant perturbation
which remains weak in the \RG\ flow.  Because in the weak-impurity and
weak-link limits the direction of the \RG\ flow is compatible towards
a strong barrier or weak link, respectively, Kane and Fisher connected
both perturbatively accessible limits and concluded that any barrier
becomes strong in the asymptotic limit.  This is supported by an exact
solution at $K = \frac 12$.  Conversely, an \emph{attractive}
interaction suppresses an initial backscattering to yield perfect
transmission even for a system with an impurity.  The \fRG\ has been
used successfully to reproduce the one-parameter scaling of the
conductance for $1/2 \leq K \leq 1$ \cite{Med03}.

\fig[width=0.9\linewidth]{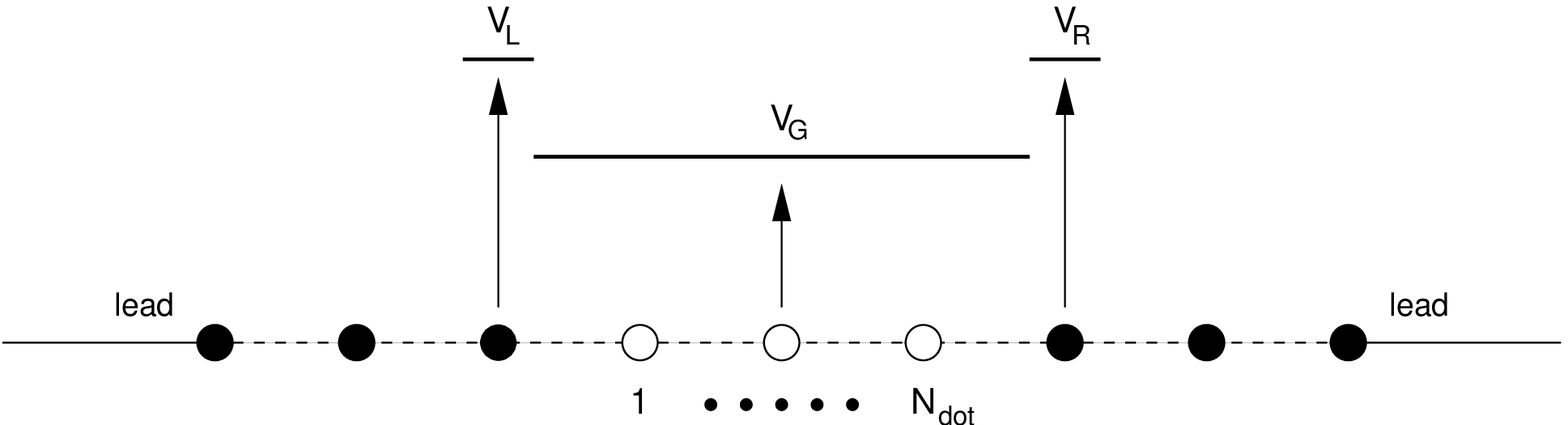}{fig:doublebar}{Model of the double
  barrier.}

In the case of a double barrier, there are resonance peaks in the
conductance $G(V_G)$ as the gate voltage $V_G$ is varied.  Using an
\RG\ analysis similar to the one for the single barrier for
asymptotically low temperatures, \cite{KF92c} derived a phase diagram
for the \emph{peak conductance} $G_\p$ depending on the barrier
strength $V$ and interaction parameter $K$.  For attractive
interaction $K > 1$ there is perfect transmission for any $V_G$, while
for repulsive $\frac 12 < K < 1$ there are sharp conductance peaks at
which perfect transmission is reached.  For $K < \frac 14$ no
transmission is possible for any $V_G$, and for $\frac 14 < K < \frac
12$ there is a line (\emph{Kosterlitz-Thouless separatrix}) of
critical barrier strengths: for stronger barriers no transmission is
possible, while for weaker barriers conductance peaks appear.  The
line shape of the resonance peaks is non-Lorentzian, the tails falling
off as $(\Delta V_G)^{-2/K}$, where $\Delta V_G$ is the detuning from
resonance.

Shortly afterwards, \cite{FN93} studied the double barrier for
arbitrary interaction by second-order perturbation theory in the
barrier strength $V \ll 1$.  At low temperatures, the deviation of the
resonant $G_\p(T)$ from $e^2/h$ scales as $T^{2K}$.  As the
temperature is increased, perturbation theory in the inverse barrier
strength $1/V \ll 1$ and a Master-equation approach \cite{Fur98} yield
a regime of \emph{uncorrelated sequential tunneling} (\UST)
characterized by peaks of height $G_\p(T) \sim T^{\ab-1}$ and width
$w(T) \sim T$, for $T\ll \Delta_\tdot$, where $\Delta_\tdot :=
\pi\vf/N_\tdot$ is the level spacing of the dot.  For $\Delta_\tdot
\ll T \ll B$, where $B$ is the bandwidth, $G_\p(T)$ increases as
$T^{2\ab}$ for increasing $T$.

However, recent experiments on carbon nanotubes \cite{Pos01} have
reported for the temperature range $\Gamma \ll T \ll \Delta_\tdot$
that both the resonant peak height $G_\p(T) \sim T^{0.7}$ and $G(V_G)$
peak width $w(T) \sim T$ decrease with decreasing temperature.  Here,
$\Gamma$ is the width of a resonance in the transmission probability
$\T(\epsilon)$ given by equation \eqref{eq:trans}, which for strong
barriers is related to the tunneling rate into and out of the dot.
While also \cite{FN93,Fur98} found power laws with a positive power of
$T$, they were expected to occur only for very strong repulsive
interactions, where $\ab$ is at least twice as large as the value $\ab
\approx 0.6 \dotsc 1.0$ observed in carbon nanotubes.  \cite{Pos01}
suggested that \emph{correlated sequential tunneling} (\CST) with
exponent $2\ab - 1 > 0$ dominates over \UST\ with exponent $\ab -1 <
0$ in the temperature range $\Gamma \ll T \ll \Delta_\tdot$.
Subsequently, \cite{Tho02,TEG04} argued that cotunneling processes of
second order in the end-tunneling local density of states (each with
exponent $\ab$) are the leading contribution to the peak conductance
(hence the $2\ab$), and claimed perfect agreement with the
experimental data of \cite{Pos01}.  In contrast, \cite{Fur98} had
found that cotunneling via virtual intermediate states dominated the
\emph{tails} of the conductance peak away from the resonance.  This
led to a renewed interest in transport through double barriers, with
several new methods available to advance beyond asymptotically low
temperatures.

Shortly after, \cite{PG03,NG03} studied the double barrier at
arbitrary temperature and impurity strength but weak interaction using
the leading-log resummation of \cite{MYG93,YGM94}.  They considered
the \RG\ flow of the energy-dependent scattering amplitudes, based on
a resummation of perturbation-theory diagrams with leading logarithmic
divergences $\ab^n \ln^n(\frac{1}{\abs{k-\kf}d})$, where $d$ is the
spatial range of the interaction.  Inelastic processes are sub-leading
at weak interaction.  Their results for a single resonant level agreed
with \cite{FN93} for $T<\Delta_\tdot$, finding \UST\ but not \CST.
\cite{PG03} extended the \RG\ study to a multi-level dot and weak
barriers (closer to experimental parameters) but again confirmed \UST.

Recently, \cite{KG03} found an exact solution for a special model at
interaction strength $K = \frac 12$.  However, in that model the
conductance scales as $T^{-1}$ even for $T<\Delta_\tdot$ where one
would expect \UST, as if the fermions were non-interacting.  Because
it shows no sign of either \UST\ or \CST, this result is probably not
generic and cannot help to resolve the puzzle.

\cite{HE04} computed the conductance using the Quantum Monte Carlo
(\QMC) method for interaction $K = 0.6$ and obtained data interpreted
to be consistent with \CST\ for weak barriers and sufficiently high
temperatures, which was explained by an additional transport channel
for strong interaction.

Against this background, it is desirable to have an unbiased method to
compute the conductance without restricting from the beginning---by
physical intuition---which physical processes are dominant.  The \fRG\ 
provides this with the only restriction that it is perturbative in the
\emph{renormalized} interaction but contains contributions of all
orders in the bare interaction $U$.  It confirms the \UST\ picture but
in addition allows to vary all parameters to see exactly at which
temperature, barrier strength, dot size etc.\ the universal scaling
sets in, if at all.  While the justification of the \fRG\ was
weak-coupling, we reproduce one-parameter scaling for the single
barrier at $U=2$ very accurately \cite{Med03}, and our
results for the double barrier qualitatively agree with the \QMC\ data
($U=\sqrt 3$), indicating that the \fRG\ is reliable up to this
interaction strength.


\subsection{Results}
\label{sec:results:cond:results}

The \fRG\ has been used before to compute the conductance through a
single impurity at zero temperature \cite{MS03a,MS03b,Med03}.  After
having developed and implemented the finite-temperature \fRG, I have
obtained results for transport through a double barrier, symmetric or
asymmetric, at or off resonance, with weak or strong barriers,
enclosing small or large dots, for repulsive interaction (see also
\cite{Med04,EMABMS04}).  I will report my findings for several
interesting regions in this large parameter space, showing agreement
with known results as well as providing clarification of a contentious
issue.

I first present results for a strong symmetric double barrier,
$V_{L,R}=10$, enclosing a dot of $N_\tdot=6$ sites.  The interacting
system size is taken to be $N=10^4$, in agreement with the size of
realistic samples of carbon nanotubes.  As the gate voltage $V_G$ is
varied across the band, there are six resonance peaks with level
spacing roughly $\Delta_\tdot$ (figure \ref{fig:gvg}).
\fig[width=7cm]{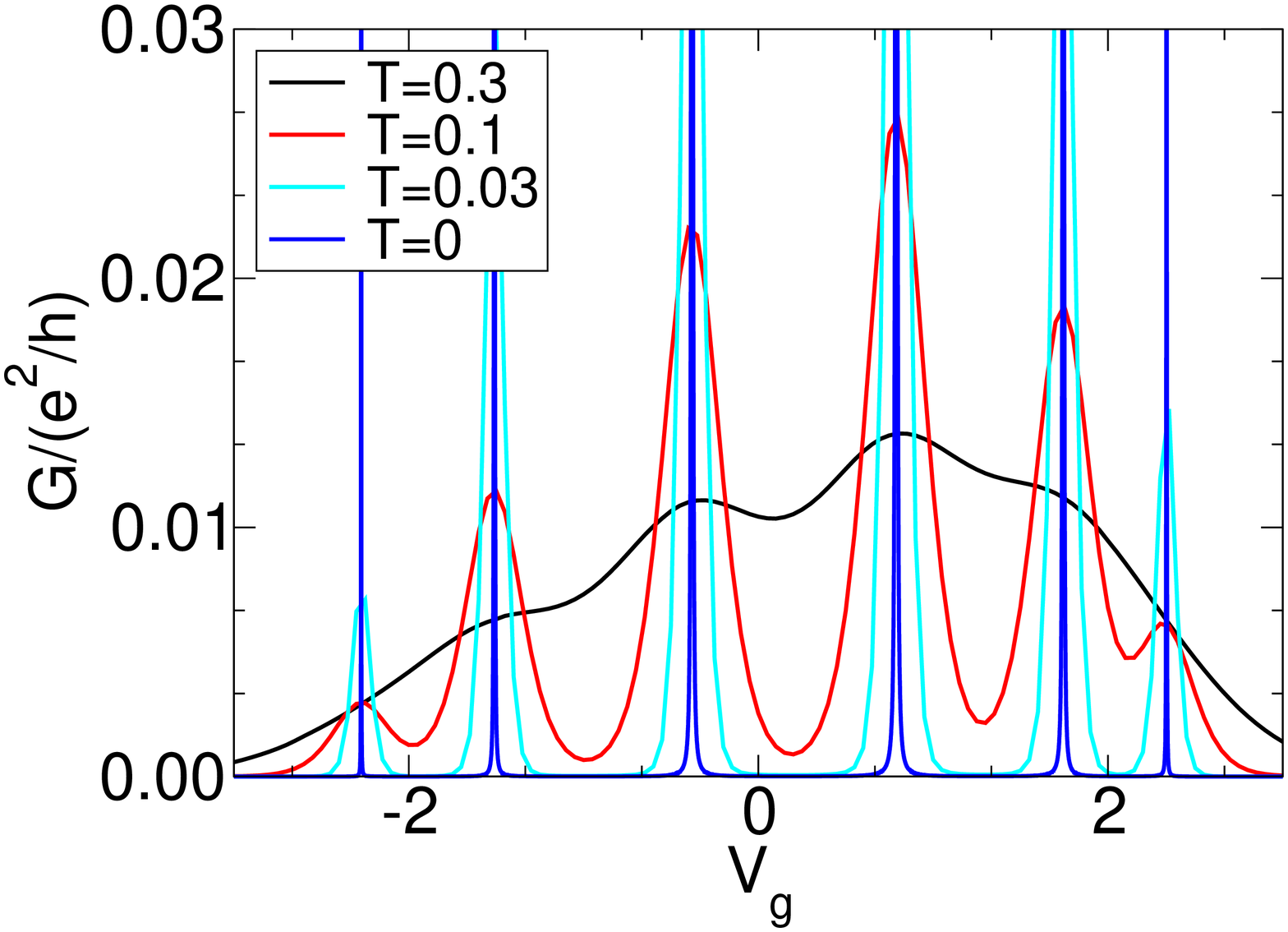}{fig:gvg}{The conductance $G(V_G)$ as a function
  of gate voltage $V_G$ for $N_\tdot = 6$, $U=0.5$, $V_{L,R}=10$,
  $N=10^4$ and different $T$.  One observes $N_\tdot$ resonance peaks
  which are widened with increasing temperature.}
In non-interacting systems the peaks have a finite width even at
$T=0$, depending on the shape of the barrier and the local density of
states of the leads at the barrier.  Repulsive interaction has a
strong influence on the line shape of the resonance peaks: they become
infinitely sharp in the asymptotic limit ($N\to\infty$, $T\to 0$) but
perfect transmission $e^2/h$ is still possible for symmetric barriers.
As the temperature is increased the peaks become wider and lower.  The
peak conductance $G_\p(T)$ as a function of temperature shows several
different power laws for appropriately chosen dot parameters.  All the
peaks in figure \ref{fig:gvg} have a different shape and size due to
band effects, however $G_\p(T)$ behaves similarly for each peak, and
in the following I will always consider the peak closest to $V_G=0$.

\fig[width=8cm]{gsymmreso}{fig:gsymmreso}{The resonant peak
  conductance $G_\p(T)$ for different dot sizes.  \emph{Upper panel:}
  $N_\tdot=2$ (circles), $6$ (squares), and $100$ (diamonds), with the
  respective level spacing $\Delta_\tdot$ indicated by the arrows.
  $U=0.5$, $N=10^4$, $V_{L,R}=10$.  The solid curve shows $G(T)/2$
  for a single barrier.  \emph{Lower panel:} Logarithmic derivative of
  $G_\p(T)$.  Solid line: $2\ab$; dashed line: $\ab-1$; dash-dotted
  line: $2\ab-1$.}
Figure~\ref{fig:gsymmreso} shows the peak conductance $G_\p(T)$ for
three different dot sizes $N_\tdot$ in the upper panel, while in the
lower panel the logarithmic derivative is plotted.
\fig[scale=0.9]{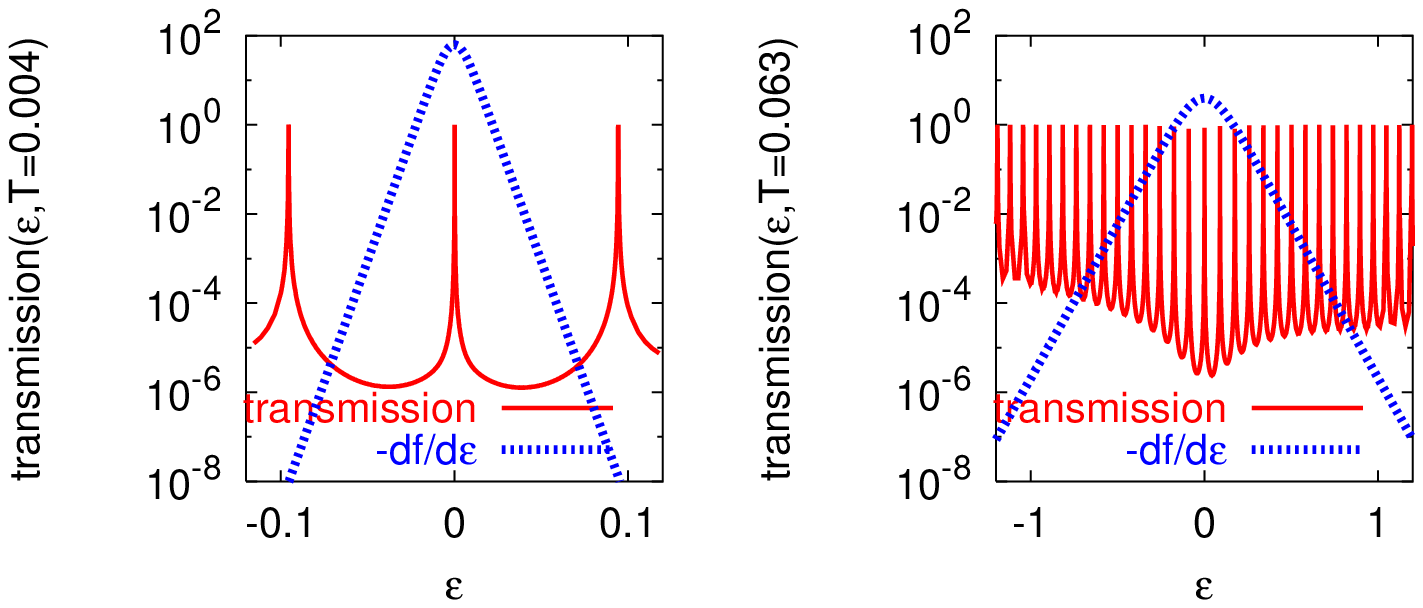}{fig:transmis}{Transmission $\T(\epsilon,T)$
  for several temperatures: \emph{left panel} $T=0.004$, \emph{right
    panel} $T=0.063$.  The dashed line is the derivative of the Fermi
  function at the respective temperature.  The dot parameters are
  $N_\tdot=100$, $U=1$, $N=10^4$ and $V_{L,R}=10$.  This shows that
  for low temperatures (\UST), a single transmission peak determines
  the conductance, while for higher temperatures (Kirchhoff) many
  peaks contribute.}
For temperatures larger than the bandwidth, the conductance scales as
$G_\p(T) \sim T^{-1}$.  This is because $f'(\epsilon)$ and the
transmission in equation \eqref{eq:cond} vary only very little over
the band, but $f'(\epsilon)$ decreases with increasing temperature as
$1/T$.  For temperatures below the dot level spacing $\Delta_\tdot$
(indicated by the arrows), the peak conductance decreases as $G_\p(T)
\sim T^{\ab-1}$ (\UST).  This exponent is marked in the lower panel by
the dashed line.  In the \UST\ regime the conductance integral is
dominated by a single peak in the transmission $\T(\epsilon,T)$ with a
width $\Gamma<T$ (cf.\ figure \ref{fig:transmis}, left panel).  As the
temperature is lowered, the \UST\ regime extends down to $T^*$ where
$T=\Gamma$.  In $G(V_G)$ the peaks are well separated and have a
width $w(T)\sim T$.

For the above parameters and dots with $N_\tdot \geq 4$, the
conductance shows a third regime for $\Delta_\tdot \lesssim T \lesssim
1$.  In this region, the conductance increases with increasing
temperature, leading to a non-monotonic overall $T$ dependence.  For
large dots $N_\tdot \gtrsim 30$, the conductance increases according
to the power law $G_\p(T) \sim T^{2\ab}$, indicated by the solid line
in the lower panel of figure \ref{fig:gsymmreso}.  Many peaks in the
transmission $\T(\epsilon,T)$ contribute to the conductance (cf.\ 
figure \ref{fig:transmis}, right panel).  This regime is most clearly
seen for strong barriers $V_{L,R} \gg 1$ and large dots $N_\tdot \gg
1$.  

If we add the resistances of both barriers separately according to
Kirchhoff's law, $1/G = 1/G_L + 1/G_R$, we obtain the solid curve in
the upper panel of figure~\ref{fig:gsymmreso}.  As our data lie on
this curve for $T\gtrsim \Delta_\tdot$, we suggest the following
interpretation: the electrons first tunnel through one barrier with a
certain probability, then incoherently through the next.  Only as the
temperature is lowered to $T\approx\Delta_\tdot$, the electron
``sees'' both barriers coherently and tunnels resonantly, such that
the conductance increases towards $e^2/h$ as $T\to 0$.

\fig[width=8cm]{gsymmoffr}{fig:gsymmoffr}{On- and off-resonance
  conductance.  \emph{Upper panel:} $G(T)$ for $U=0.5$, $N=10^4$,
  $V_{L,R}=10$, and $N_\tdot=100$.  On resonance $\Delta V_G=0$
  (circles), near resonance $\abs{\Delta V_G}=0.001$ (squares), and in
  a conductance minimum with $\abs{\Delta V_G}=0.04$ (diamonds).  The
  off-resonance curves lie on the resonance curve for $T \gtrsim
  \abs{\Delta V_G}$, then cross over to decay as $G_\p(T) \sim
  T^{2\ab}$.  \emph{Lower panel:} Logarithmic derivative of $G(T)$.
  Solid line: $2\ab$; dashed line: $\ab-1$.  At low temperature the
  double barrier off resonance behaves like a single impurity.}
It is thus only in the low-temperature range where it makes a
difference whether $V_G$ is tuned to resonance or not (cf.\
figure~\ref{fig:gsymmoffr}).  If the gate voltage is slightly off
resonance, $\Delta V_G \neq 0$ and $\abs{\delta V_G} \ll
\Delta_\tdot$, the conductance is identical to the resonant case for
$T \gtrsim \Delta V_G$, while for lower temperatures the double
barrier acts as a single impurity of strength $\abs{\Delta V_G}$,
hence the conductance scales as $G(T) \sim T^{2\ab}$.

\fig[width=8cm]{gsymmdiffu}{fig:gsymmdiffu}{Conductance for different
  interaction strength.  \emph{Upper panel:} Peak conductance
  $G_\p(T)$ for $U=0.5$ (full symbols) and $U=1$ (open symbols).
  Curves are shown for dot sizes $N_\tdot=2$ (red) and $N_\tdot=100$
  (blue).  \emph{Lower panel:} Logarithmic derivative of $G(T)$.  The
  dependence of the exponents on the interaction strength support the
  claim that the exponents are $\ab-1$ and $2\ab$, respectively.}
Up to this point, we have presented conductance data for the weak
interaction strength $U=0.5$ ($\ab=0.165$).  The numerical exponent of
$-0.835$ in the \UST\ regime was identified with $\ab-1$, and the
exponent $0.330$ in the Kirchhoff regime with $2\ab$.
Figure~\ref{fig:gsymmdiffu} shows the exponents for a different value
of $U$: at $U=1$ ($\ab=0.35$), the exponents $-0.65$ and $0.70$ can be
read off, confirming this identification within the numerical
accuracy.  A comparison of the exponents for strong and weak barriers
can be found in \cite{EMABMS04}.

For strong barriers, large dot size and weak interaction, there is no
indication of a power law with exponent $2\ab-1$ claimed by
\cite{Tho02,TEG04,HE04}.  Consider, therefore, a dot with parameters as
close as possible to those in \cite{HE04}: weak to intermediate
barriers, intermediate dot size and larger $U$ (cf.\
figure~\ref{fig:gsymmweak}).
\fig[width=8cm]{gsymmweak}{fig:gsymmweak}{Weak barriers, \emph{Upper
    panel:} $G_\p(T)$ for $U=1.5$, $N=10^4$, $N_\tdot=10$ with
  $V_{L,R}=0.8$ (circles), $N_\tdot=30$ with $V_{L,R}=1.5$
  (squares), and $N_\tdot=50$ with $V_{L,R}=0.8$ (diamonds). The
  arrows indicate $\Delta_\tdot$ for the different dot sizes.
  \emph{Lower panel:} Logarithmic derivative of $G_\p(T)$. Solid line:
  $2\ab$; dashed line: $\ab-1$.  For small dots, the $2\ab$ power law
  is not clearly developed; however the conductance agrees with the
  \QMC\ data \cite{HE04}.}
For a small dot $N_\tdot=10$ (circles), neither the exponent $\ab-1$
(dashed line) nor $2\ab$ (solid line) is clearly developed.  For
larger dots $N_\tdot=30$ (squares) and $N_\tdot=50$ (diamonds), the
$\ab-1$ exponent (\UST) is clearly visible, but the Kirchhoff exponent
$2\ab$ is still not reached.  In the \QMC\ data \cite{HE04} the
conductance is fitted to a slope of $1/3$ in a log-log plot, which is
interpreted as scaling with exponent $2\ab-1$ (\CST) for $\ab=2/3$.
Our data shows the conductance curve with a slope of $0.3\dotsc 0.7$,
depending on the dot parameters, but no fixed scaling exponent.
Therefore, we interpret this as \emph{non-universal} behavior.  

\fig[width=8cm]{gasymreso}{fig:gasymreso}{Strongly asymmetric
  barriers, \emph{Upper panel:} $G_\p(T)$ for $V_L=5$, $V_R=50$,
  $U=0.5$, $N=10^4$, $N_\tdot=2$ (circles) and $N_\tdot=20$ (squares).
  \emph{Lower panel:} Logarithmic derivative of $G_\p(T)$. Solid line:
  $2\ab$; dashed line: $\ab-1$.}
Figure~\ref{fig:gasymreso} shows the resonant peak conductance for
strongly \emph{asymmetric barriers}.  The Kirchhoff regime is
unchanged qualitatively as both barriers are very strong.  For lower
temperatures one observes a \UST\ regime which crosses over into
single-impurity scaling $G_\p(T) \sim T^{2\ab}$ for small dots at very
low temperatures.

\fig[width=6cm]{gsmall}{fig:gsmall}{Effective exponent of
  $1-G_\p(T)/(e^2/h)$ for a small dot with $N_\tdot=1$, $U=1$,
  $N=10^4$, and intermediate to weak hopping barriers $t_{l,r}$.
  Dash-dotted line: $2K$.}
For a small dot ($N_\tdot=1$) with weak barriers the conductance
approaches the perfect value $e^2/h$ according to the power law
\cite{KF92b,FN93} (cf.\ figure \ref{fig:gsmall})
\begin{align*}
  \frac{e^2}{h} - G_\p(T) \sim T^{2K}.
\end{align*}
Thus, our method yields up to four different power laws within the
same framework.


\section{Summary}
\label{sec:results:sum}

I have demonstrated that the \fRG\ is a powerful method for
multi-scale problems.  As applications I have studied Friedel
oscillations off an impurity or boundary in an interacting wire, and
transport through a double barrier.  Both problems were previously
investigated using effective field-theoretical models; in the cases
where exact results for such models are known I have obtained
quantitative agreement for weak to intermediate interactions $1/2 \leq
K \leq 1$.  For the Friedel oscillations at $T=0$, I observed the
exponent $K$ at large distances and $2K-1$ for weak impurities at
short distances \cite{AEMMSS04}.

\fig[width=11.5cm]{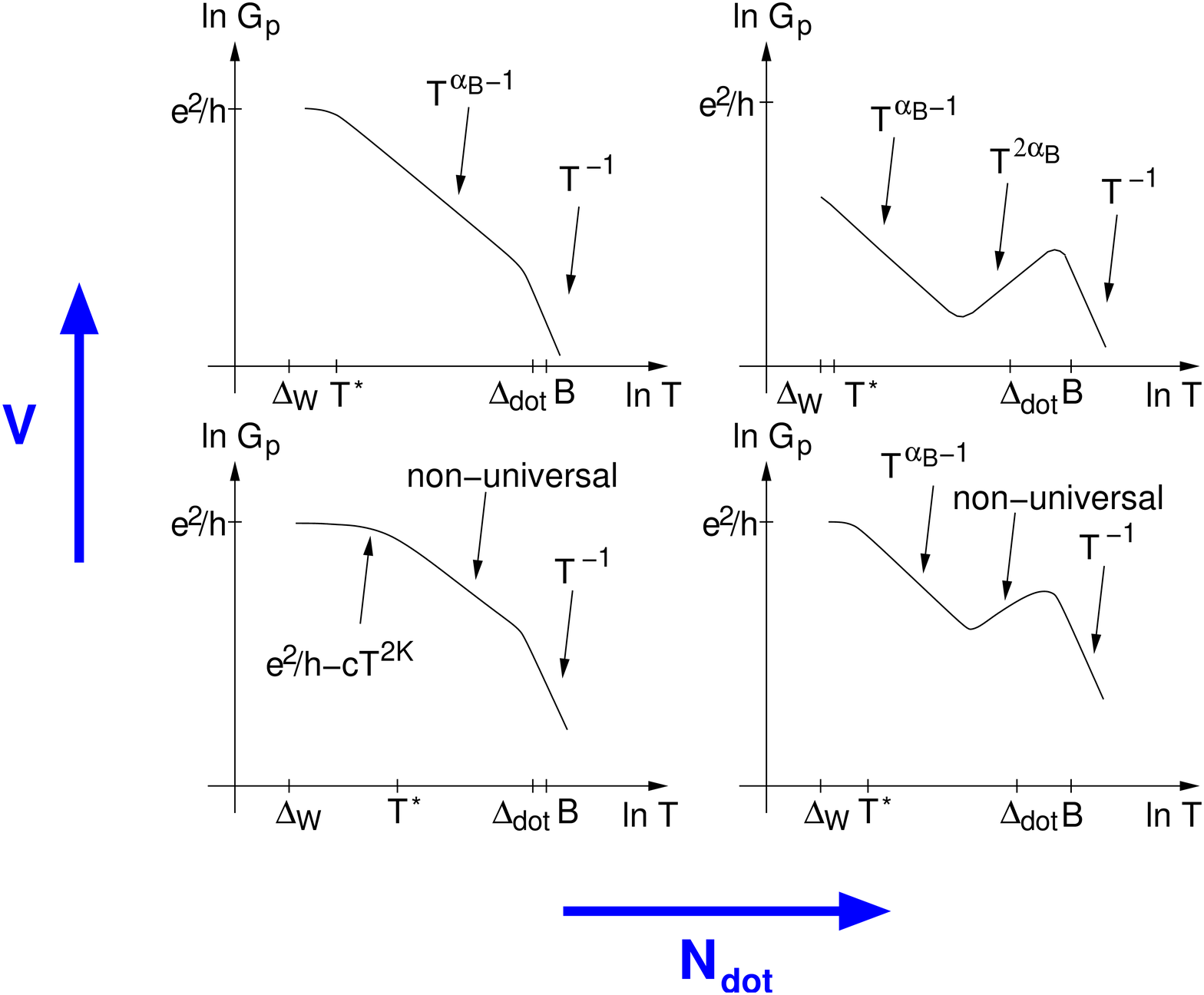}{fig:skizzereso}{Schematic plot of the
  different regimes for the scaling of the conductance $G_\p(T)$ at
  resonance for symmetric barriers found using our method.  The upper
  panels have higher barriers than the lower ones; the right panels
  have larger dot sizes than the left ones.  $\Delta_W=\pi\vf/N$ is
  the energy scale of the interacting wire.}
For resonant tunneling, depending on the parameters of the quantum dot,
I found several temperature regimes with power-law scaling as well as
non-universal behavior (cf.\ figure \ref{fig:skizzereso})
\cite{Med04,EMABMS04}.  All these temperature regimes are obtained
within the same approximation scheme.  The crossover between the
regimes can be studied in detail.  For parameters for which a
comparison is possible the results agree with the ones obtained in
lowest-order perturbation theory in the barrier height and inverse
barrier height.  I did not find any indications of a \CST\ regime with
the exponent $2\ab-1$ predicted from an approximate Master-equation
approach, and seemingly supported by \QMC\ data.  If it were present,
our method should be able to reveal such a regime since its scaling
exponent differs from the \UST\ exponent already at leading order in
the interaction.



\chapter{Conclusions and outlook}
\label{sec:conc}

In this thesis I investigate the role of symmetries and conservation
laws in the functional renormalization-group formalism, and study
specifically Friedel oscillations and transport in \oneD\ correlated
electron systems.

In the \emph{functional renormalization group} (\fRG) method an energy
cutoff scale is introduced in the bare propagator as a flow parameter
(Chapter~\ref{sec:rg}).  By solving coupled differential flow
equations for the Green functions, the effective behavior on all
energy scales can be computed for a given microscopic model.  Because
the full flow-equation hierarchy can be solved exactly only in a few
special cases, for instance the Luttinger model \cite{Schuetz04}, in
most practical applications one has to truncate the hierarchy by
setting the flow of, say, the three-particle and higher Green
functions to zero.  This approximation is justified perturbatively in
the renormalized interaction or some other small parameter.  For
applications such as the \twoD\ Hubbard model and \oneD\ impurity
problems, different basis sets of Green functions have proven to be
particularly useful.

The presence of continuous \emph{symmetries} in the bare action leads
by the Noether theorem to conservation laws and Ward identities
relating Green and response functions (Chapter~\ref{sec:ward}).  The
solution of the infinite flow-equation hierarchy preserves the
symmetry once the cutoff is removed.  The truncated flow equations
with a momentum cutoff, however, generally violate the Ward
identities.  After discussing previous results from the high-energy
physics literature, I show that if a manifestly gauge-invariant
construction is possible as, for instance, in the temperature-flow
scheme, the Ward identities between Green and response functions can
be satisfied exactly despite truncations.

The related property of \emph{self-consistency} is satisfied by
construction in the conserving approximations \cite{BK61}, and it
would be desirable if it were also satisfied in truncated \fRG\ flows.
However, I show that the commonly used truncations generally violate
self-consistency.  For special reduced models it has been shown that
truncated flow equations can be modified to satisfy self-consistency
and even yield the exact mean-field solution \cite{Kat04,SHML04}, so
the hope remains that this may be generalized to arbitrary
interaction.  However, in the \oneD\ lattice model of the Luttinger
liquid, the truncated \fRG\ is surprisingly successful and
self-consistency does not appear to play an important role on the
level of our approximation.

\emph{Correlated electron systems} in one dimension show
\emph{Luttinger-liquid\kern1pt} behavior with a strong interplay
between interaction and impurities.  The \fRG\ method is particularly
apt to compute the single-particle and transport properties of a
lattice model of spinless fermions with one or two impurities at all
energy scales (Chapter~\ref{sec:lutt}).  For systems with up to 1000
sites the results agree well with the numerically exact \DMRG.  For
large systems the \fRG\ reproduces the asymptotic, universal power
laws known from thermodynamic Bethe ansatz and bosonization, as well
as various field-theoretical methods, and in addition indicates the
onset of and crossover between different power-law regimes.

I have concentrated on two \emph{observables}: the spatial density
profile for Friedel oscillations and the conductance.  In the exact
solution the density profile can be obtained from the full propagator;
within our approximation it is much more accurate to instead treat the
density-response vertex as a separately flowing composite operator.
On the other hand, the conductance is completely determined by the
full propagator because current-vertex corrections do not arise in our
approximation in accordance with Ward identities.  The practical
usefulness of the method relies on the fast computation of loop
integrals (appendix~\ref{sec:app:loop}); a little-known mathematical
trick is used to develop a new algorithm linear instead of quadratic
in the system size which allows to treat lattices as large as $10^7$
sites very accurately.

The Friedel oscillations off an impurity or boundary
(Chapter~\ref{sec:results}) obey the characteristic power laws and
show a crossover to the linear-response regime, both for repulsive and
attractive interaction.  The measured exponents agree with the exact
values to linear order in the interaction; however the deviations are
small up to $U\approx 1.5$ because we have incorporated the flow of
the interaction vertex \cite{AEMMSS04,AndThesis}.  A double barrier
features several distinct power laws in the conductance as a function
of temperature \cite{Med04,EMABMS04}.  We reproduce the exponents for
the limiting cases treated in earlier works but in addition observe
non-universal behavior in the intermediate parameter and temperature
regions.  This serves to clarify a contentious issue in the literature
where two different universal scaling laws have been claimed for a
certain parameter range: our data agree qualitatively with the
numerically exact \QMC\ \cite{HE04} but suggest an interpretation in
terms of a non-universal crossover regime.  In conclusion, it is
remarkable that our simple approximation captures on an equal footing
effects which originate from very different physical processes.

As an \emph{outlook} there are several promising extensions of the
current scheme.
$(i)$ The electron spin should be included to make the model more
realistic \cite{AndThesis}.  Preliminary results indicate that for the
typical size of carbon nanotubes ($N \sim 10^4$), the results are even
farther from the asymptotic behavior than in the spinless case, so the
accurate treatment of the non-universal behavior is even more
important.
$(ii)$ X.~Barnab\'e-Th\'eriault studied junctions and rings pierced by
a flux, and connected to several interacting leads
\cite{BSMS04,BSMS05}.  He showed that the interpretation of the
complicated behavior of such systems is greatly simplified if one
considers not the (real) conductance but the (complex) Green function
at the interfaces to the leads as the relevant observable: if plotted
parametrically for different impurity strengths and energy scales, one
arrives at a simple flow diagram in the complex plane.
$(iii)$ The investigation of non-equilibrium phenomena using a Keldysh
variant of the \fRG\ has been started \cite{Jak04}.
$(iv)$ There are interesting Luttinger-liquid properties which appear
only at two-loop order in the interaction, related to inelastic
scattering.  To take them into account one would need to include the
frequency dependence of the interaction vertex or two-loop diagrams
into the flow equation.

The \fRG\ provides a powerful tool to compute properties of \oneD\ 
lattice models where all microscopic parameters can be flexibly
modeled.  It captures the effects on many energy scales, yielding
universal scaling as well as non-universal behavior.


%

\appendix

\chapter{Heat equation}
\label{sec:app:heat}

I hope to develop a better intuition of the formal procedures and
transformations involved in the functional formalism by repeating them
on a simple, well-known example: the heat equation in one dimension.
Given a temperature distribution $u_0(x)$ at initial time $t=0$, the
heat flow governed by the equation
\begin{align}
  \label{eq:heat}
  \partial_t \, u_t(x) & = \partial_x^2\, u_t(x),
  & u_{t=0}(x) & = u_0(x),
\end{align}
determines the temperature distribution $u_t(x)$ at any later time
$t\geq 0$.

In the left column below, we start with the solution of
\eqref{eq:heat} in integral form, a convolution with the Green
function, and transform it in several steps into the differential
form.  At the same time in the right column, we start with the
functional integral definition \eqref{eq:vdef} of the effective
interaction $\V$ because the generating functional $e^{-\V}$ is the
one formally most closely related to the temperature distribution
$u_t(x)$ at fixed time $t=1$:
\begin{align*}
  u_t(x) & = \int_{-\infty}^{\infty} dy\, G_t(x-y)\, u_0(y)
  & e^{-\V[\chi,\bar\chi]} & = e^{\inner{\bar\chi}{Q\chi} -
    \G[\eta=Q\chi,\bar\eta=Q^t\bar\chi]} \\
  & = \int_{-\infty}^{\infty} dy\, \frac{1}{\sqrt{4\pi t}}\,
  e^{-\frac{(x-y)^2}{4t}}\, u_0(y)
  & & = \frac{1}{Z_0} \int [d\psi\bar\psi]\, 
  e^{\inner{[\bar\psi-\bar\chi]}{Q[\psi-\chi]}} \\
  &
  & & \quad \times e^{-V_0[\psi,\bar\psi]} \\
\intertext{measure: $d\mu_t(x) = \mathcal{N}\cdot e^{-\frac{x^2}{4t}}
  dx, \int d\mu_t(x) = 1$}
  & = \int d\mu_t(y-x)\, u_0(y)
  & & = \int \dmux{Q}{\psi-\chi}{\bar\psi-\bar\chi}\,
  e^{-V_0[\psi,\bar\psi]}
\intertext{shift of variables:}
  & = \int d\mu_t(y)\, u_0(y+x)
  & & = \int \dmu{Q}{\psi}\, e^{-V_0[\psi+\chi,\bar\psi+\bar\chi]} \\
  & = u_0(\ds) \int d\mu_t(y)\, e^{(y+x)s} \at_{s=0}
  & & = e^{-V_0[\delta_{\bar\varphi},\delta_\varphi]} \\
  &
  & & \!\!\!\!\! \times \int \dmu{Q}{\psi}\,
  e^{\inner{\bar\varphi}{\psi+\chi}-\inner{\bar\psi+\bar\chi}{\varphi}}
  \at_{\varphi=0} \\
\intertext{completing the square: $\mathcal{N} \int
  e^{-\frac{y^2}{4t}+ys}\, dy = e^{ts^2}$}
  & = u_0(\ds)\, e^{ts^2}\, e^{xs} \at_{s=0}
  & & = e^{-V_0[\delta_{\bar\varphi},\delta_\varphi]}\,
  e^{\inner{\bar\varphi}{C\varphi}}\, 
  e^{\inner{\bar\varphi}{\chi}-\inner{\bar\chi}{\varphi}}
  \at_0 \\
  & = u_0(\ds)\, e^{t\partial_x^2}\, e^{xs} \at_{s=0}
  & & = e^{-V_0[\delta_{\bar\varphi},\delta_\varphi]}\,
  e^{\inner{\delta_\chi}{C\delta_{\bar\chi}}}\, 
  e^{\inner{\bar\varphi}{\chi}-\inner{\bar\chi}{\varphi}}
  \at_0 \\
  & = e^{t\Delta}\, u_0(x)
  & & = e^{\Laplace{C}}\, e^{-V_0[\chi,\bar\chi]} \;.
\end{align*}
The solution of the heat equation is most conveniently formulated in
Fourier space where the Laplacian $\Delta=-k^2$ is diagonal.  However,
on the field-theory side perturbation theory in the interaction $V_0$
leads to an expansion of $e^{-V_0}$ in powers of $\chi$, which
corresponds to an expansion of $u_0(x)$ in powers of position $x$, not
momentum $k$.  Therefore, we will try to solve the heat equation in
real space to make the analogy clearer.  On the way we shall obtain a
generalization of Hermite polynomials as basis functions and see how
they are obtained from a generating function, the \oneD\ analog of
generating functionals.  In the left column, we will first expand
$u_0(x)$ in powers of $x$ and then apply $e^{t\Delta}$ to it; in the
right column, we proceed in the opposite order, first applying
$e^{t\Delta}$ and then expanding in powers of $x$:
\begin{align}
  u_t(x) & = e^{t\Delta}\, u_0(x) \\
  & = e^{t\Delta} \Big[ u_0(\ds)\, e^{xs} \Big]_{s=0} & \quad
  & = \Big[ u_0(\ds)\, e^{ts^2+xs} \Big]_{s=0} \\
  & = e^{t\Delta} \sum_{k=0}^\infty \tfrac{u_0^{(k)}(0)}{k!}
  \Big[ \ds^k\, e^{xs} \Big]_0 &
  & = \sum_{k=0}^\infty \tfrac{u_0^{(k)}(0)}{k!}
  \Big[ \ds^k\, e^{ts^2+xs} \Big]_0 \\
  & = \sum_{k=0}^\infty \tfrac{u_0^{(k)}(0)}{k!}\, e^{t\Delta}\, x^k &
  & = \sum_{k=0}^\infty \tfrac{u_0^{(k)}(0)}{k!}\, H_k^t(x).
\end{align}
In the last line, we have used ``rescaled Hermite polynomials''
$H_k^t(x)$ defined by
\begin{align}
  H_k^t(x) & = e^{t\partial_x^2}\, x^k,
  & e^{ts^2+xs} & = \sum_k \frac{1}{k!} H_k^t(x)\, s^k
\intertext{which are related to (usual) Hermite polynomials $H_k(x)$,}
  H_k(x) & = e^{-\partial_x^2}\, (2x)^k,
  & e^{-s^2+2xs} & = \sum_k \frac{1}{k!} H_k(x)\, s^k
\end{align}
by
\begin{align}
  H_k^t(x) & = (-t)^{k/2} H_k(x/\sqrt{-4t}).
\end{align}
Thus, once we have expanded the initial condition into modes
$u_0^{(k)}(0)$, they evolve in time independently of each other
according to the $t$ dependence of $H_k^t(x)$.  As there are
infinitely many modes in $u_0(x) \sim e^{-V_0[\chi,\bar\chi]}$,
however, it turns out that other parametrizations are more efficient
in our applications.


\chapter{Efficient computation of tridiagonal loops in O(N)}
\label{sec:app:loop}


\section{Propagator}
\label{sec:app:loop:prop}

For the frequency cutoff at $T=0$ and $\Lambda < \Lambda_0 < \infty$
the flow equation for the self energy \eqref{eq:sigmaflowt0} can be
written as
\begin{align}
  \label{eq:appc:sigmareal}
  \dl \Sigma_{1',1}^\Lambda =
  - \frac{1}{2\pi} \sum_{2,2'} \Gamma_{1',2';1,2}^\Lambda 
  \, 2 \, \Re \left[
    \tilde G_{2,2'}^\Lambda(i\Lambda) \right] \; .
\end{align}
In order to compute its right-hand side, one needs to invert the
tridiagonal matrix
\begin{align}
  \label{eq:appc:tridiag}
  T = G_0^{-1}(i\Lambda) - \Sigma^\Lambda \; ,
\end{align}
where $T$ is complex symmetric (\emph{not} hermitean) with diagonal
elements $a_i := i\Lambda + \mu - \Sigma_{i,i}^\Lambda$,
$i=1,\ldots,N$, and first off-diagonal elements $b_i := t -
\Sigma_{i,i+1}^\Lambda$, $i=1,\ldots,N-1$.  Note that $\Im(a_i) =
\Lambda > 0$ such that $T$ is non-singular and its inverse
well-defined.

The inverse $\tilde G^\Lambda(i\Lambda) = T^{-1}$ is not tridiagonal
but a full matrix which can be computed by standard methods in
$\mathcal O(N^2)$ time.  However, for an interaction that does not
extend beyond nearest neighbors on the lattice, only the tridiagonal
part of $\tilde G$ is required, which can be computed in $\mathcal
O(N)$ time, such that much larger lattices can be treated.  We shall
first explain how this is done and then present the resulting
algorithm that can directly be incorporated into a computer program.

Under certain assumptions (see below), a matrix can be uniquely
factorized into a lower unit triangular matrix $L$, a diagonal matrix
$D$, and an upper unit triangular matrix $U$ (``\abb{LDU}
factorization''): $T=L D U$ \cite{Pre86}.  For a tridiagonal matrix
$T$ the unit triangular matrices $L$ and $U$ are in fact unit
bidiagonal: their matrix elements are unity on the diagonal, and only
the first off-diagonal is nonzero.  Since our $T$ is symmetric we have
$L=U^T$.  Thus we obtain a factorization of the form
\begin{align*}
  T = U^{+T} D^+ U^+
  = \begin{pmatrix}
    1 & & & \\
    U_1^+ & 1 & & \\
    & U_2^+ & 1 & \\
    & & \ddots & \ddots
  \end{pmatrix}
  \begin{pmatrix}
    D_1^+ & & & \\
    & D_2^+ & & \\
    & & D_3^+ & \\
    & & & \ddots
  \end{pmatrix}
  \begin{pmatrix}
    1 & U_1^+ & & \\
    & 1 & U_2^+ & \\
    & & 1 & \ddots \\
    & & & \ddots
  \end{pmatrix}
\end{align*}
where the label ``$+$'' distinguishes this factorization from another
one used below.  The prescription to compute the elements $D_i^+$ and
$U_i^+$ is well known and can be found for example in \cite{Pre86}.
Starting in the upper left corner one proceeds to increasing row and
column numbers until one arrives at the lower right corner of $T$:
\begin{align}
  \label{eq:appc:ldu}
  D^+_1:=a_1, \quad
  U^+_i:=b_i/D^+_i, \quad
  D^+_{i+1}:=a_{i+1}-b_i U^+_i \quad
  (i=1,\ldots,N-1) \; .
\end{align}
This works well since in our case $\Im(D^+_i) \geq \Lambda > 0$, such
that one never divides by zero.

To compute the inverse $\tilde G=T^{-1}$, one could directly calculate
$(U^+)^{-1} (D^+)^{-1} (U^{+T})^{-1}$.  It is however easier and more
accurate to find the inverse by solving the linear system of equations
$T \tilde G = \id$, where $\id$ is the identity matrix, by ``back
substitution''. To be specific, consider the $i^{\mathrm{th}}$ column
vector $\tilde G_{\cdot,i}$ of $\tilde G$:
\begin{align}
  \label{eq:appc:backsubst}
  e_i = T \tilde G_{\cdot,i} = U^{+T} (D^+ U^+ \tilde G_{\cdot,i})
  = U^{+T} g_i,\quad
  U^+ \tilde G_{\cdot,i} = (D^+)^{-1} g_i
\end{align}
where $e_i$ is the $i^{\text{th}}$ unit vector.  The first step is to
solve the linear system $U^{+T} g_i = e_i$ for $g_i$, and the second
step to solve $U^+ \tilde G_{\cdot,i} = (D^+)^{-1} g_i$ for $\tilde
G_{\cdot,i}$.  To solve a tridiagonal linear system for one vector
takes $\mathcal O(N)$ time, so solving for the full inverse matrix
$\tilde G$ takes $\mathcal O(N^2)$ time.

Now we shall derive an algorithm to compute the elements of $g_i$ and
$\tilde G_{\cdot,i}$.  Begin with the last column $i=N$: $U^{+T}
g_N=e_N$ can be solved from the first to the last row and gives
$g_N=e_N$.  Next $U^+ \tilde G_{\cdot,N} = (D^+)^{-1} e_N$ can be
solved starting from the last row, $\tilde G_{N,N} = 1/D_N^+$.  From
there one can work upwards by back substitution, $\tilde G_{j,N} =
-U_j^+ \tilde G_{j+1,N}$ ($j=1,\dotsc,N-1$).  For the other columns
$i<N$, one cannot take the shortcut and has to solve both linear
systems for $g_i$ and $\tilde G_{\cdot,i}$.  But it is now important
to realize that for any column vector $\tilde G_{\cdot,i+1}$, if we
somehow know the diagonal element $\tilde G_{i+1,i+1}$, the next
element above the diagonal is
\begin{align}
  \label{eq:appc:fromdiag}
  \tilde G_{i,i+1} = -U_i^+ \tilde G_{i+1,i+1} \quad 
  (i=1,\ldots,N-1) \; .
\end{align}
Thus, we have a prescription how to go up one row in $\tilde G$.
Together with the symmetry of $\tilde G$, \ie, $\tilde G_{i,i+1} =
\tilde G_{i+1,i}$, which follows from the symmetry of $T$, we get the
first off-diagonal element one column to the left \emph{without}
solving the two linear systems in \eqref{eq:appc:backsubst}. Hence, it
is possible to compute directly the tridiagonal part of the inverse.
However, there is another algorithm which is much more accurate for
near-singular matrices at the end of the \RG\ flow: the double
factorization \cite{Meu92}.  It does not rely on the symmetry of
$\tilde G$ but uses the complementary ``\abb{UDL}'' factorization
\begin{align}
  \label{eq:appc:udlfactor}
  T = U^- D^- L^- = U^- D^- U^{-T} \; ,
\end{align}
where the matrix elements are obtained as
\begin{align}
  \label{eq:appc:udl}
  D^-_N:=a_N, \quad
  U^-_i:=b_i/D^-_{i+1}, \quad
  D^-_i:=a_i-b_i U^-_i \quad
  (i=N-1,\ldots,1) \; .
\end{align}
We proceed as for the \abb{LDU} factorization above and get
\begin{align}
  \label{eq:appc:backsubst0}
  \tilde G_{1,1} & = 1/D^-_1 \\
  \label{eq:appc:todiag}
  \tilde G_{i,i+1} & = -U^-_i \, \tilde G_{i,i} \; .
\end{align}
We can combine equations \eqref{eq:appc:fromdiag} and
\eqref{eq:appc:todiag} to relate consecutive diagonal elements:
\begin{align}
  \label{eq:appc:backsubst2}
  \tilde G_{i+1,i+1} = -\tilde G_{i,i+1}/U^+_i =
  \tilde G_{i,i}\, U^-_i/U^+_i =
  \tilde G_{i,i}\, D^+_i/D^-_{i+1} \; .
\end{align}
Thus, we start with \eqref{eq:appc:backsubst0} and use the $U^-D^-L^-$
decomposition to go one matrix element to the right in the inverse
matrix, from the diagonal to the first off-diagonal
\eqref{eq:appc:todiag}, while the $L^+D^+U^+$ decomposition allows to
go down by one, back to the next diagonal element
\eqref{eq:appc:fromdiag}. There is no need to compute the full inverse
matrix.

One can implement the algorithm without knowing the derivation by
using equations \eqref{eq:appc:ldu} and
\eqref{eq:appc:udl}--\eqref{eq:appc:backsubst2}. One can further
eliminate the $U$'s using equations \eqref{eq:appc:ldu} and
\eqref{eq:appc:udl} and implement the algorithm such that only the
input vectors $a_i$, $b_i$ and the output vectors $\tilde G_{i,i}$,
$\tilde G_{i,i+1}$ enter the temporary storage.  This double
factorization is numerically accurate to more than 10 significant
digits (using double precision) even for large lattices ($10^6$ sites)
and almost singular matrices with $\abs{a_i} \sim 10^{-15}$ which
appear at the end of the flow for half filling.


\section{Bubble}
\label{sec:app:loop:bubble}

Even the \RHS\ of the flow equations \eqref{eq:Rflow} for the
density-response vertex $R^\Lambda$ and the flow of the self energy
for $T>0$ in equation \eqref{eq:sigmaflowtg0} can be computed in
$\mathcal O(N)$ time \cite{Enss04a}.  In both cases the \RHS\ of the
flow has the form of a trace of a bubble with two vertices and two
propagators,
\begin{align}
  \label{eq:bubbletrace}
  \tr(U G_1 T G_2) = \tr(U M)
\end{align}
where we have defined the product
\begin{align}
  M = G_1 T G_2
\end{align}
with $U$, $T$ tridiagonal matrices (``vertices'') and $G_1$,
$G_2$ inverse tridiagonal matrices (``propagators'').  In order to
compute the trace in the end and also as a useful intermediate result
in the flow equation, we need the tridiagonal part of $M$, \ie, 
$M_{ii}$, $M_{i,i+1}$ and $M_{i+1,i}$.

According to \cite{Meu92} the inverse of a tridiagonal matrix
generally has the following structure: the upper triangle is spanned
by two vectors $x_i$, $y_i$, while the lower triangle is spanned by
two vectors $u_i$, $v_i$:
\begin{align}
  G_{1;ik} & = 
  \begin{cases}
    u_{1i} v_{1k} & i\geq k \\
    x_{1k} y_{1i} & i\leq k \\
    \Delta_{1i}   & i=k
  \end{cases}
  & G_{2;kj} & = 
  \begin{cases}
    u_{2k} v_{2j} & k\geq j \quad \text{(lower triangle)} \\
    x_{2j} y_{2k} & k\leq j \quad \text{(upper triangle)} \\
    \Delta_{2k}   & k=j     \quad \text{(diagonal)}
  \end{cases}
\end{align}
On the diagonal, the upper and lower triangles agree, and we define an
abbreviation $\Delta_i:=x_iy_i=u_iv_i$ for it.  The four vectors have
the following interpretation: $u_i$ (multiplied by $v_1$) is the first
column, $v_i$ (multiplied by $u_N$) the last row; $x_i$ (multiplied by
$y_1$) the first row, and $y_i$ (multiplied by $x_N$) the last column.
Each of these vectors can be computed via the $L^+D^+U^+$ and
$U^-D^-L^-$ decompositions which relate one row or column to the next
or previous:
\begin{align}
  \frac{u_{i+1}}{u_i} & = -L_i^- & \frac{v_i}{v_{i+1}} = -L_i^+ \\
  \frac{x_{i+1}}{x_i} & = -U_i^- & \frac{y_i}{y_{i+1}} = -U_i^+
\end{align}
Consider first the diagonal elements $M_{ii}$, then the off-diagonal
elements $M_{i,i+1}$ and $M_{i+1,i}$ will be slight variations of it.
The tridiagonal matrix $T=(a,b,c)$ has diagonal elements $a_i$, above
the diagonal $b_i$ and below the diagonal $c_i$:
\begin{align*}
  M_{ii}
  & = \sum_k \left[ G_{1;ik} a_k G_{2;ki} + G_{1;ik} b_k G_{2;k+1,i}
    + G_{1;i,k+1} c_k G_{2;ki} \right] \\
  & = \Bigl\{ \sum_{k\leq i} u_{1i} v_{1k} a_k y_{2k} x_{2i}
    - \Delta_{1i} a_i \Delta_{2i}
    + \sum_{k\geq i} y_{1i} x_{1k} a_k u_{2k} v_{2i} \Bigr\} \\
  & + \Bigl\{ \sum_{k<i} u_{1i} v_{1k} b_k y_{2,k+1} x_{2i} 
    \qquad \qquad
    + \sum_{k\geq i} y_{1i} x_{1k} b_k u_{2,k+1} v_{2i} \Bigr\} \\
  & + \Bigl\{ \sum_{k<i} u_{1i} v_{1,k+1} c_k y_{2k} x_{2i}
    \qquad \qquad
    + \sum_{k\geq i} y_{1i} x_{1,k+1} c_k u_{2k} v_{2i} \Bigr\} \\
  & = Q_i^+ - \Delta_{1i} a_i \Delta_{2i} + Q_i^-
\end{align*}
where $Q_i^+$ are the terms on the left side inside each curly bracket
(lower $k$ indices) and $Q_i^-$ those on the right side (higher $k$
indices), so up to now we have just split the terms in the $k$ sum
into two groups.  The point of defining the $Q$'s like this is the
following: $Q_{i+1}^+$ can be computed from $Q_i^+$ in $\Ord(1)$ time
and likewise $Q_i^-$ from $Q_{i+1}^-$, so going from $M_{ii}$ to
$M_{i+1,i+1}$ is an $\Ord(1)$ operation.  Thus we can compute the whole
diagonal of $M$ in $\Ord(N)$ time!

The recursion relation for $Q_i^+$ is read off from the above
partition:
\begin{align*}
  Q_i^+ & = u_{1i} x_{2i} \sum_{k<i} \left( v_{1k} a_k y_{2k}
    + v_{1k} b_k y_{2,k+1} + v_{1,k+1} c_k y_{2k} \right)
  + \Delta_{1i} a_i \Delta_{2i} \\
  Q_{i+1}^+ & = u_{1,i+1} x_{2,i+1} \sum_{k<i+1} \left( v_{1k} a_k y_{2k}
    + v_{1k} b_k y_{2,k+1} + v_{1,k+1} c_k y_{2k} \right)
  + \Delta_{1,i+1} a_{i+1} \Delta_{2,i+1} \\
  & = \frac{u_{1,i+1}}{u_{1i}} \frac{x_{2,i+1}}{x_{2i}} Q_i^+
  + u_{1,i+1} v_{1i} b_i y_{2,i+1} x_{2,i+1}
  + u_{1,i+1} v_{1,i+1} c_i y_{2i} x_{2,i+1} \\
  & \quad + \Delta_{1,i+1} a_{i+1} \Delta_{2,i+1} \\
  & = L_{1i}^- Q_i^+ U_{2i}^-
  - L_{1i}^- \Delta_{1i} b_i \Delta_{2,i+1}
  - \Delta_{1,i+1} c_i \Delta_{2i} U_{2i}^-
  + \Delta_{1,i+1} a_{i+1} \Delta_{2,i+1}
\intertext{and likewise for $Q_i^-$:}
  Q_i^- & = y_{1i} v_{2i} \sum_{k\geq i} \left( x_{1k} a_k u_{2k}
    + x_{1k} b_k u_{2,k+1} + x_{1,k+1} c_k u_{2k} \right) \\
  Q_i^- & = U_{1i}^+ Q_{i+1}^- L_{2i}^+
  - \Delta_{1i} b_i \Delta_{2,i+1} L_{2i}^+
  - U_{1i}^+ \Delta_{1,i+1} c_i \Delta_{2i}
  + \Delta_{1i} a_i \Delta_{2i}.
\end{align*}
In the same way the off-diagonal elements $M_{i,i+1}$ and $M_{i+1,i}$
are determined:
\begin{align*}
  M_{i,i+1} & = \sum_k \left[ G_{1;ik} a_k G_{2;k,i+1}
    + G_{1;ik} b_k G_{2;k+1,i+1} + G_{1;i,k+1} c_k G_{2;k,i+1} \right] \\
  & = \Bigl\{ \sum_{k\leq i} u_{1i} v_{1k} a_k y_{2k} x_{2,i+1}
    \qquad \qquad \qquad \qquad \qquad
    + \sum_{k>i} y_{1i} x_{1k} a_k u_{2k} v_{2,i+1} \Bigr\} \\
  & + \Bigl\{ \sum_{k<i} u_{1i} v_{1k} b_k y_{2,k+1} x_{2,i+1}
    + u_{1i} v_{1i} b_i y_{2,i+1} x_{2,i+1}
    + \sum_{k>i} y_{1i} x_{1k} b_k u_{2,k+1} v_{2,i+1} \Bigr\} \\
  & + \Bigl\{ \sum_{k<i} u_{1i} v_{1,k+1} c_k y_{2k} x_{2,i+1}
    + u_{1i} v_{1,i+1} c_i y_{2i} x_{2,i+1}
    + \sum_{k>i} y_{1i} x_{1,k+1} c_k u_{2k} v_{2,i+1} \Bigr\} \\
  & = - Q_i^+ U_{2i}^- + (\Delta_{1i} b_i \Delta_{2,i+1})
  + (U_{1i}^+ \Delta_{1,i+1} c_i \Delta_{2i} U_{2i}^-)
  - U_{1i}^+ Q_{i+1}^-
  \intertext{and}
  M_{i+1,i} & = \sum_k \left[ G_{1;i+1,k} a_k G_{2;ki}
    + G_{1;i+1,k} b_k G_{2;k+1,i} + G_{1;i+1,k+1} c_k G_{2;ki} \right] \\
  & = \Bigl\{ \sum_{k\leq i} u_{1,i+1} v_{1k} a_k y_{2k} x_{2i}
    \qquad \qquad \qquad \qquad \qquad
    + \sum_{k>i} y_{1,i+1} x_{1k} a_k u_{2k} v_{2i} \Bigr\} \\
  & + \Bigl\{ \sum_{k<i} u_{1,i+1} v_{1k} b_k y_{2,k+1} x_{2i}
    + u_{1,i+1} v_{1i} b_i y_{2,i+1} x_{2i}
    + \sum_{k>i} y_{1,i+1} x_{1k} b_k u_{2,k+1} v_{2i} \Bigr\} \\
  & + \Bigl\{ \sum_{k<i} u_{1,i+1} v_{1,k+1} c_k y_{2k} x_{2i}
    + u_{1,i+1} v_{1,i+1} c_i y_{2i} x_{2i}
    + \sum_{k>i} y_{1,i+1} x_{1,k+1} c_k u_{2k} v_{2i} \Bigr\} \\
  & = - L_{1i}^- Q_i^+
  + (L_{1i}^- \Delta_{1i} b_i \Delta_{2,i+1} L_{2i}^+)
  + (\Delta_{1,i+1} c_i \Delta_{2i}) - Q_{i+1}^- L_{2i}^+.
\end{align*}

After the full derivation, let us summarize the algorithm.  We observe
that certain combinations of terms appear frequently so we define the
following abbreviations:
\begin{align}
  \label{eq:bubbleabbrev}
  & A_i := \Delta_{1i} a_i \Delta_{2i} \\
  & B_i := \Delta_{1i} b_i \Delta_{2,i+1} \\
  & C_i := \Delta_{1,i+1} c_i \Delta_{2i}.
\end{align}
At the beginning of the algorithm, $Q_i^+$ and $Q_i^-$ need to be
computed via the recursion formulas
\begin{align}
  \label{eq:qplusrecur}
  & Q_{i+1}^+ := L_{1i}^- Q_i^+ U_{2i}^-
  - L_{1i}^- B_i - C_i U_{2i}^- + A_{i+1} \\
  & Q_i^- := U_{1i}^+ Q_{i+1}^- L_{2i}^+
  - B_i L_{2i}^+ - U_{1i}^+ C_i + A_i
\end{align}
with initial conditions
\begin{align}
  & Q_1^+ = A_1 \\
  & Q_N^- = A_N.
\end{align}
Then the tridiagonal components of $M$ can be computed in any order:
\begin{align}
  \label{eq:bubbleresult}
  & M_{i,i} := Q_i^+ - A_i + Q_i^-
  = Q_i^+ - B_i L_{2i}^+ - U_{1i}^+ C_i + U_{1i}^+ Q_{i+1}^- L_{2i}^+ \\
  & M_{i,i+1} := B_i - Q_i^+ U_{2i}^- - U_{1i}^+ Q_{i+1}^- + U_{1i}^+
  C_i U_{2i}^- \\
  & M_{i+1,i} := C_i - L_{1i}^- Q_i^+ - Q_{i+1}^- L_{2i}^+ + L_{1i}^-
  B_i L_{2i}^+.
\end{align}
Finally, the trace \eqref{eq:bubbletrace} is
\begin{align}
  \label{eq:bubbletraceresult}
  & \tr(UM) :=
  \sum_i (U_{ii} M_{ii} + U_{i+1,i} M_{i,i+1} + U_{i,i+1} M_{i+1,i}).
\end{align}


%

\backmatter
\let\cleardoublepage\clearpage
\setsize{1.0}
\setbibpreamble{In the online \abb{PDF} file, simply click on the journal
  references and \texttt{arXiv} numbers!\\
  After each citation appear the page numbers where it is referenced.
  \bigskip}
\bibliographystyle{thesis}
{\raggedright\bibliography{thesis}}


\chapter{Publications}
\label{sec:publ}

Parts of this thesis are contained in the following publications:
\begin{itemize}
\item Enss~T, Meden~V, Andergassen~S, Barnab\'e-Th\'eriault~X,
  Metzner~W, and Sch{\"o}nhammer~K, \textit{Impurity and correlation
    effects on transport in one-dimensional quantum wires}, 
  \href{http://dx.doi.org/10.1103/PhysRevB.71.155401}
  {Phys.\ Rev.~B \textbf{71}, 155401 (2005)},
  \href{http://arXiv.org/abs/cond-mat/0411310}
  {\texttt{cond-mat/0411310}}.
\item Meden~V, Enss~T, Andergassen~S, Metzner~W, and Sch{\"o}nhammer~K,\\
  \textit{Correlation effects on resonant tunneling in one-dimensional
    quantum wires},\\
  \href{http://dx.doi.org/10.1103/PhysRevB.71.041302}
  {Phys.\ Rev.~B \textbf{71}, 041302(R) (2005)},
  \href{http://arXiv.org/abs/cond-mat/0403655}
  {\texttt{cond-mat/0403655}}.
\item Honerkamp~C, Rohe~D, Andergassen~S, and Enss~T,\\
  \textit{Interaction flow method for many-fermion systems},\\
  \href{http://dx.doi.org/10.1103/PhysRevB.70.235115}
  {Phys.\ Rev.~B \textbf{70}, 235115 (2004)},
  \href{http://arXiv.org/abs/cond-mat/0403633}
  {\texttt{cond-mat/0403633}}.
\item Andergassen~S, Enss~T, Meden~V, Metzner~W, Schollw{\"o}ck~U, and
  Sch{\"o}nhammer~K,\\
  \textit{Functional renormalization group for Luttinger liquids with
    impurities},\\
  \href{http://dx.doi.org/10.1103/PhysRevB.70.075102}
  {Phys.\ Rev.~B \textbf{70}, 075102 (2004)},
  \href{http://arXiv.org/abs/cond-mat/0403517}
  {\texttt{cond-mat/0403517}}.
\item Enss~T, \textit{Loops of tridiagonal and inverse tridiagonal
    matrices in O(N)},\\
  in preparation, 2005.
\end{itemize}


\setsize{1.1}

\chapter{Acknowledgments}
\label{sec:ack}

\begingroup
\setlength{\parskip}{1.5ex}
\setlength{\parindent}{0ex}

First of all I am indebted to Walter Metzner for giving me the
opportunity to write my \PhD\ thesis at the Max-Planck-Institut in
Stuttgart.  It has been a great pleasure to work in his theory group.
I wish to thank Walter Metzner for proposing this intriguing and
challenging subject, always having time for discussions, reading my
notes quickly and very carefully, and for giving me the opportunity to
travel to such interesting and far-away places as Bras\'ilia.  I am
grateful to Manfred Salmhofer for many discussions and suggestions on
the more formal and mathematical aspects of the \fRG.  I wish to thank
Siegfried Dietrich for co-examining the thesis.

The close work with Sabine Andergassen on the \oneD\ systems has
been a great pleasure indeed, with discussions on virtually every
detail of the calculations.  I am thankful for a quick proofreading
of the manuscript.

I wish to thank Kurt Sch\"onhammer and Volker Meden in G\"ottingen
for encouragement and for being an inexhaustible fountain of
interesting physical questions and parameter regions to investigate.

Xavier Barnab\'e-Th\'eriault was a great source of motivation, fun
and programming tricks, and had an admirable desire to understand
the physics and not to get caught up in technical details.  He met a
sudden and untimely death by a tragic traffic accident on August 15,
2004.

Daniel Rohe always has new ideas for variations on the scheme and
deserves my thanks for many discussions and quickly proofreading the
whole manuscript.

I wish to thank Julius Rei\ss{} for insightful discussions on the
foundations of different \RG\ schemes, and for organizing the fine
cineastic evenings in the institute.

Carsten Honerkamp proposed several worthwhile applications of our
machinery and always conveys the fun of doing physics.

I am indebted to the computer service group under Armin Burkhardt,
and Daniel Rohe for providing a very fine \abb{IT} infrastructure and
responding very quickly and flexibly to my individual wishes.

I extend my thanks to all the wonderful people in the theory group,
Dmitry Aristov, Heinz Barentzen, Sergej Brener, Luca Dell'Anna,
Martin Feldbacher, Roland Gersch, Karsten Held, Peter Horsch, Andrej
Katanin, Dirk Manske, Matthias Mayr, Hiroyuki Yamase, Yi-Feng Yang,
and Roland Zeyher for support and stimulating discussions about
physics and everything else, and to Mrs Knapp for help in all
organizational matters.

I thank the \abb{ICCMP} in Bras\'ilia and the
Erwin-Schr\"odinger-Institut in Vienna for the hospitable
environment where many discussions on this thesis took place.

Last but not least I am indebted to my wife Carmen for never-ending
encouragement, motivation and typographical expertise.

\endgroup



\chapter{Curriculum Vitae}

\hspace{5mm}
\begin{tabular}[t]{ll}
\parbox[t]{3.6cm}{Name} & Tilman Enss\\
Nationality & German\\
Date of birth & 8 October 1975\\
Place of birth & Bielefeld, Germany\\
Marital status & Married \\
\end{tabular}

\subsubsection*{Studies}
\hspace{5mm}
\begin{tabular}[t]{ll}
\parbox[t]{3.6cm}{02/2002 -- 02/2005} & \PhD\ --- with distinction --- \\
 & Max-Planck-Institut f\"ur Festk\"orperforschung, Stuttgart\\
 & \textit{Renormalization, conservation laws and transport in}\\
 & \textit{correlated electron systems}, with Prof.\ Walter Metzner\\[1.6ex]
10/2000 -- 01/2002 &
Diploma --- with distinction --- \\
 & Ludwig-Maximilians-Universit\"at, Munich\\
 & \textit{Transfer-matrix renormalization group far from the}\\
 & \textit{statistical equilibrium}, with Dr.\ Ulrich Schollw\"ock\\[1.6ex]
10/1999 -- 07/2000 &
studies of physics, Hebrew University Jerusalem, Israel\\
 & Conformal field theory, with Prof.\ Eliezer Rabinovici\\[1.6ex]
10/1998 -- 07/1999 &
studies of physics, Ludwig-Maximilians-Universit\"at, Munich\\[1.6ex]
02/1997 -- 08/1998 &
three one-month research visits, Princeton University\\
 & Magnetic resonance imaging, with Prof.\ Warren S. Warren \\[1.6ex]
05/1998 &
``Vordiplom'' degree in physics, grade 1.0\\[1.6ex]
10/1996 -- 07/1998 &
studies of physics, Friedrich-Schiller-Universit\"at, Jena\\[1.6ex]
07/1995 -- 07/1996 &
civilian service, Diakoniepflegestation Aachen\\[1.6ex]
06/1995 &
Abitur, Kaiser-Karls-Gymnasium Aachen\\
\end{tabular}

\subsubsection*{Scholarships}
\hspace{5mm}
\begin{tabular}[t]{ll}
\parbox[t]{3.6cm}{11/1999 -- 01/2002} &
Studienstiftung des deutschen Volkes\\[1.6ex]
10/1999 -- 06/2000 &
\abb{DAAD} scholarship for Israel\\
\end{tabular}


\end{document}